%% file: main_stripped.tex
\documentclass[aps,physrev,reprint,superscriptaddress,nofootinbib]{revtex4-2}

\usepackage{amsmath, amssymb, bm}
\usepackage{graphicx}%
\usepackage{subcaption}
\usepackage{array,booktabs,multirow}
\usepackage{dcolumn}%
\usepackage{acronym}
\usepackage{xspace}
\usepackage[dvipsnames]{xcolor}
\usepackage[shortlabels]{enumitem}
\usepackage{hyperref}
\usepackage[noabbrev,capitalize]{cleveref}
\usepackage{accents}

\AddToHook{cmd/appendix/before}{\crefalias{section}{appendix}}

\newcolumntype{C}[1]{>{\centering\arraybackslash}m{#1}}
\graphicspath{fig}

\input{commands}
\input{acronyms}

\newcommand{\changed}[1]{{\color{black} #1}}

\begin{document}

\rule{0pt}{0pt} %

\title{A unified harmonic framework for dark siren cosmology}

\author{April Qiu Cheng}
\email{aqcheng@princeton.edu}
\affiliation{Department of Astrophysical Sciences, Princeton University, Princeton, New Jersey 08544, USA}
\affiliation{Max Planck Institute for Gravitational Physics (Albert Einstein Institute), Am Mühlenberg 1, 14476 Potsdam, Germany}

\author{Jonathan Gair}
\affiliation{Max Planck Institute for Gravitational Physics (Albert Einstein Institute), Am Mühlenberg 1, 14476 Potsdam, Germany}

\date{\today}

\begin{abstract}
The galaxy catalog dark siren method aims to infer cosmological parameters from gravitational waves (GWs) without an electromagnetic counterpart by statistically marginalizing over possible host galaxies. 
The cross-correlation of GW sources and galaxies is a promising avenue for cosmological inference without requiring observed host galaxies, by leveraging 2-point statistics.
We provide a detailed guide to the cross-correlation method, clarifying its relationship to standard dark siren techniques as well as the assumptions necessary to be able to use this formalism on GW data.
We show that a \changed{2x2 point cross-correlation analysis} \changed{extends the} angular part of the galaxy catalog method \changed{transformed to harmonic space, in which the unknown galaxy field is marginalized over rather than treated as fixed and uniform.} Combined with the spectral sirens method, which encodes information from the \ac{GW} rate evolution, mass distribution, and selection effects, one can perform an inference that leverages the joint constraining power of all dark siren methods.
We also present a strategy to rigorously fold GW measurement errors into the likelihood. 
Using this method, we show that with a 2 Einstein Telescope + 1 Cosmic Explorer setup, the GW--galaxy cross-correlation part alone can jointly measure $H_0$ and \Omn to 1\% and 5\% precision with just 2 years of data, demonstrating its potential as a precise and scalable inference technique in the next generation of \ac{GW} and galaxy surveys. This is in contrast with canonical \changed{hierarchical Bayesian} population inference techniques, which are known to scale poorly with the precision and catalog size expected of next-generation \ac{GW} experiments.
\changed{Because the cross-correlation method implicitly requires large-number statistics, a \ac{GW}--galaxy cross-correlation measurement is not possible with current GW catalogs.}
\end{abstract}

\maketitle

\section{Introduction\label{sec:intro}}

The present-day expansion rate of the universe $H_0$ is a key observable of $\Lambda$CDM cosmology. There is currently a large ($\gtrsim 4\sigma$) tension between early-universe and late-universe measurements of $H_0$ with electromagnetic probes \cite{2016A&A...594A..13P,2020A&A...641A...6P,2022ApJ...934L...7R,2019NatAs...3..891V,2021CQGra..38o3001D}, motivating alternate avenues of cosmological inference. 
The observation of \acp{GW} sourced from merging compact objects is one such independent avenue, as one can directly measure their luminosity distance from the waveform amplitude and phase evolution without any need for a distance ladder
\cite{1986Natur.323..310S,2005ApJ...629...15H,2006PhRvD..74f3006D}. Hence, \ac{GW} sources are often dubbed \textit{standard sirens}.

The goal of \ac{GW} cosmology can be cast as an inference problem on the redshifts of the \ac{GW} sources, which allows one to probe the $d_L-z$ relation. The source redshift can be measured directly with the observation of an electromagnetic counterpart and subsequent host galaxy identification  \cite{2017Natur.551...85A,2018Natur.562..545C,2021ApJ...908L...4C}. However, the vast majority of \ac{GW} sources have no electromagnetic counterpart, and a variety of methods have been developed and applied in the literature to use these \textit{dark sirens} for cosmological inference. 
\changed{In this manuscript, we use the term ``dark sirens'' to refer to any \ac{GW} source without an electromagnetic counterpart for precise localization; dark siren methods then refer to the broad class of methods for extracting redshift information from the population without host galaxy identifications.}

\changed{One} method of this inference is known as the \textit{spectral sirens method}, which leverages features in the \ac{GW} mass distribution to infer the redshifts of the \ac{GW} sources \cite{1993AAS...183.6705F,2012PhRvD..86b3502T,2022PhRvL.129f1102E,2025ApJ...980...85M,2021PhRvD.104f2009M,2025PhRvD.111j3012M,2022JCAP...09..012L,2012PhRvD..85b3535T,2025ApJ...978..153F,2026A&A...709A.197T}.
One can also use a galaxy catalog to statistically marginalize over possible host galaxies on an event-by-event basis, accounting for the catalog completeness \cite{2012PhRvD..86d3011D,2019ApJ...871L..13F,2023PhRvD.108d2002M,2023JCAP...12..023G,2023AJ....166...22G,2020PhRvD.101l2001G,2023RNAAS...7..250B,2024MNRAS.528.3249A,2023ApJ...943...56P,2020ApJ...900L..33P,2019ApJ...876L...7S,2023ApJ...949...76A,2025arXiv250904348T,2024ApJ...964..191B}. 
\changed{While initially viewed as distinct methods, it is now understood that the galaxy catalog method ought to be carried out as an  extension of the spectral sirens method, in which the spatial population prior is upgraded to one informed by the observed galaxy field \cite{2021PhRvD.104f2009M,2023JCAP...12..023G,2023PhRvD.108d2002M}}.
Both of these methods fall within the standard hierarchical Bayesian formalism for inferring population properties of \acp{GW}, in which one infers the hyperparameters of an assumed population model by marginalizing over the possible source parameters of the observed strain data of each event \cite{2019MNRAS.486.1086M,2022hgwa.bookE..45V,2019PASA...36...10T}.

One of the primary limitations of the galaxy catalog method is the incompleteness of galaxy catalogs at moderate to high redshifts \cite{2026A&A...709A.197T,2020PhRvD.101l2001G,2025PASA...42..149C,2025ApJ...979....9H,2025A&A...698A.128P}. Because the galaxy catalog method typically assumes a uniform distribution for the unobserved galaxies, little to no information is gained on the redshift of the \ac{GW} source at these distances. 
This ignores the clustering of galaxies, as in principle the observed galaxy field can be used to inform the field of missing galaxies. 
On the other hand, techniques which directly infer cosmology from the cross-correlation statistics of \acp{GW} and galaxies have been proposed as a complementary method to dark siren techniques. These \textit{cross-correlation methods} generally aim to constrain cosmology by cross-correlating \acp{GW} in luminosity distance space with galaxies in redshift space, most straightforwardly via the 3x2pt analysis of the \ac{GW}--\ac{GW}, \ac{GW}--galaxy, and galaxy--galaxy power spectra \cite{2016PhRvD..93h3511O,2025arXiv250410482P,2026PhRvD.113j3532P}. Alternate methods have also been proposed, including \changed{cross-correlations with} weak gravitational lensing \changed{\cite{2018PhRvD..98h3524O,2024JCAP...10..074B}}, inference with the \ac{GW}--galaxy power spectra conditioned on the observed galaxy overdensity field \changed{\cite{2021PhRvD.103d3520M,2022MNRAS.511.2782C,2024ApJ...975..189M,2020MNRAS.494.1956M}}, and using 2-point correlations to predict local galaxy overdensities near observed \acp{GW} \cite{2024IAUGA..32P1303G}. 
There is also significant interest in the \ac{GW} clustering itself, both to inform the \ac{LSS} of the universe as well as to infer the \ac{GW} tracer bias, which could provide insights on the host environments and formation origins of \ac{GW} sources \cite{2016PhRvD..94b4013N,2018arXiv181011915Z,2023PhRvD.108j3017V,2021JCAP...02..035L,2023ApJ...959...35K,2025JCAP...04..056D}. \changed{\ac{GW} clustering alone can also constrain cosmology with \ac{3G} detectors, as first proposed by ref. \cite{2016PhRvL.116l1302N}.}

Because the cross-correlation method uses the estimated cross-correlation of the \ac{GW} field as the observable, they bypass the standard hierarchical Bayesian formalism for \ac{GW} population analysis, obscuring the question of how selection effects and individual event localizations should be rigorously treated. 
In particular, it is not clear how to construct the \ac{GW} overdensity field in the first place when the \ac{GW} sky localizations are not directly measured, but rather are a Bayesian posterior obtained from the measured strain data\changed{, unlike electromagnetic observations}.
Most forecasts of the cross-correlation method use a Fisher matrix formalism \cite{2016PhRvD..93h3511O,2024JCAP...10..074B,2024OJAp....7E.110G,2025MNRAS.537.1912Z,2025arXiv250410482P,2026PhRvD.113j3532P}, bypassing any explicit form of a hierarchical likelihood on the strain data. 
Cross-correlation analyses that do perform Bayesian inference on real or mock \ac{GW} catalog typically use the \ac{GW} posterior localization ellipses themselves to construct the overdensity field \cite{2022MNRAS.511.2782C,2024ApJ...975..189M,2025JCAP...04..008F}. 
This localization uncertainty is either ignored by assuming the cross-correlation signal is unaffected at the \changed{angular scales probed in the analysis} \cite{2024ApJ...975..189M,2016PhRvD..93h3511O,2024JCAP...10..074B}, or more commonly by modeling it as a convolution of the map of the true \ac{GW} positions with a single circular Gaussian kernel, which can be neatly folded into the theoretical prediction analytically \cite{2022MNRAS.511.2782C,2025arXiv250410482P,2026PhRvD.113j3532P}.
Given that \acp{GW} have a wide range of localization uncertainties that are often neither isotropic nor Gaussian, there is reason to scrutinize the robustness of these assumptions.
Furthermore, a convolution of the true source position with a Gaussian kernel is an ellipse centered at the true source position, which is a suitable model for the point spread function of a telescope, but not necessarily the posterior measurement of \ac{GW} strain data.

This motivates us to cautiously examine some of the optimistic forecasts of the cross-correlation method in the literature. In at least one case, the cross-correlation method has been used to place tighter constraints on cosmological parameters than the standard galaxy catalog method with real \ac{GW} data \changed{\cite{2024ApJ...975..189M}}. A possible explanation is that the cross-correlation method gains information from 2-point statistics. 
Nonetheless, the cross-correlation method uses a compressed form of the \ac{GW} data as its observable (i.e., the cross-correlation power spectra) and imposes weaker requirements, namely that \ac{GW} sources need only to trace the same underlying \ac{LSS} rather than necessarily being hosted in galaxies. Moreover, the cross-correlation method makes no use of the information from the source-frame mass distribution and rate evolution, which is used by the spectral sirens method.
Further complicating the picture are various techniques in the middle, which extend the galaxy catalog method to use some form of galaxy clustering to help inform the distribution of unobserved galaxies \cite{2024JCAP...12..013L,2026JCAP...01..013L,2024JCAP...02..024D,2026JCAP...01..034D}. 
This motivates a clarification of the relationship between the galaxy catalog and cross-correlation methods.
Given the same two datasets---a \ac{GW} catalog and a galaxy catalog---how does one maximize the constraining power of the inference? What information is lost or gained by the galaxy catalog method with respect to the cross-correlation method, and vice versa?

With these questions in mind, we are motivated to provide a rigorous derivation of the cross-correlation method from first principles, with the simultaneous goal of acting as a pedagogical resource for those newer to 2-point statistical analyses, especially those from the \ac{GW} population inference side. Moreover, we derive \changed{novel} expressions for modeling \ac{GW} localization errors within the machinery of cross-correlation analyses, making explicit our model of the \ac{GW} measurement process and its underlying assumptions. 
Because this paper is written with the aim of bridging cross-correlation and dark siren population analyses, we limit the scope of this paper to Gaussian 2-point statistics in the linear regime. In particular, we ignore peculiar velocities and relativistic effects in this work, including lensing, magnification, \acp{RSD} and \acp{LSD}. We refer the reader to references \cite{2024JCAP...10..087B,2023JCAP...08..050F,2024JCAP...05..095Z,2021JCAP...01..036N,1998ApJ...498L...1S,2011PhRvD..84f3505B} and references therein for the theoretical treatment of these effects.

This paper is thus organized as follows. In \cref{sec:notation}, we summarize for the reader's reference the notation used in this work. In \cref{sec:theory}, we \changed{present a pedagogical review of} the formalisms for dark siren population analyses, \acp{GRF}, and multi-tracer cross-correlations in both Cartesian and spherical coordinates. In \cref{sec:unified}, we describe the relationship between the standard dark siren and cross-correlation methods by outlining a continuous perturbation in the population prior from the former to the latter. In \cref{sec:noise}, we describe how to treat \ac{GW} and galaxy measurement noise, with an emphasis on \ac{GW} angular localizations. We summarize our theoretical results by presenting a unified hierarchical Bayesian posterior in \cref{sec:posterior}; a reader interested in the main theoretical results but not in the details can jump ahead here. Finally, we demonstrate our cross-correlation formalism on simulated catalogs of galaxies and \acp{BBH}, outlining our methodology in \cref{sec:implementation} and presenting the inference results in \cref{sec:results}. We conclude and summarize in \cref{sec:conclusions}.

\changed{While \cref{sec:theory} consists primarily of an overview of existing theory, the results in \cref{sec:unified} and onward are novel and the main results summarized as follows. 1) The angular part of the galaxy catalog population prior is shown to be a special case of the cross-correlation population prior in which the missing %
galaxies are uniformly distributed and uncorrelated with the observed galaxies. Thus, the cross-correlation method generalizes the angular part of the galaxy catalog method up to the binning resolution, insofar as the galaxy and \ac{GW} tracer fields are Gaussian. 2) Previous works on \ac{GW} cross-correlations have borrowed from the \ac{CMB} literature to model localization errors as uniform and isotropic point-spread functions, even though \ac{GW} localization errors arise from an inherently unique noise process and arise from Bayesian \ac{PE}. We propose a new method to fold in localization errors, valid under certain assumptions. 3) The cross-correlation method as it is presented here is not valid in the small catalog limit ($N \lesssim \mathcal{O}(100s)$), and will have poor constraining power until \ac{3G} detectors are operational. 4) A \ac{3G} network composed of two \acp{ET} and one \ac{CE} can achieve percent-level precision in measuring $H_0$ and $\Omn$ with 2 years of data.}

\section{Notation\label{sec:notation}}

\begin{table*}[t]
\centering
\begin{tabular}{@{} c c l @{}}
\toprule[.1em]
 & ~ Symbol ~ & Definition \\
\midrule
\multirow{6}{2cm}{\centering population inference} & $p$, $\mathcal{L}$ & probability distribution, likelihood \\
& $\Pi$ & prior distribution \\
& $\mathcal{P}$ & posterior distribution \\
                         & $\vecd$ & data vector \\
                         & $\veclam$ & event-level parameters \\
                         & $\vecLam$ & population parameters \\
                         
\midrule
\multirow{7}{2cm}{\centering spatial quantities} & $\vecr, \veck$ & 3D position vector, wave-vector \\
                         & $r$ & arbitrary radial distance measure \\ 
                         & $d_L$ & luminosity distance \\
                         & $z$ & redshift \\ 
                         & $\chi$ & radial comoving distance \\
                         & $\vecOm$ & 2D sky position vector \\
                         & $\theta$ & Angular distance, $\cos\theta = \vecOm \cdot \vecOm^\prime$\\
\midrule
\multirow{9}{2cm}{\centering GRFs} & $P(k)$ & 3D power spectrum \\
 & $C_\ell$ & 2D angular power spectrum \\
 & $\tilde{C}_\ell$ & 2D angular power spectrum estimator \\
 & $n^X$ & number density field of tracer $X$ \\
 & $\bar{\rho}^X$ & probability density field of tracer $X$ \\
 & $\rho$ & matter density field \\
 & $\delta$ & 3D overdensity field \\
 & $\Delta$ & 2D overdensity field \\
 & $s_{\{\ell, k\}}$ & (angular, 3D) power spectrum shot noise \\[0.5ex]
 \bottomrule[.1em]
\end{tabular}
\caption{Summary of key symbols and notation used throughout this work.}
\label{tab:symbols}
\end{table*}

In this section, we define for the reader's reference the basic notation used throughout this manuscript; these quantities will be defined in detail in \cref{sec:theory,sec:noise}. \cref{tab:symbols} summarizes the symbols used to describe various quantities. In general, superscripts will be used to denote the tracer, while subscripts will denote the position vector or wavenumber in real or Fourier space. For example, $\delta^g_{\veck}$ denotes the galaxy overdensity field in Fourier space. This is not to be confused with $\delta^{(D)}$ and $\delta^{(K)}$, the Dirac and Kronecker delta functions. 

Power spectra with two superscripts (e.g., $C_\ell^{XY}$) indicates a cross-power spectrum between fields $X$ and $Y$. In this text, we will often use $X$ or $Y$ to denote arbitrary tracers, and substitute in $G$ (gravitational wave source) or $g$ (galaxy) as needed. We use greek letters $\alpha,\beta$ to index bins of any tracer. For brevity and consistency with the notation of \ac{GW} population analyses, individual event-level parameters such as $d_L, z,$ etc. refer to \ac{GW} event parameters, unless otherwise specified by a superscript to belong to a different tracer (e.g., $z^g$ for galaxies). We will do the same for probability distributions, using $p(\veclam |\vecLam)$ to refer to by default the \ac{GW} population prior. We will also use \ac{GW}, \ac{GW} source, and \ac{BBH} somewhat interchangeably to refer to \ac{GW} sources as tracers of \ac{LSS}, as \acp{BBH} make up the vast majority of \ac{GW} sources and their population modeling is currently more robust.

As we will describe in \cref{sec:theory}, \ac{GW} sources can be modeled as a Poisson realization of the galaxies, which in turn are a Poisson realization of the underlying dark matter field.
Thus, to distinguish between probability density fields and Poisson-realized density fields, and between the presence and absence of measurement noise, we introduce the following notation:
\begin{itemize}
    \item An overline denotes the \textit{probability} density field. For example, $\overline{\rho}^X(k)$ denotes the probability density field of tracer $X$, which is proportional to the dark matter density field.
    \item A plain symbol, such as $a_{\ell m}^X$ or $P^X(k)$, indicates the \textit{realized} tracer overdensity field and its angular power spectrum without measurement noise. For the power spectrum $P^X(k)$, this is $\overline{P}^X(k)$ plus a Poisson shot noise term.
    \item A hat ( $\hat{}$ ) indicates a measured quantity. For example, the measured \ac{GW} map $\hat{a}_{\ell m}^G$ can differ from the true \ac{GW} map $a_{\ell m}^G$ of the observations due to localization errors. All tracer overdensity fields (both of the true positions $a_{\ell m}^X$ and the observed positions $\hat{a}_{\ell m}^X$) are constructed from finite observations, i.e. with selection effects implicitly folded in. 
    \item A tilde ( $\tilde{}$ ) denotes an \textit{estimator}. Because we only observe one realization of the universe, it is impossible to access the true power spectrum. We estimate the power spectrum with observations by computing an estimator $\tilde{C}_\ell$, with $\langle \tilde{C}_\ell \rangle = C_\ell$ for an unbiased estimator.
\end{itemize}

Because some quantities such as $C_\ell^{\alpha\beta}$ are symmetric in $\alpha \leftrightarrow \beta$ by construction, we will denote all unique pairings with $[\alpha, \beta]$, which enforces $\alpha \geq \beta$. $\{C_\ell^{[\alpha\beta]}\}$ thus refers to the set of all unique cross-correlation power spectra.

In this work, we will refer to the galaxy catalog method with a heuristic (e.g., uniform) completion of missing galaxies as the ``standard galaxy catalog method", and the full 3x2pt analysis of the power spectra of \acp{GW} and galaxies in radial tomographic bins as the ``cross-correlation method". Although these are the methods we focus on bridging, we will make some references to alternate flavors of galaxy catalog and cross-correlation methods throughout this work.

\section{Theory\label{sec:theory}}

\subsection{The canonical dark siren formalism} \label{subsec:darksiren}

In this section, we will review the standard methods for population inference with \ac{GW} observations without electromagnetic counterparts. For more in-depth reviews, see, e.g., \cite{2024AnP...53600180M,2026enap....5..557P,2023AJ....166...22G}.

Given a set of \changed{$N^G$} \ac{GW} observations $\{\vecd\}$, the Bayesian posterior on a set of \changed{population} parameters $\vecLam$ is
\begin{equation}
    \mathcal{P}(\vecLam | \{\vecd\}, \changed{N^G}) \propto \Pi(\vecLam) \mathcal{L}(\{\vecd\}, \changed{N^G} | \vecLam) 
\end{equation}
Under the presence of selection effects, these probabilities become implicitly conditioned on the fact that the data are detected. If we denote the detection of each observation as $D$, then the posterior can be given in terms of an inhomogeneous Poisson likelihood \cite{2019MNRAS.486.1086M,
2019PASA...36...10T,2022hgwa.bookE..45V,2024ApJ...962..169E,
2025PhRvD.111j3012M}
\begin{equation} 
\begin{split}
    \mathcal{P}(\vecLam | \{\vecd\}, \{D\}, N^G) &\propto \Pi(\vecLam) \mathcal{L}(\{\vecd\} | N^G, \{D\}, \vecLam) p(N^G |  \vecLam)\\
    \propto &~\Pi(\vecLam) \frac{ \mathcal{L}(\{\vecd\} | \vecLam)}{\xi(\vecLam)^{N^G}} e^{-N_\text{exp}(\vecLam)} \left[N_\text{exp}(\vecLam)\right]^{N^G}
\end{split}
\end{equation}
where the fact that the likelihood conditioned on detection $\mathcal{L}(\{\vecd\} | N^G, \{D\}, \vecLam)$ is equal to the unconditioned likelihood up to a renormalization $\mathcal{L}(\vecd_i | \vecLam) / \xi(\vecLam)$ follows from the fact that detection is a property of the observed data only, $P(D_i | \vecd_i) = 1$.
Here, $\xi(\vecLam)$ is the fraction of detectable \changed{mergers marginalized over a given a population model}
\begin{equation}
\changed{ \xi(\vecLam)=\iint_{d \text { detectable }} \di  d \, \di \veclam \,p(d | \veclam) p(\veclam | \vecLam) }
\end{equation}
and $N_\text{exp}(\vecLam) = N_\text{tot}(\vecLam)\xi(\vecLam)$ is the expected number of observed events in the spacetime 4-volume, given the rate of events in the Universe is $N_{\rm tot}(\vecLam)$. Therefore, the Bayesian posterior under selection effects is 
\begin{equation} \label{eq:poisson_ll}
    \mathcal{P}(\vecLam |\{\vecd\}, \{D\}\changed{, N^G} ) \propto \Pi(\vecLam) e^{-N_\text{exp}(\vecLam)} \changed{N_\text{tot}^{N^G}(\vecLam)} \,\mathcal{L}(\{\vecd\}| \vecLam).
\end{equation}
If the parameters of interest $\vecLam$ are population-level parameters that govern a probability distribution \Ppop on the individual event-level parameters $\{\veclam\}$, the likelihood on the data is expressed as a marginalization over $\{\veclam\}$
\begin{equation} \label{eq:standardpop}
    \mathcal{L}(\{\vecd\} | \vecLam) = \prod_{i=1}^{N^G} \int \di \veclam_i \,\Ppop(\veclam_i | \vecLam)\,\mathcal{L}(\vecd_i | \veclam_i, \vecLam).
\end{equation}
Because of this, the population parameters $\vecLam$ are often referred to as ``hyperparameters'', as opposed to the (event-level) parameters $\veclam$.

The goal of \ac{GW} cosmology is to infer cosmological parameters $\vecLam_c$ such as the Hubble parameter \Hn and the present-day matter density parameter \Omn as our population parameters of interest. Unlike electromagnetic standard candles, the luminosity distance to the \ac{GW} source can be inferred directly from the amplitude and phase evolution of the waveform, bypassing any sort of calibrated distance ladder. On the other hand, the redshift is degenerate with the source-frame mass in the waveform, and therefore constraining the astrophysical source-frame mass distribution will also constrain the $d_L-z$ relation, and therefore the cosmology \cite{1986Natur.323..310S,2005ApJ...629...15H}. Thus, the relevant event-level parameters are \changed{the source-frame masses and luminosity distance} $\veclam = \{m_1, m_2, d_L\}$, and the population prior can be most generally written as
\begin{align} \label{eq:GW_Ppop}
    \Ppop(\veclam | \vecLam) = \Ppop(m_1, m_2 | d_L, \vecLam_m, \vecLam_c) \Ppop(d_L | \vecLam_c, \vecLam_r)
\end{align}
\changed{where the mass population model may in general be allowed to be redshift-dependent, and $\vecLam_m$ and $\vecLam_r$ refer to hyperparameters which govern the source-frame mass distribution and rate evolution, respectively.}
$\Ppop(d_L | \vecLam) = \deriv{z}{d_L}(\vecLam_c) \Ppop(z | \vecLam_c, \vecLam_r)$ is usually parameterized in terms of the redshift $z$ as a source-frame rate $\mathcal{R}(z)$ per comoving volume
\begin{equation} \label{eq:GW_rate} 
    \changed{\Ppop(z | \vecLam) \propto \frac{\mathcal{R}(z|\vecLam_r)}{1+z} \deriv{V_c}{z}\big(z|\vecLam_c\big)}.
\end{equation}
The factor of $1+z$ comes from the conversion of the source-frame rate to detector-frame rate.  
Assuming \cref{eq:GW_rate} in \cref{eq:GW_Ppop} is known as the spectral sirens method, and uses no supplementary electromagnetic data in the inference. All methods in dark siren cosmology can be understood as generalizations of this method with the inclusion of additional information. 
\changed{While the galaxy catalog method was initially carried out independently of the spectral sirens method by assuming a fixed population model for the mass spectrum, doing so can lead to significant biases \cite{2021PhRvD.104f2009M} because of correlations between parameters in \ac{PE}. Current implementations of the galaxy catalog method carry out a joint inference of the mass spectrum of the \ac{GW} sources alongside the 3D spatial positions \cite{2023JCAP...12..023G,2023PhRvD.108d2002M,2026arXiv260527227T}.}

Assuming that \acp{GW} are hosted in galaxies, the spatially uniform redshift prior $\deriv{V_c}{z}$ in \cref{eq:GW_rate} can be upgraded to a 3D spatial prior for galaxies $p\gtot(z, \vecOm)$,
\begin{equation} \label{eq:GW_rate_gen} 
    \Ppop(z, \vecOm | \vecLam. \dg) \propto \frac{\mathcal{R}(z|\vecLam_r)}{1+z} p\gtot(z, \vecOm | \vecLam_c, \dg) 
\end{equation}
in which case \cref{eq:GW_rate} is the statement that the redshift prior for galaxies is uniform in comoving volume in the absence of galaxy data. The galaxy catalog method improves the constraining power of \cref{eq:GW_rate} by using a galaxy catalog \dg to inform $p\gtot(z, \vecOm)$. 

If we knew the positions of all the galaxies of the universe, then the redshift prior would be a sum\footnote{A more general model modifies \cref{eq:gcatcomplete} to a weighted sum, rather than assuming that \acp{GW} are drawn uniformly from the galaxies. In practice, this replaces our galaxy field with an effective weighted galaxy field, with an accordingly modified detection probability (since the weight may be a function of e.g., the galaxy luminosity). 
Therefore, without loss of generality, we will proceed in the rest of the paper with the uniform weights in \cref{eq:gcatcomplete}.} of Dirac-delta functions at the galaxy positions
\begin{equation} \label{eq:gcatcomplete}
    p\gtot(z, \vecOm \mid \vecLam_c, \dg) =  \frac{1}{N\gtot} \sum_{i=1}^{N\gtot} \delta^{(D)}(z-z_i)\delta^{(D)}(\vecOm-\vecOm_i)
\end{equation}
or equivalently,
\begin{equation} \label{eq:gcat_basic}
\frac{\di N^G}{\di z \di \vecOm} \propto  \frac{\mathcal{R}(z)}{1+z} \frac{\di N\gtot}{\di z \di \vecOm}.
\end{equation}
Therefore, the general goal is to infer $\frac{\di N\gtot}{\di z\,\di \Omega}$, from which we can infer the \ac{GW} field $\frac{\di N^G}{\di z\,\di \Omega}$ from which the observed \acp{GW} are drawn. 

In practice, galaxy surveys are flux-limited, and therefore the underlying galaxy density field $\frac{\di N\gtot}{\di z\,\di \vecOm}$ usually cannot be recovered. We can rewrite the total underlying galaxy field in \cref{eq:gcatcomplete} as a weighted sum of the observed and unobserved galaxy fields: 
\begin{equation} \label{eq:Pgal}
    \begin{split}
        \frac{1}{N\gtot} \frac{\di N\gtot}{\di z \di \vecOm} = \frac{1}{N\gtot} 
        \Bigg[ \sum_{i=1}^{N\gobs}\, \delta^{(D)}(z-z_i)\delta^{(D)}(\vecOm-\vecOm_i)& \\ 
        + \sum_{i=1}^{N\gmiss} \delta^{(D)}(z-z_i)\delta^{(D)}(\vecOm-\vecOm_i)& \Bigg] \\
        =  f\gobs \frac{1}{N\gobs}\frac{\di N\gobs}{\di z \di \vecOm} + (1 - f\gobs)\frac{1}{N\gmiss}\frac{\di N\gmiss}{\di z \di \vecOm}&
    \end{split}
\end{equation}
where $f\gobs = \frac{N\gobs}{N\gtot}$ is the detectable fraction of galaxies over the integration volume. In Bayesian language, this is a marginalization of whether ($D^g$) or not ($\undertilde{D}^g$) the host galaxy of a \ac{GW} is detected in the catalog \cite{2023JCAP...12..023G,2023PhRvD.108d2002M,2024ApJ...964..191B}:
\begin{equation} \label{eq:Pgal_Bayes}
    \begin{split}
        p\gtot(z, \vecOm \mid \vecLam_c, \dg) = p(D^g | \vecLam_c) \,&\Pgal(z, \vecOm | \vecLam_c, \dg, D^g) + \\
         \left[ 1 - p(D^g | \vecLam_c) \right] \,&\Pgal(z, \vecOm | \vecLam_c, \dg, \undertilde{D}^g)
    \end{split}
\end{equation}
where $p(D^g | \vecLam_c) = \fgdet$ is the host galaxy detection probability and the in-catalog term $P(z, \vecOm \mid \vecLam_c, \dg, D^g)$ is the observed galaxy field \cref{eq:gcatcomplete}. The out-of-catalog term $P(z, \vecOm \mid \vecLam_c, \dg, \undertilde{D}^g)$ is typically taken to be completely uninformative
\begin{equation} \label{eq:homogeneous_comp}
    p(z, \vecOm \mid \vecLam_c, \dg, \undertilde{D}^g) \propto [1 - p(D^g|z, \vecLam_c)] \frac{dV_c}{dz}(z|\vecLam_c).
\end{equation}
where the redshift-dependent prefactor reflects the fact the fact that the galaxy completeness is not uniform over the survey volume.
Alternatively, one may assume that the missing galaxies trace the observed galaxies, such that 
\begin{equation} \label{eq:mult_comp}
\begin{split}
    p(z, \vecOm \mid \vecLam_c, \dg, \undertilde{D}^g) &\propto \frac{1 - p(D^g|z, \vecLam_c)}{p(D^g|z, \vecLam_c)} \times \\
    &\times p(z, \vecOm \mid \vecLam_c, \dg, D^g)
\end{split}
\end{equation}
These two methods, dubbed the ``homogeneous completion" and ``multiplicative completion" methods \cite{2024ApJ...964..191B} represent two extremes of modeling the distribution of the missing galaxies\changed{, neither of which are necessarily realistic.} In either case, the low completeness of galaxy catalogs compared to the sensitivity depth of \ac{GW} observations is a primary bottleneck of the standard galaxy catalog method. This method also requires modeling electromagnetic selection effects for the galaxy detection probability $p(D^g|\vecLam_c)$, which is dependent on the assumed galaxy luminosity function.

\subsection{Gaussian Random Fields} \label{subsec:GRF}

In this section, we will review the basic formalism of modeling the underlying matter field and its Poisson tracers as a \ac{GRF}, beginning with the 3D power spectrum in \cref{subsubsec:3D_GRF} and changing to spherical coordinates in \cref{subsubsec:2D_GRF}.  For simplicity we will assume perfect localizations and complete sky coverage, and leave treating noise for \cref{sec:noise}. Much of the information in this section can be found in standard cosmology texts \cite{1981lssu.book.....P,2020moco.book.....D} .

\subsubsection{The 3D power spectrum} \label{subsubsec:3D_GRF}

It is well-known that the primordial dark matter field which formed the initial seeds of \ac{LSS} is a \ac{GRF}. That is, the random realization of initial dark matter fluctuations can be modeled as a draw from its characteristic power spectrum. In linear perturbation theory, the matter density field at late times can still be considered to be Gaussian on sufficiently large scales, allowing one to write down the probability distribution of an observed field given a power spectrum. 

For a homogeneous and isotropic survey volume $V$ with periodic boundary conditions, the power spectrum of a \ac{GRF} $\delta(\vecr)$ is defined as
\begin{equation} \label{eq:Pk_defn}
    \langle \delta^{*}_{\veck} \,\delta_{\veck^\prime} \rangle \equiv \frac{1}{V} P(k)\,\delta^{(D)}(\veck - \veck^\prime),
\end{equation}
describing the power concentrated across comoving scales $k \equiv |\veck|$.
Here, $\delta_{\vecr}$ denotes the field in real space whereas $\delta_{\veck}$ denotes its Fourier transform; $\delta^{(D)}$ is the Dirac-delta function.

In perturbation theory, the field of interest is the matter overdensity field
\begin{equation}
    \delta^m({\vecr}) \equiv \frac{\rho^m(\vecr) - \langle \rho^m \rangle}{ \langle \rho^m \rangle}.
\end{equation}
where $\rho^m(\vecr)$ is the matter density field and $\langle \rho^m \rangle$ is its expectation value. The underlying matter fields $\rho^m$ and $\delta^m$ are not directly observable; however, bright astrophysical objects such as galaxies, galaxy clusters, and even \ac{GW} sources trace the underlying matter field. These tracers are generally modeled as a biased Poisson realization of the matter field:
\begin{equation} \label{eq:linear_bias_poisson}
    \overline{\delta}^X(\vecr) \equiv \frac{\bar{\rho}^X(\vecr) - \langle \bar{\rho}^X \rangle}{ \langle \bar{\rho}^X \rangle} = b^X\, \delta^m(\vecr).
\end{equation}
where $\bar{\rho}^X(\vecr)$ is now the Poisson probability density field of tracer $X$, and $b^X$ is its bias parameter. Here, we use the overline $\overline{\delta}$ to indicate that this is the overdensity field of the tracer probability (i.e. Poisson rate), whereas a plain $\delta$ denotes the Poisson-realized overdensity field, as we describe below.

The relevant observable for tracers is not the Poisson probability, but rather the realized number counts field $n^X(\vecr)$ and corresponding overdensity field
\begin{equation} \label{eq:3D_overdensity}
    \delta^X({\vecr}) \equiv \frac{n^X(\vecr) - \langle n^X \rangle}{ \langle n^X \rangle}.
\end{equation}
We can write the shot noise as an additive component $s$
\begin{equation}
\begin{split}
    \delta^X(\vecr) &= \overline{\delta}^X(\vecr) + s^X(\vecr) \\
    \delta^X(\veck) &= \overline{\delta}^X(\veck) + s^X(\veck).
\end{split}
\end{equation}
where $s^X(\vecr)$ is defined as the difference between the expected and realized overdensity fields.
Assuming homogeneity and isotropy and substituting the Poisson variance $\langle s^X(\vecr) s^X(\vecr^\prime) \rangle = \frac{\delta^{(D)}(\vecr - \vecr^\prime)}{\langle n^X(\vecr) \rangle}$ and the definition of the power spectrum in \cref{eq:Pk_defn} yields\footnote{Here, we assume that the shot noise is decoupled from the expected tracer field $\overline{\delta}^X(\vecr)$, which is true in the Gaussian limit of Poisson noise ($\langle n^X \rangle \gg 1$). Thus, we can average the shot noise over realizations of the field as we have done above (see e.g. \cite{2010MNRAS.406...60J}).} 
\begin{equation} \label{eq:Pk_auto_with_shot}
    P^X(k) = \overline{P}^X(k) + \frac{1}{\langle n^X \rangle} = (b^X)^2P^m(k) + \frac{1}{\langle n^X \rangle},
\end{equation}
i.e., the power spectrum of the number counts field is the power spectrum of the probability field plus a scale-independent shot noise.
Applying the central limit theorem ($n^X V \gg 1$), the observed number overdensity field $\delta^X$ also approximately follows a Gaussian, with an effective power spectrum of $P^X(k)$. 

Writing down the probability distribution for a \ac{GRF} is simple, especially in Fourier space. 
The Fourier modes are discrete for a finite rectangular survey volume, and homogeneity and isotropy imply that they are independent:
\begin{equation} \label{eq:P_veck}
p(\{ \delta_\veck^X \}) = \prod_{\veck \geq 0} \frac{1}{\sqrt{2\pi \langle |\delta_\veck^X|^2 \rangle}} \exp\left(-\frac{1}{2} \frac{|\delta_\veck^X|^2}{\langle |\delta_\veck^X|^2 \rangle} \right).
\end{equation}
We take the product for modes  $k_{\{x,y,z\}}\geq0$ because the reality condition implies that $\delta_\veck = \delta^*_{-\veck}$, reducing the number of independent modes.
The elements of the diagonal covariance matrix, $\langle |\delta_\veck^X|^2 \rangle$, are given by the 3D power spectrum (up to a volumetric constant dependent on the Fourier convention), as in \cref{eq:Pk_defn}. Note that $\langle \delta^X \rangle = 0$ by the definition of the overdensity field. The power spectrum characterizes the clustering patterns of a tracer.

\cref{eq:P_veck} is straightforwardly generalized to multiple tracers. 
The joint Gaussian distribution on the overdensity fields of two tracers $X$ and $Y$ is
\begin{equation} \label{eq:P_veck_multi}
p(\{ \delta_\veck^X \},\{ \delta_\veck^Y \}) = \prod_{\veck \geq 0}
    \mathcal{N}\left[ \bm{\delta}_\veck; 0,  \covC_\veck \right] 
\end{equation}
where
\begin{equation} 
\mathcal{N}\left[ \bm{\delta}_\veck; 0,  \covC_\veck \right] \equiv 
 \frac{1}{\sqrt{(2\pi)^2 \det\covC_\veck}} \exp\left(-\frac{1}{2} \bm{\delta}_\veck^\dagger \covC_\veck^{-1} \bm{\delta}_{\veck} \right)
\end{equation}
is a complex multivariate Gaussian on the field vector
\begin{equation}
    \bm{\delta}_\veck \equiv 
    \begin{pmatrix}
        \delta_\veck^X \\
        \delta_\veck^Y
    \end{pmatrix}.
\end{equation}
with mean $0$ and covariance matrix
\begin{equation}
    \covC_\veck \equiv 
 \begin{pmatrix}
    \langle |\delta^X_\veck|^2 \rangle & \langle \delta_\veck^{X\,*} \delta_\veck^Y \rangle\\
    \langle \delta_\veck^{X} \delta_\veck^{Y\,*} \rangle & |\langle \delta^Y_\veck|^2 \rangle 
\end{pmatrix} = V 
\begin{pmatrix}
    P^X(k) & P^{XY}(k) \\
    P^{XY}(k) & P^{Y}(k)
\end{pmatrix}.
\end{equation}
This defines the cross-correlation power spectrum $P^{XY}(k)$.

One can similarly derive an expression for the cross shot noise
\begin{equation} \label{eq:ang_cross_shot}
    \left\langle s^X(\vecr)  s^Y (\vecr) \right\rangle = \frac{\langle n^X n^Y \rangle}{\langle n^X \rangle \langle n^Y \rangle}
\end{equation}
where $\langle{n^X n^Y}\rangle$ is equal to the number density of coincident tracers. Thus,
\begin{equation} \label{eq:Pk_cross_with_shot}
    P^{XY}(k) = b^X b^Y P^m(k) + \frac{\langle{n^X n^Y}\rangle}{\langle n^X \rangle \langle n^Y \rangle}
\end{equation}
The shot noise of the cross-power spectrum is typically taken to be $0$, since for two distinct fields the coincident number density is $0$. 
Nonetheless, the cross shot noise plays an important role in relation to the standard galaxy catalog method because it encodes the \ac{GW}--galaxy coincident number density, i.e., how often one expects a \ac{GW} host galaxy to be in the catalog. 
In the limit of galaxy catalog completeness, the cross shot noise is the same as the galaxy shot noise, $1/\langle n^g \rangle$.

Generally, both the power spectrum and bias parameters are functions of $z$, characterizing the growth of \ac{LSS} throughout cosmic history. 
Additionally, the simple constant linear bias in \cref{eq:linear_bias_poisson} only works on linear scales where $|\delta| \ll 1$. On non-linear scales ($\sim 10\Mpc$), the matter density field is highly non-linear with steep overdensities in the form of clusters and halos. 
In the quasi-linear regime, one can continue the perturbation theory series expansion and fit a polynomial bias (e.g., \cite{2007MNRAS.378..852P})
\begin{equation} \label{eq:poly_bias}
    b(k) = b + \sum_{n=1} a_n\left(\frac{k}{k_*}\right)^n.
\end{equation}
for some suitable pivot scale $k_*$.
In the fully non-linear regime, one would need a halo model to describe the 2-point (and higher order) statistics of the matter field \cite{2002PhR...372....1C,2003MNRAS.341.1311S,2012ApJ...761..152T}. 
We do not consider this further in this work.

\subsubsection{The angular power spectrum} \label{subsubsec:2D_GRF}

The 3D Fourier prescription is usually insufficient for modern surveys, which are deep enough to probe the evolution of the power spectrum as a function of redshift and wide enough such that the survey area can no longer be approximated as a Cartesian box.
Many galaxy surveys have turned to a spherical harmonic approach to the problem \cite{1973ApJ...185..413P,1994MNRAS.266..219F,1995MNRAS.275..483H,2002ApJ...572..140D,2012MNRAS.427.1891A}, analyzing the data in tomographic radial slices. This approach has the added benefit of allowing the separate characterization of angular and radial sources of error, which we will discuss in \cref{sec:noise}.
Moreover, it allows us to analyze maps of tracers with different radial distance measures, such as galaxies in redshift space and \acp{GW} in luminosity distance space, without assuming a cosmology \textit{a priori}. 

Let us bin tracer $X$ into $M$ radial bins using radial window functions $w^{X_\alpha}(r)$ for $\alpha = 1, 2, \ldots, M$. The projected angular number counts density field of each bin is
\begin{equation}
    n^{X_\alpha}(\vecOm) = \int\,\di r\,w^{X_\alpha}(r) \frac{\di N^X_\text{obs}}{\di r \, \di \vecOm} 
\end{equation}
where
\begin{equation}
     \frac{\di N^X_\text{obs}}{\di r \, \di \vecOm} = p^X_\text{obs}(r) \frac{\di N^X_\text{obs}}{\di \vecOm} \bigg|_r
\end{equation}
is the \textit{observed} radial distribution of tracer $X$, i.e., with selection effects folded in\footnote{\changed{For \ac{GW} sources, this is implicitly marginalized over the other source parameters, as we bin the tracers radially irrespective of their binary parameters.}}. The 3D overdensity field at an infinitesimal slice at distance $r$ is\footnote{Henceforth,  3D number densities will be referred to explicitly as $n_\text{3D}(\vecr)$, and $n(\vecOm)$ will refer to angular number densities.}
\begin{equation}
\begin{split}
    \delta^X\left(\vecr=(r, \vecOm)\right) &= \frac{n^X_\text{3D}(\vecr)}{\langle n^X_\text{3D}(\vecr) \rangle} - 1 \\
    &= \frac{\frac{1}{\chi(r)^2}\frac{\di N^X_\text{obs}}{\di \chi \di \vecOm} \big|_{\chi(r)}}{\left\langle \frac{1}{\chi(r)^2}\frac{\di N^X_\text{obs}}{\di \chi \di \vecOm} \big|_{\chi(r)} \right\rangle} - 1 \\
    &= \frac{n^X(\vecOm|r)}{\langle n^X(\vecOm|r) \rangle} - 1.
\end{split}
\end{equation}
where $\chi$ is the radial comoving distance.
Thus, if we define the effective window function 
\begin{equation} \label{eq:eff_radial_window}
\phi^{X_\alpha}(r) \equiv \frac{w^{X_\alpha}(r) p^X_\text{obs}(r)}{\int \di r\,w^{X_\alpha}(r) p^X_\text{obs}(r)},
\end{equation}
the 2D-projected overdensity field $\Delta^X(\vecOm)$ can be expressed as a weighted integral of the 3D overdensity field 
\begin{equation} \label{eq:windowfunc}
\Delta^{X_\alpha}(\vecOm) \equiv \frac{n^{X_\alpha}(\vecOm) - \langle n^{X_\alpha} \rangle}{\langle n^{X_\alpha} \rangle} = \int \di r \,\phi^{X_\alpha}(r) \,\delta^X(\vecr).
\end{equation}
Because linear operations preserve Gaussianity, the radially projected 2D density field can also be described by a \ac{GRF}. Many of the results for the 3D power spectrum have analogous angular counterparts, with the \ac{SHT} as the spherical analog of the Fourier transform. 
Let $\alm^{X_\alpha}$ be the \ac{SHT} of $\Delta^{X_\alpha}(\vecOm)$:
\begin{align} 
    \Delta^{X_\alpha}(\vecOm) &= \sum_{\ell=0}^\infty \sum_{m=-\ell}^\ell a_{\ell m}^{X_\alpha} Y_{\ell m}(\vecOm) \label{eq:SHTdef} \\ 
    a_{\ell m}^{X_\alpha} &= \int \di \vecOm\, \Delta^{X_\alpha}(\vecOm) Y_{\ell m}^*(\vecOm). \label{eq:SHTdefback}
\end{align}
Analogously to the 3D power spectrum, the $M$ overdensity fields of tracer $X$ are characterized by their angular power spectra 
\begin{equation}
    C_\ell^{X_\alpha X_\beta} \delta^{(K)}_{\ell \ell^\prime} \delta^{(K)}_{m m^\prime} \equiv \left\langle \left(\alm^{X_\alpha}\right)^* a_{\ell^\prime m^\prime}^{X_\beta} \right\rangle,
\end{equation}
which is diagonal in $\ell$ and $m$ due to the spherical symmetry of the fields; $\delta^{(K)}$ denotes the Kronecker delta function. Note that radial Fourier modes couple different angular shells together. Similarly, the angular cross-power spectrum describes the coupling between the binned density fields of two tracers $X$ and $Y$:
\begin{equation} \label{eq:angcrosspower}
    C_\ell^{X_\alpha Y_\beta} \delta^{(K)}_{\ell \ell^\prime} \delta^{(K)}_{m m^\prime} \equiv \left\langle \left(\alm^{X_\alpha}\right)^* a_{\ell^\prime m^\prime}^{Y_\beta} \right\rangle.
\end{equation}

\cref{eq:windowfunc} expresses the projected angular density field as a weighted integral of the 3D density field over the bin's effective window function. Using this and the definition of the angular power spectrum in \cref{eq:angcrosspower}, one can show that the angular power spectrum is itself an integral of the 3D power spectrum over the window functions.
Specifically, given tracers $X, Y$ with probability density fields characterized by a 3D power spectrum $\overline{P}^{XY}$, the cross-correlation of the angular probability overdensity fields $\overline{C}_\ell^{\alpha\beta}$ between bin $\alpha$ of tracer $X$ and bin $\beta$ of tracer $Y$ is
\begin{equation} \label{eq:limber}
\begin{aligned}
\overline{C}_{\ell}^{\alpha\beta} = \int_0^{\infty} & \frac{\di  \chi}{\chi^2} \, \left(\frac{\di r}{\di \chi}\right) \left(\frac{\di r^\prime}{\di \chi}\right) \, \phi_i^\alpha \left(r(\chi)\right) \, \phi_j^\beta \left(r^\prime(\chi)\right) \\ 
& \times b^X(k, z)\,b^Y(k,z) \,P^m\left(k=\frac{\ell+1 / 2}{\chi}, z(\chi)\right)
\end{aligned}
\end{equation}
where $r$ is the radial coordinate for tracer $X$ and $r^\prime$ is the radial coordinate for tracer $Y$, which are not necessarily the same (as is the case for \acp{GW} and galaxies). 
Recall that the 3D power spectrum of the probability density fields is given by $\overline{P}^{XY}(k,z) = b^X(k,z) \,b^Y(k,z) \,P^m(k,z)$ (where $b$ are the bias parameters and $P^m$ is the matter power spectrum); it has gained a $z$ dependence because working in spherical slices allows us to probe the evolving \ac{LSS} of the universe as we go back farther in $z$.
\cref{eq:limber} is a first-order approximation in $(\ell+1/2)^{-1}$ known as the Limber equation \footnote{The exact formula is a complicated integral of the unequal-time power spectrum $P(k, z(r), z(r^\prime))$ over $k$, $r$, and $r^\prime$. The \textit{Limber approximation} assumes the power spectrum is slowly-varying with respect to the sharply peaked spherical Bessel function $j_\ell(k\chi)$ at $k \chi(r) = k\chi(r^\prime) \approx \ell + 1/2$, which greatly simplifies the expression.} \cite{1953ApJ...117..134L,2008PhRvD..78l3506L}. 
The Limber approximation has its limitations; it breaks down at large angular scales ($\ell \lesssim 20$), and also if the relative bin width $\delta\chi/\chi$ is too small \cite{2007A&A...473..711S}.
Furthermore, while clustering in the radial direction is encoded in the coupling between radial shells, it is a higher-order term that is ignored in the Limber approximation. Nonetheless, it is in principle possible to use the exact integral form, or include higher orders in the approximation \cite{2008PhRvD..78l3506L}.

At this point, we will drop the tracer label for brevity (as the tracer and bin label always appear together) and use labels $\alpha, \beta \in \{1, 2, \ldots, M+M^\prime\}$, such that $1 \leq \alpha, \beta \leq M$ corresponds to the $M$ bins of tracer $X$ and $M+1 \leq \alpha, \beta \leq M+M^\prime$ corresponds to the $M^\prime$ bins of tracer $Y$.

Analogously to \cref{eq:Pk_cross_with_shot}, we can also derive the shot noise for the 2D number counts field. If we model each infinitesimal volume $\di V$ in space as a Poisson point process, then for any given radial bin the number count per pixel $\di \vecOm$ integrated over the bin is also Poisson. 
Thus, the power spectrum of the number counts field can again be written as a sum of the power spectrum of the probability field and a scale-independent shot noise
\begin{align} 
    \alm^\alpha &= \overline{a}_{\ell m} + s_{\ell m} \label{eq:alm_with_shot} \\
    C_\ell^{\alpha\beta} &= \overline{C}_\ell^{\alpha\beta} + \frac{\langle n^{\alpha}n^{\beta} \rangle}{\langle n^{\alpha} \rangle \langle n^{\beta} \rangle}
\end{align}
If the coincident number density is negligible (e.g. different tracers or non-overlapping bins), the shot noise is given by
\begin{equation} \label{eq:Cl_with_shot}
    C_\ell^{\alpha\beta} = \overline{C}_\ell^{\alpha\beta} + \frac{\delta^{(K)}_{\alpha\beta}}{\langle n^{\alpha}\rangle},
\end{equation}
which is its usual form in the literature. 
As before, modeling the shot noise term as Gaussian requires a sufficiently large number of objects expected per radial bin.

To write down the probability distribution for $M$ and $M^\prime$ overdensity fields of tracers $X$ and $Y$, respectively, we define for each $\ell, m$ the vector and corresponding covariance matrix
\begin{align}
    \mathbf{a}_{\ell m} & = 
        \begin{pmatrix}
            \alm^{1}, \alm^{2}, \ldots\alm^{M+M^\prime}
        \end{pmatrix}^\mathrm{T} \\
    \mathbf{\Sigma}_{\ell, \alpha\beta} &= \left\langle \,\mathbf{a}_{\ell m}^*[\alpha] \,\mathbf{a}_{\ell m}[\beta]  \,\right\rangle = C_\ell^{\alpha\beta}. \label{eq:angular_cov}
\end{align}
Thus, the probability of realizing these $M+M^\prime$ overdensity fields given a power spectrum $C_\ell^{\alpha\beta} $ is then a complex multivariate Gaussian diagonal in $\ell$ and $m$:
\begin{equation} \label{eq:Gaussian_alm_ll}
    p(\{ \alm^X \}, \{ \alm^Y \} \mid \{C_\ell^{\alpha \beta}\}) = \prod_{\ell=\ell_{\min}}^{\ell_{\max}} \prod_{m=0}^{\ell} \mathcal{N}\left[ \mathbf{a}_{\ell m}; 0,  \mathbf{\Sigma}_{\ell}\right] \\
\end{equation}
where the covariance matrix $\mathbf{\Sigma}_{\ell}$ is given by \cref{eq:angular_cov}.
Note that $\ell_{\min}$ and $\ell_{\max}$ are typically set by either the Limber approximation or the survey size and resolution, and that the product in \cref{eq:Gaussian_alm_ll} runs from $m=0$ to $m=\ell$ because the reality condition requires $\alm = a_{\ell -m}^*$. 

\subsection{An optimal data compression step} \label{subsec:optimal_compression}

If we now let $X=G$ (\ac{GW} sources, $r=d_L$) and $Y=g$ (galaxies, $r^\prime=z$), then \cref{eq:Gaussian_alm_ll} defines a joint probability distribution for the observed \ac{GW} and galaxy overdensity fields given a theoretical power spectrum. \cref{eq:Gaussian_alm_ll}, however, is not the population prior used in the cross-correlation method. This is because \cref{eq:Gaussian_alm_ll} is effectively a joint distribution of all voxels in the volume, which quickly becomes unwieldy with even modest resolutions; a map with a  $1^\circ$ resolution will have a $ \mathcal{O}(10^4) \times (M+M^\prime)^2$ size covariance matrix \textit{after} factoring in the decoupling of different harmonic modes!

The problem of how to compress the data stored in these maps without losing information on the cosmological parameters of interest was studied extensively for the efficient analysis of high-resolution \ac{CMB} maps. Much of the following discussion is taken from \cite{1997PhRvD..55.5895T}; we review the core pieces here. 

\cite{1997PhRvD..55.5895T} defined the notion of a \textit{lossless compression} as one in which ``any set of cosmological parameters can be measured just as accurately from the compressed data set as from the original data set''. In particular, for some unknown parameters $\vecLam$, the Cramer-Rao inequality sets a lower bound on the error of any unbiased estimator constructed from the data \vecd
\begin{equation} \label{eq:cramer}
    \Delta \Lambda_i(\vecd) \geq (\vecF^{-1})_{ii}^{1/2}
\end{equation}
where \vecF is the Fisher information matrix. Therefore, if the variance of an unbiased estimator $\hat{\vecLam}(\tilde{\vecd})$ constructed from the compressed data $\tilde{\vecd}$ achieves the bound set by the Cramer-Rao inequality on the original dataset $\vecd$, then the compression $\tilde{\vecd}$ can be considered lossless. 

In the problem at hand, our uncompressed data are the maps $\{\Delta^\alpha(\vecOm)\}$, or equivalently, its \ac{SHT} $\{a_{\ell m}^\alpha\}$; the parameters of interest $\vecLam$ are the cosmological parameters $\vecLam_c$. Because all the cosmological information in the Gaussian distribution of the fields in \cref{eq:Gaussian_alm_ll} is encoded in the elements of the covariance matrix (i.e., the angular power spectra $\{C_\ell^{\alpha\beta}\}$), it is sufficient to show that a compression is lossless on the power spectra to show that it is lossless on any cosmological parameters derived from the power spectrum. That is, for this exercise we can effectively take $\{C_\ell^{\alpha\beta}\}$ itself to be the parameters of interest $\vecLam$.

Using \cref{eq:Gaussian_alm_ll}, the inverse Fisher matrix for data $\{a_{\ell m}^\alpha\}$ and parameters $\{C_\ell^{\alpha\beta}\}$ is \cite{2022JCAP...08..073A,2022JCAP...04..013A}
\begin{equation} \label{eq:Cl_covmat}
    \vecF^{-1} = \left\langle C_{\ell}^{\alpha \beta} C_{\ell^\prime}^{\mu \nu}\right\rangle = \frac{\delta^{(K)}_{\ell\ell^\prime}}{2 \ell+1}\left[C_{\ell}^{\alpha \mu} C_{\ell}^{\beta \nu}+C_{\ell}^{\alpha \nu} C_{\ell}^{\beta \mu}\right].
\end{equation}

Next, we want to construct an unbiased estimator for the power spectrum that can saturate this bound. The most straightforward estimator, which can be shown to be minimum-variance with isotropic noise and sky coverage, is constructed by simply averaging over the modes for each $\ell$:
\begin{equation} \label{eq:naive_estimator}
    \tilde{C}_\ell^{\alpha\beta} = \frac{1}{2\ell + 1} \sum_{m=-\ell}^\ell \alm^{\alpha\,*} \alm^\beta 
\end{equation}
Under the assumption of Gaussianity of the field, the error of this estimator can be computed with Wick's theorem and also shown to be equal to \cref{eq:Cl_covmat}, thus saturating the Cramer-Rao bound on the uncompressed data, i.e., the fields themselves. 

Therefore, we can take the power spectrum estimator of \cref{eq:naive_estimator} itself to be our compressed data; $\{\tilde{C}_\ell^{\alpha\beta}\}$ compresses $\{\alm^\alpha\}$ from $\sim (M+M^\prime)\ell_{\max}^2$ numbers to $\sim (M+M^\prime)^2 \ell_{\max}/2$ numbers without losing any information on the angular power spectra $\{C_\ell^{\alpha\beta}\}$\footnote{Of course, this is only a compression if $\ell_{\max} > (M+M^\prime)/2$, which is to say that this compression becomes less efficient if we try to use a large number of bins while having poor resolution (low $\ell_{\max}$).}. This ought to make sense intuitively: due to isotropy, the astrophysically relevant information is encoded in the total power in different angular scales $\ell$ rather than the phase information in each $m$. In the case of \acp{GW} and galaxies, for example, the cross power spectrum of a given $z$-shell and $d_L$-shell is encoded in how similarly the galaxies and \acp{GW} cluster together averaged over the entire sphere, rather than in any specific direction. 
This is analogous to how in standard \ac{GW} population analyses, the product of individual event likelihoods probes $p^g(z, \vecOm)$ at event positions over the entire sky for each $\di z$. 

This motivates the use of the cross-correlation estimator $\{\tilde{C}_\ell^{\alpha\beta}\}$ as our observable, rather than the fields themselves. Because the cross-correlation estimators are symmetric in $\alpha \leftrightarrow \beta$ by construction, we use the notation $\{\tilde{C}_\ell^{[\alpha\beta]}\}$ to enforce $\alpha\geq\beta$, referring to the non-redundant set of cross-correlation power spectra. 
For $\alpha, \beta$ spanning two tracers $X, Y$, this is a ``3x2pt" analysis, referring to the 2-point statistics of 3 sets of power spectra: $C_\ell^{XX}$, $C_\ell^{XY}$, and $C_\ell^{YY}$.

Strictly speaking, $\{\tilde{C}_\ell^{[\alpha\beta]}\}$  is a complex Wishart distributed, since it is the sample covariance of a multivariate-Gaussian distributed variable. However, one can use the central limit theorem to approximate the distribution as Gaussian
\begin{equation} \label{eq:Clhat_Gaussian_ll}
\begin{split}
    p&\left(\left\{\tilde{C}_\ell^{[\alpha\beta]}\right\} \mid \left\{C_\ell^{[\alpha\beta]}\right\}\right) = 
    \prod_{\ell=\ell_{\min}}^{\ell_{\max}}
    \frac{1}{\sqrt{(2\pi)^{r}\det\mathbf{\Sigma}_\ell}} \\& \exp\left[-\frac{1}{2} \sum_{\alpha \leq \beta} \sum_{\mu \leq \nu} \sum_{\ell} \Delta C_{\ell}^{\alpha\beta} (\covS_\ell)^{-1}_{\alpha\beta, \mu\nu} \, \Delta C_{\ell}^{\mu \nu}\right],
\end{split}
\end{equation}
where $r=\frac{(M+M^\prime)(M+M^\prime+1)}{2}$ is the number of power spectra per $\ell$ for $1 \leq \beta \leq \alpha \leq M+M^\prime$, the covariance matrix $\covS_\ell$ was given in \cref{eq:Cl_covmat}, and
\begin{equation}
\Delta C_{\ell}^{\alpha\beta} \equiv \tilde{C}_\ell^{\alpha\beta} - C_\ell^{\alpha\beta}
\end{equation}
is the difference between the power spectrum estimator and our model for the power spectrum. 
Note that the central limit theorem does not work for small $\ell$ \cite{1997ApJ...480...22T}. Imposing $\ell \gtrsim 20$, for example, ensures at least $\sim 40$ independent modes. 
Another caveat is that \cref{eq:naive_estimator} is a cosmic variance estimator, which means it estimates the power spectrum by assuming the different $m$ modes sample independent realizations of the power spectrum; the Gaussianity of \cref{eq:Clhat_Gaussian_ll} also relies on the independence of the different modes via the central limit theorem.
This requires the number of observations per bin $N \gg 2\ell+1$, the number of degrees of freedom per $\ell$; if $N$ is small, then the shot noise is highly non-Gaussian and the variance of the $\tilde{C}_\ell$ no longer satisfies \cref{eq:Cl_covmat}.
Probing scales up to $\ell\sim60$ ($\sim1^\circ$ resolution), for example, would require at least hundreds of events per bin.

In summary, a probability distribution on \acp{GRF} $\{\alm^\alpha\}$ can be compressed into a distribution on the cross-correlation estimator $\{\tilde{C}_\ell^{\alpha\beta}\}$; this compression is both optimal and lossless. The distribution is Gaussian if $\ell$ is not small and, for a Poisson-realized tracer field, $N \gg 2\ell+1$.

\section{Towards a unified population prior} \label{sec:unified}

In order to construct a method that can leverage the combined information from the canonical galaxy catalog and cross-correlation methods, it is useful to first examine qualitatively where each method derives its cosmological constraining power.

The \ac{GW} and galaxy data can be effectively considered as measurements $\{d_L, \vecOm, m_1, m_2 \ldots\}$ and $\{z^g, \vecOm^g, \ldots\}$, respectively, for $N^G$ events and $N^g$ galaxies.
As described in \cref{subsec:darksiren}, dark siren methods probe cosmology by directly measuring the luminosity distance and redshifted masses of the source merger, thus allowing one to probe the $d_L-z$ relation by modeling the source-frame mass and redshift distributions.
The galaxy catalog method uses a 3D spatial prior informed by the positions of observed galaxies 
\begin{equation} \label{eq:gcat_spatial_prior}
\begin{split}
    \Ppop(z, \vecOm | \dg, \vecLam) &= \Ppop(z | \vecLam, \dg) \,\Ppop(\vecOm | z, \dg, \vecLam) \\
    &\propto \frac{\mathcal{R}(z|\vecLam)}{1+z}\deriv{V_c}{z}(\vecLam) \,\Ppop(\vecOm | z, \vecLam, \dg) \\
\end{split}
\end{equation}
where in going from the first line to the second line, we have approximated the distribution of galaxies as uniform in comoving volume averaged over all sky directions. One can in more generality use a $\Ppop(z | \vecLam, \dg)$ informed by the radial distribution of the observed galaxies, but given the incompleteness of galaxy surveys at typical \ac{GW} luminosity distances, the information we get from galaxies after factoring out the angular dependence is negligible.
Thus, \cref{eq:gcat_spatial_prior} splits the spatial distribution into a radial part, encoding information from the vanilla spectral sirens method (or in greater generality the radial part of the galaxy catalog method), and a conditional angular part, which encodes the additional information on field overdensities from a galaxy catalog.

The cross-correlation method, on the other hand, models a population prior on the \ac{GW} and galaxy \textit{fields}, rather than individual events. The cosmological information in the cross-correlation method is contained entirely in the covariance matrix of the Gaussian fields and therefore the angular power spectrum $\overline{C}_\ell^{\alpha\beta}$. If we revisit \cref{eq:limber}, we can see that are three ways in which cosmological parameters can enter $\overline{C}_\ell^{\alpha\beta}$:
\begin{itemize}
    \item The $d_L-z$ relation, as we integrate over radial window functions in two different coordinates $r=d_L$ and $r^\prime=z$ 
    \item The comoving angular diameter distances $\chi$ to each shell, which leverages features of the power spectrum at scales $P^m(k\approx\ell/\chi)$ as standard rulers 
    \item $P^m(k)$ itself, which is sensitive to the cosmology
\end{itemize}
Much of the cross-correlation literature focuses on the $d_L-z$ relation as the source of constraining power, as the cross-correlation matches shells of the \ac{GW} angular density field in $d_L$ space to shells of the galaxy angular density field in $z$-space \cite{2025arXiv250410482P,2025JCAP...04..008F}. However, as noted in \cite{2025arXiv250410482P}, the amplitude and scale of these correlations are coupled to the shape of the power spectrum and therefore also to galaxy ($C_\ell^{gg})$ and \ac{GW} ($C_\ell^{GG}$) auto-correlations. The constraining power of these auto-power spectra are highly nontrivial, as they form the basis of experiments such as DESI \cite{2016arXiv161100036D,2025JCAP...09..008A,2025JCAP...04..012A}, Euclid \cite{2022A&A...662A.112E}, SKA \cite{2016arXiv161100036D}, and LSST \cite{2009arXiv0912.0201L,2019ApJ...873..111I}. 
We will demonstrate an example of this later in \cref{fig:corner}, where the combined 3x2pt analysis of power spectra $C_\ell^{GG}, C_\ell^{Gg}, C_\ell^{gg}$ sets significantly stronger constraints than an analysis of galaxy--galaxy or \ac{GW}--galaxy cross-correlations alone.

Upon comparing the dark siren and cross-correlation methods, we can identify information that is used by one method but not the other. The cross-correlation method does not use any kind of information on the rate evolution of \acp{GW} in $\Ppop(z|\vecLam)$. It is also agnostic to the completeness of the galaxy catalog, which the galaxy catalog method makes use of to predict how often \acp{GW} and galaxies are coincident. In fact, the cross-correlation method does not even require \ac{GW} and galaxies to be coincident at all, only that they trace (with possibly different biases) the same underlying matter field.
On the other hand, canonical dark siren methods do not use any 2-point statistics, effectively assuming a flat power spectrum. Nonetheless, there are similarities: both methods derive information by coupling the luminosity distance of \acp{GW} to the redshift measurements of galaxies.

One can imagine a unified population prior that utilizes both the spectral sirens rate information and the full Gaussian 2-point statistics of the cross-correlation method, from both galaxies and \acp{GW}. The goal of this section is to describe such a method.

\subsection{What is our model conditioned upon?} \label{subsec:independence}

The most obvious difference between a standard \ac{GW} population analysis and the cross-correlation method is that the former models the probability of the positions of individual \ac{GW} events and galaxies, while the latter models the distribution of \ac{GW} and galaxy density fields jointly at all points in space. A standard hierarchical \ac{GW} population prior (i.e., \cref{eq:standardpop}) assumes that \acp{GW} are 
\ac{IID} (implicit in the factorization of the event-level likelihoods), and thus does not allow for any 2-point correlations in the spatial field.

This assumption of independence is a subtle one. Physically, it must be true: whether or not a \ac{BBH} merges in one galaxy must be independent of what is happening in another galaxy. 
Given the underlying galaxy field from which the \ac{GW} sources are drawn, the location of \ac{GW} sources are conditionally independent.
However, if we only have a probabilistic description of the galaxy field, then the observed \acp{GW} are no longer independently distributed because one must marginalize over its possible realizations. 
In any realization of the universe, the positions of galaxies are correlated and its 2-point statistics are given by the matter power spectrum which sources the galaxy probability field.
In other words, the \ac{GW} tracer density field is a \textit{doubly stochastic process}, because the underlying probability density field sourcing the Poisson process (the galaxy field) is itself random, and thus the tracers cannot be considered to be \ac{IID} if the underlying probability density field is not known.

In the case of dark siren cosmology, we do not have information on the missing galaxies. Therefore, one must either condition upon the observed galaxy density field and marginalize over all possible realizations of the missing galaxies according to the power spectrum, or jointly model the observed \acp{GW} and galaxy fields. As we will describe below, the latter method lends itself to the cross-correlation method. 

\subsection{From dark sirens to cross-correlations} \label{subsec:bridge}

We begin by recasting the dark siren galaxy prior $p^g(z, \vecOm)$ in terms of the overdensity field. Consider the angular overdensity field of some tracer $X$ binned in an infinitesimal shell at radius $r$:
\begin{equation}
    1 + \Delta^{X_r}(\vecOm) = \frac{n^{X_r}(\vecOm)}{\langle n^{X_r}\rangle} =  \frac{\di N^X}{\di r\, \di \vecOm} \left[\frac{1}{4\pi} \frac{\di N^X}{\di r}(r)\right]^{-1}
\end{equation}
The conditional angular distribution of some tracer $p^X(\vecOm | r)$ can be written via Bayes' theorem as
\begin{equation} \label{eq:bayes_ang_cond}
\begin{split}
    p^X(\vecOm | r) &= \frac{p(r, \vecOm)}{p(r)} \\
    &= \frac{\di N^X}{\di r \, \di\vecOm} \left(\frac{\di N^X}{\di r}\right)^{-1}
\end{split}
\end{equation}
and hence the angular overdensity field is directly related to the angular distribution $P(\vecOm | r)$:
\begin{equation} \label{eq:delta_POm_relation}
    \frac{1}{4\pi}\left[1+\Delta^X(\vecOm | r)\right] = p^X(\vecOm | r).
\end{equation}
\cref{eq:bayes_ang_cond} also implies from \cref{eq:gcat_basic}
\begin{equation}
    p^G(\vecOm | z) = p\gtot(\vecOm | r)
\end{equation}
or equivalently
\begin{equation}
    \bar{\Delta}^G(\vecOm | z) = \Delta\gtot(\vecOm|z),
\end{equation}
i.e., the \ac{GW} rate evolution $\mathcal{R}(z)$ is normalized out. Note that $\bar{\Delta}^G$ indicates that the \ac{GW} field is itself a stochastic realization of $\Delta\gtot(\vecOm|z)$.

The angular distribution of galaxies at a given redshift $z$ is the weighted sum of the distributions of observed and unobserved galaxies
\begin{equation} \label{eq:pgal_ang}
\begin{split}
    p\gtot(\vecOm|z, \dg) = p(D^g|z,\vecLam_c) \,&p\gobs(\vecOm|z, \dg, \vecLam) + \\
    (1 -p(D^g|z,\vecLam_c)) \, &p\gmiss(\vecOm|z, \dg, \vecLam).
\end{split}
\end{equation}
where \dg is the galaxy data, and $D^g$ ($\undertilde{D}^g$) indicates whether (or not) the galaxy is detected. 
This is equivalent to the galaxy catalog method population prior derived in \cref{eq:Pgal_Bayes}.
The detection coefficient in \cref{eq:pgal_ang} is different than \cref{eq:Pgal_Bayes} because the former conditions on the redshift.

\cref{eq:pgal_ang} can be written in terms of overdensity fields via \cref{eq:delta_POm_relation}
\begin{equation}
\begin{split} \label{eq:pgal_ang_fields}
    \Delta\gtot(\vecOm|z) = \, &p(D^g|z,\vecLam_c)\Delta\gobs(\vecOm|z) + \\
    (1 - \, &p(D^g|z,\vecLam_c)) \Delta\gmiss(\vecOm|z).
\end{split}
\end{equation}

As we will show, the differences between the cross-correlation and standard galaxy catalog method lie in how the missing galaxies $\Delta\gmiss(\vecOm | z)$ are treated, as well as what data we are conditioning on. 
We will now put everything together and describe four different approaches to dealing with the missing galaxies, representing a continuous deformation from the standard galaxy catalog method to the cross-correlation method.

For the rest of this manuscript, we will abbreviate the superscript $g, \text{obs} \to g$, such that a lone $g$ implicitly refers to the observed galaxies.

\subsubsection{The canonical galaxy catalog method}
The simplest way to deal with the missing galaxies is just to ignore their 2-point statistics altogether and take $\Delta\gmiss=0$. It is easy to see from \cref{eq:delta_POm_relation} that this is equivalent to taking the distribution of the missing galaxies to be uniform, such that
\begin{equation} \label{eq:angdist}
\begin{split}
    \Ppop(\vecOm | z, \vecLam, \dg) =  & p(D^g|z,\vecLam_c)p^g(\vecOm|z, \vecLam, \dg) + \\
    &\frac{1 -p(D^g|z,\vecLam_c)}{4\pi} 
\end{split}
\end{equation}
which, once we put back in $p\gtot(z) \propto \di V_c / \di z$, is just the ``homogeneous completion" of missing galaxies in the standard galaxy catalog method.

Alternatively, one can assume that the missing galaxies trace the observed galaxies $\Delta\gmiss=\Delta^g$, which is the ``multiplicative completion" of missing galaxies in the standard galaxy catalog method. 
In either case, one conditions on a fixed distribution as the ground truth for the distribution of missing galaxies, which we by definition have no knowledge of.

\subsubsection{Marginalizing over the missing galaxies}
A correction would be to marginalize over all possible realizations of the missing galaxies, given the observed galaxies and a power spectrum:
\begin{widetext}
\begin{equation} 
\begin{split}
    \Ppop(\vecOm | z, \vecLam, \{\Delta^g\}) \propto p(D^g|\vecLam_c)&\left[1 + \Delta^g(\vecOm| z)\right] \\
    + \left(1 - p(D^g|\vecLam_c)\right) &\bigg[1 + \int \di \{\Delta\gmiss\}\, \Delta\gmiss(\vecOm | z)  p \left(\{\Delta\gmiss\} \mid \{\Delta^g\}, \{z^g\}, \vecLam_b, \vecLam_c \right)\bigg]
\end{split}
\end{equation}
\end{widetext}
Here, the possible angular distributions of the missing galaxies $\{\Delta\gmiss\}$ can be taken to be a vector of values on a discretized set of voxels.
The conditional distribution $p \left(\{\Delta\gmiss\} \mid \{\Delta\gobs\},\ldots \right)$ is a conditional multivariate Gaussian, which can be evaluated from the full multivariate Gaussian in \cref{eq:Gaussian_alm_ll} when transformed to harmonic space. 
The theoretical power spectrum can be computed from the cosmology, bias parameters $\vecLam_b$, and shot noise.

An equivalent approach is to make a change of variables in \cref{eq:pgal_ang_fields} and marginalize over the total galaxy field. %
This approach has been taken by recent works such as \cite{2024ApJ...964..191B,2024JCAP...02..024D,2026JCAP...01..013L}, and can be considered a hybrid approach that directly incorporates large-scale structure to inform the underlying $p\gtot(z, \vecOm)$ in the dark siren method. The downside of this method is that it can be quite computationally expensive; one must marginalize over all possible realizations of the entire galaxy field, and thus the dimensionality of the parameter space is proportional to the survey volume at fixed resolution.

\subsubsection{From coordinates to fields} \label{subsubsec:field_ll}

Gaussian 2-point statistics are much more easily expressed if we write the probability distribution in terms of the \ac{GW} density field $n^G(\vecr)$, rather than the individual event parameters $\{z, \vecOm\}$. 
Besides the voxel resolution, the only information that is lost in this change of variables is the distinguishability of the different objects, which is irrelevant for evaluating the population prior, as the ordering of the product in \cref{eq:standardpop} does not matter. Nonetheless, the labelling of different objects is generally important in \ac{GW} population analyses because one must account for the different measurement errors of different events, which we will revisit in \cref{sec:noise}.

For \acp{GW}, the desired change of variables for thin radial shells $\bar{r}$ and pixels $\bar{\vecOm}$ is
\begin{equation} \label{eq:change_to_alm}
    \{r, \vecOm\} \to \{r\}, \Delta^G(\bar{\vecOm} | \bar{r}) \leftrightarrow \{r\}, \mathbf{a}_{\ell m}^G(\bar{r})
\end{equation}
which factors the information on the 3D distribution of the tracer into a radial distribution $\{r\} \to n^G(\bar{r})$ and a conditional angular distribution $\Delta^G(\bar{\vecOm} | \bar{r})$.
This is sensible, since we showed earlier that the angular distribution $p(\vecOm|r)$ is equivalent to the overdensity field $\bar{\Delta}^G(\vecOm|r)$. 
Note that just as the pixel resolution sets a finite truncation of the field in real space, the pixel resolution also sets a $\ell_{\max} \sim 1/\delta\theta$, such that our data vector is finite.

Under this change of variables, the \ac{GW} population prior $\Ppop(\vecOm|z)$ is replaced by a prior on the overdensity field $\alm^G[\bar{z}]$ at each redshift shell $\bar{z}$
\begin{equation}
    p(\{\vecOm\} | \{z\}, \dg, \vecLam) \to p(\{\alm^G[\bar{d}_L]\}|\{\alm^g[\bar{z}]\},\{d_L\},\{z^g\}, \vecLam)
\end{equation}
which is again a complex multivariate Gaussian given in \cref{eq:Gaussian_alm_ll}. Here, we have changed coordinates from $z$ to $d_L$ for \acp{GW} (binning the \acp{GW} in luminosity distance space); the conditional dependence on the $\{d_L\}$ and galaxy redshifts $\{z^g\}$ remains in order to compute the expected shot noise $1/\langle n^{\{G, g\}}[\bar{r}]\rangle$. 
As is standard in survey cosmology, we approximate $ \langle n^\alpha \rangle$ with the sample mean for each bin $\alpha$, which allows us to compute the overdensity field without a simultaneous inference of the overall rate. This is satisfied if the Poisson fluctuations per bin $1/\sqrt{N^\alpha} \ll 1$, requiring at least $N^\alpha \sim \mathcal{O}(10^2)$ events per bin for a 10\% error.

\changed{Since this change in variables is invertible up to the resolution truncation, a natural validation is to show that the galaxy catalog population prior can be recovered as a special case of this \ac{GW} field population prior $p(\{\alm^G[\bar{d}_L]\}|\{\alm^g[\bar{r}]\},\{d_L\},\{z^g\}, \vecLam)$. We show this explicitly in \cref{app:gcat_field_bridge}.} In particular, the probability distribution on the \ac{GW} field conditioned on the \ac{GW} redshifts and \textit{total} galaxy field recovers $p(\vecOm|z, \vecLam)$ of the standard galaxy catalog method (i.e., \cref{eq:pgal_ang}), as the standard galaxy catalog assumes that the distribution of missing galaxies is known. \changed{The standard ``homogeneous completion" of the galaxy catalog method is given by fixing the overdensity field of the missing galaxies to be identically $0$, i.e. uniformly distributed throughout space.}

\changed{This result may be surprising because the power spectrum does not enter the galaxy catalog method at all. The fact that the galaxy catalog method can be recovered as a special case of the cross-correlation method is because the galaxy catalog method implicitly conditions upon knowledge of the entire galaxy field, including observed and missing galaxies.} In general, for any tracer $X$ that directly traces another tracer $Y$, the probability distribution of the field $\alm^X$ conditioned on the field $\alm^Y$ will have no 2-point correlations. This is as we discussed earlier in \cref{subsec:independence}: we can model the events as independently distributed if and only if the underlying probability distribution is known. 
This does not necessarily imply that the standard galaxy catalog method is incorrect, only that whatever is assumed for the distribution of missing galaxies constitutes an uninformative prior guess.

\subsubsection{A joint GW -- galaxy population prior}

The galaxy data is both noisy (from spectroscopic or photometric redshift errors) and informative on the cosmology (from features in the matter power spectrum). A natural extension is therefore to jointly model the \acp{GW} and galaxies, which was also noted by \cite{2023AJ....166...22G} as the exactly correct approach to accounting for galaxy measurement errors.

Our angular conditional population prior now includes the observed galaxy field
\begin{equation}
    \Ppop(\{ \alm^G[\bar{d}_L] \}, \{ \alm^g[\bar{z}] \} \mid \{d_L\}, \{z^g\}, \vecLam)
\end{equation}
which is now the full joint Gaussian probability distribution in \cref{eq:Gaussian_alm_ll}.

A population prior on the \ac{GW} positions conditioned on the observed galaxy positions asks what cosmology best matches the \acp{GW} to the galaxies; a population prior on the observed galaxies asks what cosmology can best reproduce the observed galaxy clustering patterns, among all possible realizations of the matter density field.
A joint \ac{GW}--galaxy population model adds these two complementary pieces of information together.

\subsubsection{The cross-correlation method}

\begin{figure*}[t]
    \centering 
    \includegraphics[width=\textwidth]{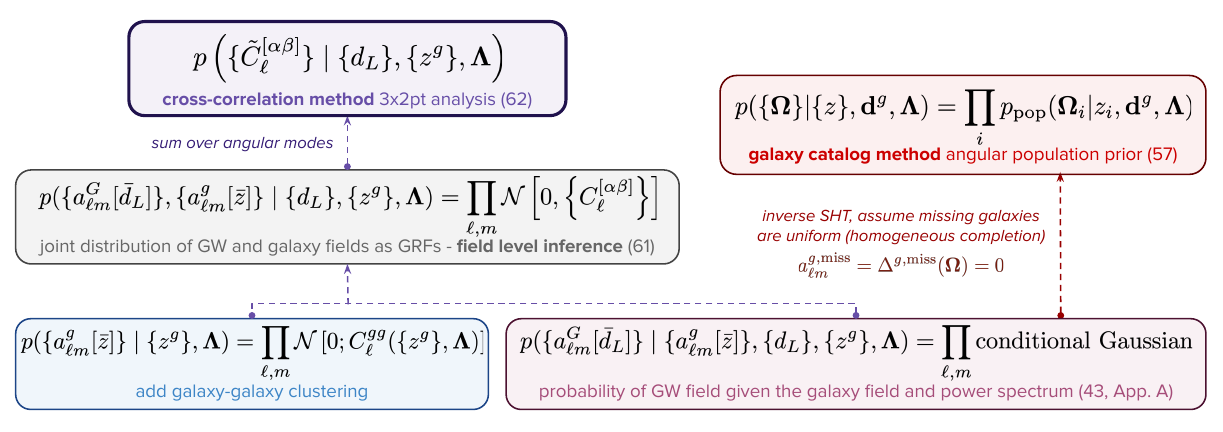}
    \caption{A flowchart summarizing the relationship between the galaxy catalog and cross-correlation methods, outlined in \cref{sec:unified}. The parentheses indicate the corresponding equation in the text.  \label{fig:unified}}
\end{figure*}

The final change of variables is \changed{a compression into 2-point statistics}. Let $\alpha, \beta$ again index the radial shells of both \acp{GW} and galaxies. We reduce the population prior on the full fields $\{ \alm^\alpha \}$ to a prior on the cross-correlation estimator $\{\tilde{C}_\ell^{\alpha\beta}\}$
\begin{equation}
    \Ppop(\{\tilde{C}_\ell^{[\alpha\beta]}\} \mid \{d_L\}, \{z^g\}, \vecLam)
\end{equation}
which is just \cref{eq:Clhat_Gaussian_ll}. \changed{As discussed in \cref{subsec:optimal_compression}, this compression is optimal and not lossy if the fields are Gaussian.}

\changed{We summarize our results up to this point in \cref{fig:unified}, where we show a flowchart visualizing the relationship between the galaxy catalog and cross-correlation methods. The angular part of the galaxy catalog population prior is the real-space analog of a field-level inference conditioned on the galaxy data. Adding galaxy--galaxy clustering and compressing the $\{\alm\}$ into cross-correlation statistics yields the full 3x2pt cross-correlation method.}

\subsubsection{The joint cross-correlation + spectral sirens method}

A probability distribution on angular overdensity fields (or equivalently, their estimated power spectra) is an angular distribution conditioned on the radial distances. Adding back the radial dependence, as well as any other \ac{GW} parameters we are modeling ($\{m_1, m_2, \ldots\}$, yields the full population prior
\begin{equation} \label{eq:Ppop_xcorr_unified}
\begin{split}
    &\Ppop(\{\tilde{C}_\ell^{[\alpha\beta]}\}, \{d_L, m_1, m_2, \ldots\} \mid \{z^g\}, \vecLam) = \\
    & \qquad \left[\prod_{i=1}^{N^G} \Ppop(d_{L, i}, m_{1, i}, m_{2, i}, \ldots | \vecLam)  \right] \\
    & \qquad \qquad\times p(\tilde{C}_\ell^{[\alpha\beta]} \mid \{d_L\}, \{z^g\}, \vecLam)
\end{split}
\end{equation}
where the middle line is just the spectral sirens population model\footnote{Recall from discussion at the beginning of \cref{sec:unified} that we take the redshift distribution of galaxies marginalized over sky direction to be uniform in comoving volume even after conditioning upon the observed galaxies, which is a good enough approximation given galaxy completeness at typical \ac{GW} distances.}.
\cref{eq:Ppop_xcorr_unified} is the unified joint population prior encoding the full joint information \changed{from the galaxy and \ac{GW} catalogs, combining the cross-correlation and spectral siren analyses}.
The fact that it is neatly factorizable into a probability on the cross-correlation statistic and the spectral sirens population prior shows that they use exactly complementary information. This is because the cross-correlation is an analysis on the overdensity fields, which are normalized to the rate at each radial bin, whereas the spectral sirens method uses information on the overall rate evolution of \acp{GW}.
As we discussed above, this decoupling is only possible if the number of events per bin is sufficiently large (at least $\mathcal{O}(10^2)$), in order for the expected shot noise to be well approximated by the sample mean for each bin. Otherwise, the expected cross-correlation would be coupled to an inference on the expected rate for each bin.

\section{Building noise into the likelihood} \label{sec:noise}

Thus far, we have only discussed the population model on the true positions of the \acp{GW} and galaxies. In canonical \ac{GW} population analyses, one writes down the likelihood on the data by marginalizing over all possible values of the true source parameters. This is typically done via Monte-Carlo integration over the posterior $p_{\rm PE}(\veclam|\vecd^G)$ for each \ac{GW} source $\vecd^G$, using samples generated from \ac{PE}, using a fixed prior $\pi_{\rm PE}(\veclam)$. The $i$-th event-level likelihood in \cref{eq:standardpop} would be computed as 
\begin{equation}
\begin{split}
\mathcal{L}_i (\vecLam) &\equiv \int \di \veclam \,\Ppop(\veclam | \vecLam) \mathcal{L}(\vecd^G_i|\veclam) \\
&= \int \di \veclam \,\Ppop(\veclam | \vecLam) \frac{p_{\rm PE}(\veclam | \vecd^G_i)}{\pi_{\text{PE}}(\veclam)} {\cal Z}_{\rm PE}(\vecd^G_i) \\
&\appropto \frac{1}{m}\sum_{k=1}^m \frac{\Ppop(\veclam_i^k|\vecLam)}{\pi_{\text{PE}}(\veclam_i^k)}
\end{split}
\end{equation}
for $m$ \ac{PE} samples per event. Here, ${\cal Z}_{\rm PE}(\vecd^G_i) \equiv \int \di\veclam \, {\cal L}(\vecd^G_i | \veclam) \pi_{\rm PE}(\veclam)$ is the Bayesian evidence for the \ac{PE} prior for each event\changed{, and the symbol $\appropto$ indicates that $\mathcal{L}_i$ is approximately proportional to the last line.}

In principle, one could do the same here and marginalize over all possible true maps $\{a_{\ell m}^G\}$ and therefore all possible $\{\tilde{C}_\ell^{\alpha\beta}\}$ by repeatedly drawing from \ac{PE} samples. Direct Monte Carlo marginalization is impractical, however, for at least three compounding reasons:
\begin{enumerate}
    \item \textbf{Exponential sample complexity.} The population prior on the $\{a_{\ell m}^G\}$ is effectively a joint prior on the positions of all $N^G$ events. Thus, unlike standard population inference, one would have to generate joint samples across all $N^G$ events, and covering this space at fixed resolution requires exponentially many samples with the number of events. In standard population inference, the product over individual event likelihoods
    \begin{equation}
        \mathcal{L} = \prod_{i=1}^{N^G}\mathcal{L}_i \sim \prod_{i=1}^{N^G} \sum_{k=1}^m \frac{\Ppop(\veclam_i^k|\vecLam)}{\pi_{\text{PE}}(\veclam_i^k)}
    \end{equation}
    implicitly explores a grid of $m^{N^G}$ samples in the joint parameter space $\{\veclam_1, \veclam_2, \ldots, \veclam_{N}\}$ for $m$ samples per event. Attempting to keep this resolution would cause the per-likelihood computational cost to grow exponentially with the number of events.

    \item \textbf{Nonlinearity}. The mapping from coordinates $\{\vecr\}$ to $\{\tilde{C}_\ell^{\alpha\beta}\}$ is highly non-linear, which means that \ac{PE} samples do not yield an unbiased estimate of the power spectra. For \acp{GW} with noisy measured positions $\{\hat{\vecr}\}$ and true positions $\{\vecr\}$ satisfying $\mathbb{E}_{\text{noise}}\left[ \hat{\vecr}_i \right] = \vecr_i$, the expectation value of the corresponding measured cross-correlation statistic $ \mathbb{E}_{\text{noise}}\left[ \hat{\tilde{C}}_\ell^{\alpha\beta}\right]$ is not equal to $\tilde{C}_\ell^{\alpha\beta}$, the cross-correlation computed from the true positions $\{\vecr\}$. As we will show in \cref{subsec:ang_loc}, angular localization errors act as a smoothing kernel that suppress power at angular scales on the order of the localization error. As a result, the fraction of \ac{PE} posterior samples consistent with the true $\{\tilde{C}_\ell^{\alpha\beta}\}$ shrinks exponentially with the number of events with localizations worse than $\sim1/\ell$.

    \item \textbf{Per-sample computational cost.} Finally, each sample requires its own evaluation of $\{\tilde{C}_\ell^{\alpha\beta}\}$, a nontrivial computation that includes sorting the \acp{GW} into radial bins, computing the \ac{SHT} of the field in each shell, and computing all auto- and cross-power spectra. Beyond computation time, memory constraints also limit the Monte Carlo sample size: for a combined total of $\sim30$ \ac{GW} and galaxy bins and $\ell$ up to $\sim1^\circ$ resolution, a single realization of $\{\tilde{C}_\ell^{\alpha\beta}\}$ takes $\sim 1$ Mb of storage! This limits us to Monte Carlo integrals with at most $10^4-10^5$ samples over a parameter space that is, as argued above, extremely high dimensional.
\end{enumerate}

Therefore, we take a different approach where we forward model the errors into the theory-level prediction, which are the mean and covariance matrices of the likelihood on the cross-correlation power spectra $\tilde{C}_\ell^{[\alpha\beta]}$. To do so, we need to be able to model the \ac{GW} posterior measurement in an analytically tractable way. This is possible in the high \ac{SNR} limit ($\gtrsim 25$), where posteriors are expected to be Gaussian \cite{2008PhRvD..77d2001V,2023MNRAS.519.2736G}. 

Our approach will be as follows. We will take our \ac{GW} sky location data to be the \ac{MAP} point of each event $\hat{\veclam} = \{\hat{d}_L, \hat{\vecOm}\}$ and their associated measurement errors $\Delta\veclam = \{\Delta d_L, \Delta \vecOm\}$. Assuming flat priors, the high \ac{SNR} limit implies that the posterior and likelihood are Gaussian:
\begin{equation}
\begin{split}
    P(\veclam |\hat{\veclam}, \Delta\veclam) &= \mathcal{N}(\veclam; \hat{\veclam}, \Delta \veclam) \\
    \mathcal{L}(\hat{\veclam}|\veclam, \Delta\veclam)  &= \mathcal{N}(\hat{\veclam}; \veclam, \Delta \veclam)
\end{split}
\end{equation}

We will make the further assumption that the \ac{GW} radial and angular localization errors can be considered to be independent (i.e., the off-diagonal components of the Fisher covariance matrix are small), allowing us to analyze them separately.

\subsection{Angular localization errors} \label{subsec:ang_loc}

The effect of a localization error on the power spectrum of a 2D map is well known from studies of the \ac{CMB}. In particular, if the observed map can be modeled by a convolution of the true map with a kernel that describes the point-spread function of the telescope, then the observed \ac{SHT} coefficients are just a linear combination of the true coefficients from the convolution theorem. For a homogeneous and isotropic kernel, the relationship between the blurred map $d_{\ell m}$ and the true map \alm is simply multiplicative:
\begin{equation} \label{eq:isotropic_beam_mult}
    d_{\ell m} = B_\ell a_{\ell m}.
\end{equation}
We will refer to $B_\ell$ as the angular beam window function. For a circular Gaussian kernel with \ac{RMS} width $\sigma$, one can show that 
\begin{equation} \label{eq:Gaussian_beam}
    B_\ell(\sigma) = e^{-\frac{1}{2}\ell(\ell+1)\sigma^2}
\end{equation}
and hence the convolution has the effect of suppressing the signal for $\ell \gtrsim 1/\sigma$.
This result has been extrapolated to model \ac{GW} localization errors in the literature, often by using the \ac{GW} angular localization posteriors themselves as the observed density field.
In doing so, one assumes that a) the posteriors are Gaussian, b) all events can be modeled as having the same localization error $\sigma$ (say, the median localization error of the radial bin \cite{2025arXiv250410482P}), and crucially c) the \ac{MAP} points are located at the true positions.
To our knowledge, \changed{the validity of these} assumptions have yet to be explicitly addressed in the \ac{GW} cross-correlation literature.
Ultimately, a \ac{GW} posterior measurement is a result of a Bayesian inference on a noisy \ac{GW} strain measurement, and cannot be naively modelled as blurred image of the true position as one can for electromagnetic maps such as the \ac{CMB}.

We discussed that the high-\ac{SNR} limit can justify the Gaussianity of the \ac{GW} measurement error. If the \ac{MAP} points are Gaussian-distributed from the true positions, then the probability distribution under measurement noise of the measured sky position $\hat{\vecOm}$ of each \textit{individual} event is a convolution of the true position with a Gaussian kernel the size of its localization error. This allows us to apply \cref{eq:Gaussian_beam} on an individual event basis. A more rigorous derivation of this result is given in \cref{app:beam}. 

The number counts overdensity field constructed from the true positions of $N^\alpha$ \acp{GW} in radial bin $\alpha$ is given by
\begin{equation}
\begin{split}
\delta^{\alpha}(\vecOm) &= \frac{1}{n^{\alpha}}\sum_{i=1}^{N^{\alpha}} \delta^{(D)}(\vecOm - \vecOm_i) \\
\alm^{\alpha} &= \frac{1}{n^{\alpha}} \sum_{i=1}^{N^{\alpha}} Y_{\ell m}^*(\vecOm_i) \equiv \frac{1}{n^\alpha} \sum_{i=1}^{N^{\alpha}}  \alm^{(i)}
\end{split}
\end{equation}
where $n^{\alpha} = \frac{N^{\alpha}}{4\pi}$ is the mean angular density of bin $\alpha$, and $\alm^{(i)} \equiv Y_{\ell m}^*(\vecOm_i)$ is the \ac{SHT} of a Dirac-delta localization at the $i$th event's true position $\vecOm_i$. For brevity, we omit the negative offset term that sets $\langle \delta^\alpha \rangle = 0$, since it only has the effect of setting $a_{00}=0$.

The \ac{MAP} position $\hat{\vecOm}_i$ of each event is drawn from a Gaussian probability distribution $p^{(i)}(\vecOm)$ centered on the true position $\vecOm_i$. Each $p^{(i)}$ is given by a convolution of the true position with the Gaussian kernel given in \cref{eq:Gaussian_beam} corresponding to its localization error $\sigma_i$, which in harmonic space is
\begin{equation} \label{eq:beam_window}
d_{\ell m}^{(i)} = B_\ell(\sigma_i) \alm^{(i)}
\end{equation}
where $d_{\ell m}^{(i)}$ is the \ac{SHT} of $p^{(i)}$. 
Note that if we were to take the posterior distributions themselves as the map, this would correspond to an \textit{additional} Gaussian convolution, since the \ac{MAP} measurements are offset from the true positions! This would have the effect of doubly suppressing the signal at small scales.

The total noise probability overdensity field is would therefore be
\begin{equation}
    d_{\ell m}^{\alpha} = \frac{1}{n^{\alpha}} \sum_{i=1}^{N^\alpha} d_{\ell m}^{(i)} = \frac{1}{n^{\alpha}} \sum_{i=1}^{N^\alpha} B_\ell(\sigma_i) \alm^{(i)}.
\end{equation}
This is the expectation value over noise of the measured overdensity field $\hat{a}_{\ell m}^\alpha$, i.e. 
\begin{align}
    \E_\text{noise}\left[ \hat{a}^{\alpha}_{\ell m} \right] &= d_{\ell m}^{\alpha} \\
    \E_\text{noise}\left[ \hat{a}^{(i)}_{\ell m} \right] &= d_{\ell m}^{(i)}.
\end{align}
which we will use later in \cref{eq:beam_i=j}. 

\changed{For clarity, \alm is the true overdensity field composed of a sum of Dirac delta functions at the true source positions, $d_{\ell m}$ is the probability density field of the measured \ac{MAP} sky locations of the sources, and $\hat{a}_{\ell m}$ is the measured overdensity field, which is} a sum of Dirac delta functions drawn from each $p^{(i)}(\vecOm)$. 

$\hat{a}^{(i)}_{\ell m}$ is \changed{therefore} the \ac{SHT} of the delta function draw from event probability distribution $d_{\ell m}^{(i)}$. The \ac{SHT} of this measured overdensity field is therefore 
\begin{equation}
    \hat{a}^\alpha_{\ell m} = \frac{1}{n^\alpha}\sum_{i=1}^{N^\alpha} \hat{a}^{(i)}_{\ell m} = \frac{1}{n^\alpha}\sum_{i=1}^{N^\alpha} Y_{\ell m}^*(\hat{\vecOm}_i).
\end{equation}

We are interested in finding the auto- and cross-power spectra corresponding to these measured maps. The \ac{GW} auto-power spectrum is given by
\begin{equation} \label{eq:measured_auto_power_spectra}
\left\langle \hat{a}^{\alpha\,*}_{\ell m} \hat{a}^\alpha_{\ell^\prime m^\prime} \right\rangle = \frac{1}{\left(n^\alpha\right)^2} \sum_{i, j}^{N^\alpha} \left\langle \hat{a}^{(i)\,*}_{\ell m} \hat{a}^{(j)}_{\ell^\prime m^\prime} \right\rangle 
\end{equation}
for some luminosity distance slice $\alpha$.
This expectation value now averages over both measurement noise realizations and realizations of the universe, i.e.,
\begin{equation}
    \left\langle Z \right\rangle = \E_\text{cosmo}\Big[ \, \E_\text{noise}\left[ Z \right] \Big]
\end{equation}
for any quantity $Z$.
Since the $\hat{a}^{(i)}_{\ell m}$ are drawn from $d_{\ell m}^{(i)}$,
\begin{equation}
    \E_\text{noise}\left[ \hat{a}^{(i)}_{\ell m} \right] = d_{\ell m}^{(i)} = \int \di \vecOm \, p^{(i)}(\vecOm) Y_{\ell m}^*(\vecOm)
\end{equation}
and therefore
\begin{equation} \label{eq:measured_map_exp}
\begin{split}
\E_\text{noise}&\left[ \hat{a}^{(i)\,*}_{\ell m} \hat{a}^{(j)}_{\ell^\prime m^\prime} \right] = \E_\text{noise}\left[ Y_{\ell m}^*(\hat{\vecOm}_i) \,Y_{\ell^\prime m^\prime}(\hat{\vecOm}_j) \right] \\
=&
\begin{cases}
    \int \di \vecOm \, p^{(i)}(\vecOm) Y_{\ell m}^*(\vecOm) \,Y_{\ell^\prime m^\prime}(\vecOm) \quad & (i = j) \\
    \int \di \vecOm \, \di \vecOm^\prime \, p^{(i)}(\vecOm) p^{(j)}(\vecOm^\prime) Y_{\ell m}^*(\vecOm) \,Y_{\ell^\prime m^\prime}(\vecOm^\prime) \quad & (i \neq j)
\end{cases}
\end{split}
\end{equation}
Now we need to take an expectation value over realizations of the universe. For the $i=j$ case, we have $\langle p^{(i)}(\vecOm) \rangle = \frac{1}{4\pi}$ for a homogeneous and isotropic universe, and thus
\begin{equation} \label{eq:beam_i=j}
    \sum_{i}^{N^\alpha} 
    \left\langle \hat{a}^{(i)\,*}_{\ell m} \hat{a}^{(i)}_{\ell^\prime m^\prime} \right\rangle  = \frac{N^\alpha}{4\pi} \delta^{(K)}_{\ell \ell^\prime} \delta^{(K)}_{m m^\prime}
\end{equation}

Because taking the expectation value over noise returns us to a map of probability distributions of the \ac{GW} measurements (i.e., $\E_\text{noise}\left[ \hat{a}_{\ell m} \right] = d_{\ell m}$), a further expectation value over realizations of the universe takes us to the power spectrum of the probability map $d_{\ell m}$. 
We can use this fact to deduce the sum of the $i\neq j$ terms in \cref{eq:measured_map_exp}.
Using the definition of the \ac{SHT} yields
\begin{equation} 
\begin{split}
    \left\langle d^\alpha_{\ell m} d^\alpha_{\ell^\prime m^\prime} \right\rangle &= \frac{1}{\left(n^\alpha\right)^2} \sum_{i,j} \left\langle d^{(i)\,*}_{\ell m} d^{(j)}_{\ell^\prime m^\prime} \right\rangle \\
    &= \frac{1}{\left(n^\alpha\right)^2} \sum_{i,j} \int \di \vecOm \, \di \vecOm^\prime \left\langle p^{(i)}(\vecOm) p^{(j)}(\vecOm^\prime) \right\rangle \\
    & \qquad\qquad\qquad\qquad \times  Y_{\ell m}^*(\vecOm) \,Y_{\ell^\prime m^\prime}(\vecOm^\prime)
\end{split}
\end{equation}
We can see that the $i \neq j$ terms are the same as those of \cref{eq:measured_map_exp}.
To compute these terms, we can plug in \cref{eq:beam_window}:
\begin{equation} \label{eq:blurred_map_exp}
    \frac{1}{\left(n^\alpha\right)^2} \sum_{i,j} \left\langle d^{(i)\,*}_{\ell m} d^{(j)}_{\ell^\prime m^\prime} \right\rangle = \frac{\left\langle B_\ell(\sigma)\right\rangle \left\langle B_{\ell^\prime}(\sigma)\right\rangle}{\left(n^\alpha\right)^2} \sum_{i,j}  \left\langle \alm^{(i)\,*} a_{\ell^\prime m^\prime}^{(j)} \right\rangle 
\end{equation}
The independent brackets indicate the independent draws of the localization error $\sigma$.
The summand is simply the power spectrum of the true number counts overdensity field:
\begin{equation}
\begin{split}
    \frac{1}{\left(n^\alpha\right)^2} \sum_{i,j}\left\langle \alm^{(i)\,*} a_{\ell^\prime m^\prime}^{(j)} \right\rangle &= \left\langle a_{\ell m}^{\alpha\,*} a_{\ell^\prime m^\prime}^\alpha \right\rangle \\
    &= \delta^{(K)}_{\ell \ell^\prime} \delta^{(K)}_{m m^\prime} \left[ \overline{C}_\ell^{\alpha\alpha} + \frac{1}{n^\alpha} \right].
\end{split}
\end{equation}
One can show that the shot noise $1/n^\alpha$ can be identified with the $i=j$ terms, while the $i \neq j$ terms add up to $\overline{C}_\ell^{\alpha\alpha}$. Therefore, we obtain our $i\neq j$ terms
\begin{equation}
\begin{split}
    \frac{1}{(n^\alpha)^2} \sum_{i \neq j} \left\langle \hat{a}^{(i)\,*}_{\ell m} \hat{a}^{(j)}_{\ell^\prime m^\prime} \right\rangle &= \frac{1}{(n^\alpha)^2} \sum_{i \neq j} \left\langle d^{(i)\,*}_{\ell m} d^{(j)}_{\ell^\prime m^\prime} \right\rangle \\
    &= \delta^{(K)}_{\ell \ell^\prime} \delta^{(K)}_{m m^\prime} \left\langle B_\ell(\sigma)\right\rangle^2 \overline{C}_\ell^{\alpha\alpha}.
\end{split}
\end{equation}
Finally, combined with \cref{eq:beam_i=j}, we have the auto-power spectrum under independent event-level localization errors:
\begin{equation}
    \left\langle \hat{a}^{\alpha\,*}_{\ell m} \hat{a}^\alpha_{\ell^\prime m^\prime} \right\rangle = \delta^{(K)}_{\ell \ell^\prime}  \delta^{(K)}_{m m^\prime} \left[ \left\langle B_\ell\right\rangle^2 \overline{C}_\ell^{\alpha\alpha} + \frac{1}{n^\alpha}\right].
\end{equation}
We can see that for the \ac{GW}--\ac{GW} auto-power spectrum, the signal component is suppressed by the localization beam, while the shot noise is not: each event has random localization errors, but the realized position of each event is still self-correlated. 

Deriving the cross-power spectra is a straightforward generalization. For two non-overlapping shells of \acp{GW}, it is always the case that $i \neq j$, and hence there is no shot noise. The \ac{GW}--\ac{GW} angular cross-correlation is therefore
\begin{equation} \label{eq:hatClGG_with_angloc}
    \hat{C}_\ell^{G_\alpha G_\beta} \equiv \left\langle \hat{a}^{G_\alpha\,*}_{\ell m} \hat{a}^{G_\beta}_{\ell m} \right\rangle = \left\langle B_\ell\right\rangle_\alpha \left\langle B_\ell\right\rangle_\beta  \, \overline{C}_\ell^{G_\alpha G_\beta} + \frac{\delta^{(K)}_{\alpha\beta}}{n^{G_\alpha}}.
\end{equation}
where $\left\langle B_\ell\right\rangle_\alpha $ is the expected \ac{GW} angular beam window function in bin $\alpha$.

The \ac{GW}--galaxy cross-correlation is a bit trickier, since overlapping shells can have a coincident case corresponding to the \ac{GW} host galaxy being in the galaxy catalog. Because galaxy and \ac{GW} localizations are independent, we always have $\hat{\vecOm}_i \neq \hat{\vecOm}_j$ for \ac{GW} $i$ and galaxy $j$ and therefore
\begin{equation}
\begin{split}
\E_\text{noise}&\left[ \hat{a}^{G, (i)\,*}_{\ell m} \hat{a}^{g, (j)}_{\ell^\prime m^\prime} \right] = \E_\text{noise}\left[ Y_{\ell m}^*(\hat{\vecOm}_i) \,Y_{\ell^\prime m^\prime}(\hat{\vecOm}_j) \right] \\
&= \int \di \vecOm \, \di \vecOm^\prime \, p^{G,(i)}(\vecOm) p^{g, (j)}(\vecOm^\prime) Y_{\ell m}^*(\vecOm) \,Y_{\ell^\prime m^\prime}(\vecOm^\prime).
\end{split}
\end{equation}
That is, compared to \cref{eq:measured_map_exp}, there is no special $i=j$ case. 
Taking the expectation value over realizations of the universe, we follow the same steps as above to evaluate the power spectra of the measured maps in terms of the power spectra of the true maps. The result is
\begin{equation} \label{eq:hatClGg_with_angloc}
    \hat{C}_\ell^{G_\alpha g_\beta} \equiv \left\langle \hat{a}^{G_\alpha\,*}_{\ell m} \hat{a}^{g_\beta}_{\ell m} \right\rangle = \left\langle B_\ell^a \right\rangle \left[ \overline{C}_\ell^{G_\alpha g_\beta} + \frac{\langle n^{G_\alpha g_\beta} \rangle}{\langle n^{G_\alpha} \rangle \langle n^{g_\beta} \rangle} \right].
\end{equation}
Because galaxy localization errors are negligible compared to \ac{GW} localization errors, we have taken the beam window function of the galaxy localization $B_\ell^g=1$.
We can see that the cross shot noise term is now suppressed by the localization beam, resulting from the fact that \acp{GW} and their host galaxies are localized independently.
Consistent with intuition, the information that \acp{GW} are located in host galaxies becomes diluted with poor localization errors.

Remarkably, we need only to carry the $\textit{average}$ beam window function per shell $\alpha$
\begin{equation} \label{eq:mean_beam}
    \langle B_\ell \rangle_\alpha = \int \di \sigma P^\alpha(\sigma) B_\ell(\sigma).
\end{equation}
If the number of events in the shell is large, $B_\ell$ can be directly estimated from its empirical distribution as a once-through pre-processing step
\begin{equation} \label{eq:mean_beam_approx}
    \langle B_\ell \rangle_\alpha \approx \frac{1}{N^{G_\alpha}}\sum_{i=1}^{N^{G_\alpha}} B_\ell(\sigma_i).
\end{equation}

A corollary of this is that using the mean localization error $\langle \sigma \rangle$ to compute $B_\ell$ will overestimate the degreee to which small angular scales are suppressed. That is, $B_\ell(\langle \sigma\rangle) < \langle B_\ell(\sigma) \rangle$ if and only if $\sqrt{\ell(\ell+1)} > \langle \sigma \rangle$. This follows from Jensen's inequality. We show this for a set of realistic mock data  in \cref{fig:beam}, which will be described in more detail in \cref{sec:implementation}.

In short, \cref{eq:hatClGG_with_angloc,eq:hatClGg_with_angloc} describe how to compute the theoretical expected power spectra corrected for localization errors by multiplying by a beam window function that suppresses small angular scales, which can be estimated via \cref{eq:mean_beam_approx} (given a sufficiently large number of observations per bin).
$\left\{\hat{\tilde{C}}_\ell^{\alpha\beta}\right\}$, the cross-correlation estimators of the noisy \ac{GW} data, are then an unbiased estimator of these error-corrected theoretical power spectra 
\begin{equation}
    \hat{C}_\ell^{\alpha\beta} = \left\langle \hat{\tilde{C}}_\ell^{\alpha\beta} \right\rangle
\end{equation} 
Because angular localization errors are effectively suppressing the Gaussian fluctuations at various scales, we can plug these error-corrected theoretical power spectra $\{\hat{C}_\ell^{\alpha\beta}\}$ directly into \cref{eq:Clhat_Gaussian_ll} for the covariance of the $\{\hat{\tilde{C}}_\ell^{\alpha\beta}\}$.

Finally, although we discussed only Gaussian, circular, and homogeneous localization errors, this formalism can be generalized in principle to any distribution of localization errors.
Specifically, the expression for the beam window function in \cref{eq:Gaussian_beam} will be different for non-Gaussian distributions.
In general, as we discuss further in \cref{app:beam}, the beam of a spatially varying (i.e. the localization error depends on $\vecOm$) and anisotropic kernel is given by
\begin{equation}
d^{(i)}_{\ell m} = \sum_{\ell^\prime, m^\prime} D_{\ell m, \ell^\prime m^\prime} a_{\ell^\prime m^\prime}
\end{equation}
rather than the simple multiplicative form in \cref{eq:isotropic_beam_mult}; the corresponding expressions for the expected cross-correlation power spectra above could be re-derived in this more complicated case.
Circular localizations require triangulation from multiple ($\geq 3$) sensitive detectors, while measurement errors may be spatially varying due to accumulated modulations from Earth's rotation and orbit \cite{2017ApJ...835...31C}.
This expansion will be left for future work. 

\subsection{Incomplete or anisotropic sky coverage} \label{subsec:aniso_sky}

The treatment of incomplete or anisotropic sky coverage has been developed extensively for analyses of both galaxy surveys and the \ac{CMB}, and thus we only briefly review the results here. For more details, see e.g., \cite{2019MNRAS.484.4127A} and references therein.

Galaxy surveys do not have complete coverage of the sky; at the very least, the galactic plane and other bright sources are masked out. It is also possible that the depth of either the galaxy or \ac{GW} catalog does not have uniform sensitivity across the sky, in the latter case  because of the angular beam pattern of the GW detector network. In the case of angular selection effects, the expected density $\langle n(r, \vecOm) \rangle$ depends on not only the distance but also the sky location, and thus a mean density encoding the angular selection effects must be used when computing the overdensity field.
The result is a measured overdensity field that has been multiplied by a weights map, $\hat{\Delta}^\alpha(\vecOm) = \Delta^\alpha(\vecOm) \,m^\alpha(\vecOm)$. A sky mask would simply be a binary weights map. 

This weighted map is no longer isotropic, and thus different $\ell$ modes are coupled together, making the naive estimator in \cref{eq:naive_estimator} biased. The expectation value of the naive estimator is related to the true power spectrum via a mode-coupling matrix
\begin{equation}
    \left\langle \hat{\tilde{C}}_\ell^{\alpha\beta} \right\rangle = \sum_{\ell^\prime} M_{\ell \ell^\prime} C_{\ell^\prime}^{\alpha\beta}.
\end{equation}
There are several approaches to deal with the mode-coupling matrix: one could convolve the theory prediction with $M_{\ell \ell^\prime}$, bin together the $\ell$'s that are coupled together in bandpowers \cite{2019MNRAS.484.4127A}, or use a different estimator (e.g., maximum likelihood \cite{1998PhRvD..57.2117B} or minimum-variance quadratic \cite{1997PhRvD..55.5895T}) to produce an unbiased estimate of the power spectrum in the presence of the weights map. 

In practice, if the sky-mask is well-behaved, its fractional sky coverage is not too small, and the underlying power spectrum $C_\ell$ is smooth with respect to the coupling width, then $M_{\ell \ell^\prime}$ is effectively diagonal or close to diagonal, such that $M_{\ell \ell^\prime} \approx \delta^{(K)}_{\ell\ell^\prime}$, where $\fsky^{\alpha\beta}$ is the overlapping fractional sky coverage of the masks corresponding to both fields
\begin{equation}
    \fsky^{\alpha\beta} = \frac{1}{4\pi} \int d\vecOm \, m^\alpha(\vecOm) m^\beta(\vecOm).
\end{equation}

In this case, one can use the estimator
\begin{equation} \label{eq:fsky_naive_estimator}
    \hat{\tilde{C}}_\ell^{\alpha\beta} = \sum_{\ell^\prime} M_{\ell\ell^\prime} \tilde{C}_{\ell^\prime}^{\alpha\beta} \approx \frac{1}{\fsky^{\alpha\beta}(2\ell + 1)} \sum_{m=-\ell}^\ell \hat{a}_{\ell m}^{\alpha\,*} \hat{a}_{\ell m}^\beta 
\end{equation}
where $\tilde{C}_{\ell^\prime}^{\alpha\beta}$ is the naive estimator given in \cref{eq:naive_estimator}.

The fractional sky coverage also affects how well the power spectra can be measured. On scales smaller than the sky mask, it is often assumed that the number of independent modes is decreased in proportion to the available sky area, which effectively increases the covariance of the angular power spectrum estimators (originally given in \cref{eq:Cl_covmat}) by a factor of $\fsky^{\alpha\beta\mu\nu}$, the fractional sky coverage shared between all four fields \cite{2002ApJ...567....2H,2007MNRAS.381.1347C}. That is,
\begin{equation} \label{eq:skymask_Clcovmat}
     \left\langle \hat{\tilde{C}}_\ell^{\alpha\beta} \,\hat{\tilde{C}}_{\ell^\prime}^{\mu\nu} \right\rangle \approx
     \frac{\delta^{(K)}_{\ell\ell^\prime}}{\fsky^{\alpha\beta\mu\nu}(2 \ell+1)}\left[\hat{C}_{\ell}^{\alpha \mu} \hat{C}_{\ell}^{\beta \nu}+\hat{C}_{\ell}^{\alpha \nu} \hat{C}_{\ell}^{\beta \mu}\right].
\end{equation}
As with \cref{eq:fsky_naive_estimator}, this assumption holds with diagonal or close to diagonal $M_{\ell \ell^\prime}$.

\subsection{Radial localization errors} \label{subsec:radial_loc}

In \cref{subsubsec:2D_GRF}, we defined in \cref{eq:eff_radial_window} an effective window function $\phi(r)$ that acts as the radial weighting function when integrating over the 3D power spectrum to produce the theoretical 2D power spectrum in \cref{eq:limber}. In the presence of radial localization errors, the window function $w$ we use for binning is not in the true distance $r$ but rather in the measured distance $\hat{r}$, which we must marginalize over \cite{2012MNRAS.427.1891A}:
\begin{equation} \label{eq:eff_window_func}
\phi^{X_\alpha}(r) \propto p^X_\text{obs}(r)\int \di \hat{r}\,w^{X_\alpha}(\hat{r}) \mathcal{L}^{X_\alpha}(\hat{r}|r)
\end{equation}
where $p^X_\text{obs}(r)$ is the distribution of observable (i.e. with selection effects) tracers $X$, since the window function is weighted by the number of tracers accessible at each distance $r$.

For \acp{GW}, the measurement error $\mathcal{L}(\hat{d}_L|d_L)$ is not deterministic, varying from event to event depending on the noise realization and source properties. 
Nonetheless, we can imagine the resulting localization of each \ac{GW} as arising from some distribution of errors. 
Generally, the expected localization error will depend on the measured source properties. Nonetheless, motivated by the idea that the measured strain $h \sim 1/\hat{d}_L$, the distribution of $\Delta\ln \hat{d}_L = \Delta d_L / \hat{d}_L$ evolves slowly across the width of the bin, and thus we can approximate the $\Delta\ln \hat{d}_L$ of the \acp{GW} in that bin as being drawn independently from a distribution $p^\alpha(\Delta\ln \hat{d}_L )$. If the number of \acp{GW} in each bin is large, then this distribution can be approximated directly, with e.g., a \ac{KDE} or parameterized fit. The effective window function for each \ac{GW} bin $\alpha$ (in the high-\ac{SNR} / gaussian measurement regime) is then 
\begin{equation}  \label{eq:GW_eff_window_func}
\begin{split}
\phi^{G_\alpha}&(d_L) \propto p^{G}_\text{obs}(d_L) \int \di \hat{d}_L\,w^{G_\alpha}(\hat{d}_L)  
\\& \times \int \di (\Delta\ln \hat{d}_L) \, p^\alpha(\Delta\ln \hat{d}_L ) \,
\mathcal{N}\left[\hat{d}_L; d_L, \hat{d}_L \cdot \Delta\ln \hat{d}_L\right]
\end{split}
\end{equation}
up to normalization. 

The same idea of applies to galaxies. A common model for galaxy redshift errors is a fixed relative redshift error $\Delta\ln z = \Delta z / z$, for example $\approx0.05$ for photometric redshifts and $\approx 0.01$ for spectroscopic redshifts \cite{2022A&A...662A..93E,2024ApJ...964..191B,2025arXiv250410482P}. In this case, 
\begin{equation}  \label{eq:g_eff_window_func}
\phi^{g_\beta}(z) \propto p\gobs(z) \int \di \hat{z}\,w^{g_\beta}(\hat{z}) \,\mathcal{N}\left[\hat{z}; z, z \cdot \Delta\ln z\right].
\end{equation}
for some fixed $\Delta \ln z.$

Note that both \cref{eq:GW_eff_window_func,eq:g_eff_window_func} require knowledge of the true radial distribution of the observed tracers $p^{X}_\text{obs}(r)$, despite radial localization errors. We will assume that this can be inferred separately before the cross-correlation analysis, a reasonable assumption for a large number of relatively well-localized observations.

\section{Summary of Formalism} \label{sec:posterior}

In \cref{sec:theory}, we reviewed the standard formalism for dark siren analyses and cross-correlation analyses. 
In \cref{sec:unified}, we connected these two methods and showed that the Gaussian formalism can encompass dark siren analyses, insofar as Poisson statistics tend to Gaussian with large $N$. 
In this framing, the standard galaxy catalog method infers the probability of an observed \ac{GW} field conditioned on a fixed background galaxy field, which is usually constructed by assuming a uniform distribution of missing galaxies.
On the other hand, the cross-correlation method models a joint distribution of the observed \ac{GW} \textit{and} galaxy data, conditioning only on the cosmology rather than any specific realization of the universe. 
We also showed that because we cross-correlate normalized overdensity fields, it uses complementary information to the spectral sirens method.
Thus, we are able to write down in \cref{eq:Ppop_xcorr_unified} a joint population model on the \ac{GW} and galaxy parameters, which are the \ac{GW} parameters $\veclam = \{m_1, m_2, d_L\}$ and cross-correlation estimators $\{\tilde{C}_\ell^{[\alpha\beta]}\}$. 
This population prior includes information from the spectral sirens method, \ac{GW}--galaxy cross-correlations, and galaxy--galaxy clustering.

\begin{figure*}[t]
    \centering 
    \includegraphics[width=\textwidth]{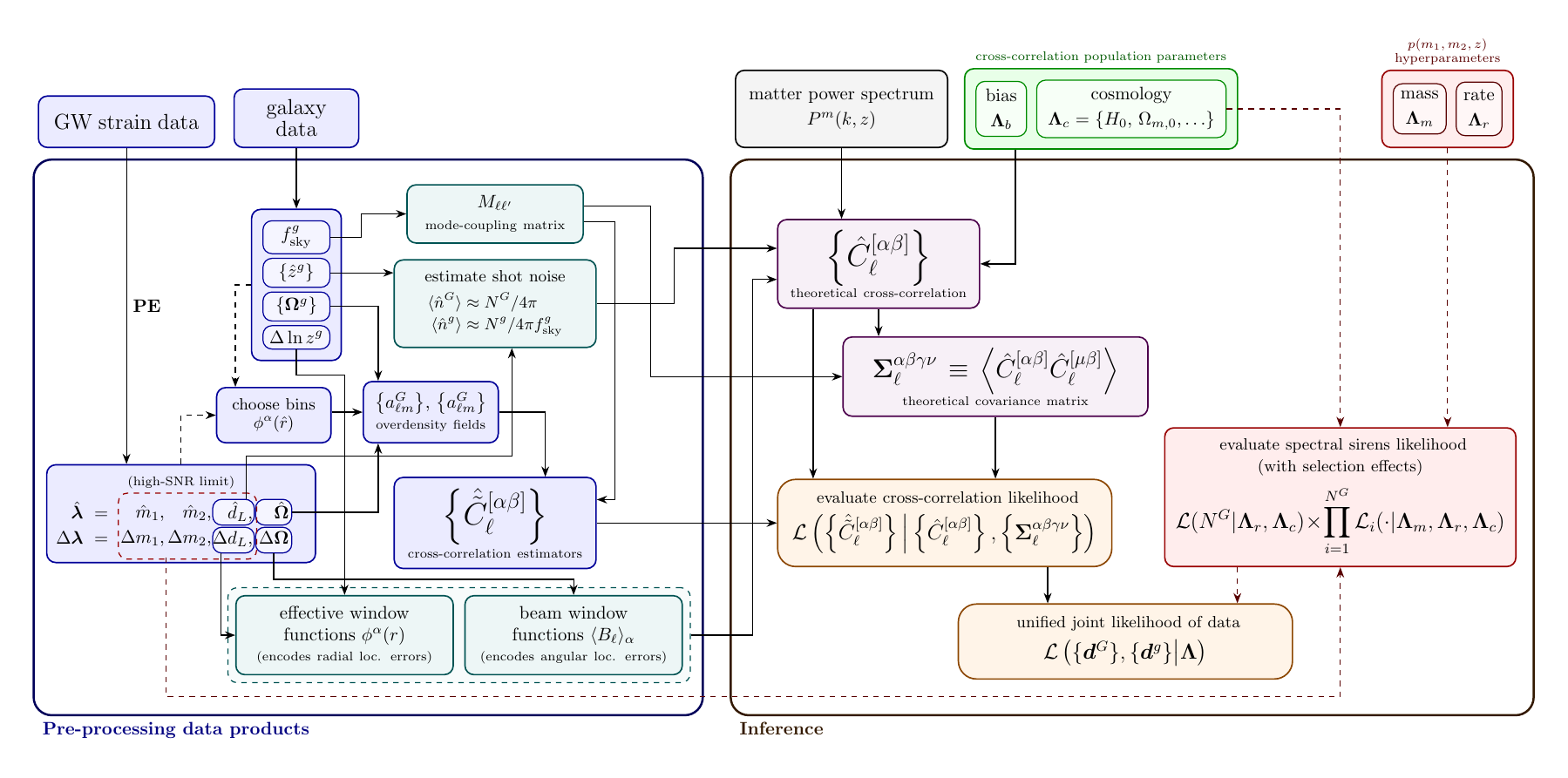}
    \caption{A flowchart summarizing our cross-correlation formalism. Pre-processing steps for the data (\cref{sec:noise}) are shown in the left box, which are used to compute the theoretical cross-correlation and covariance matrix during inference (right box). Boxes are colored roughly as follows: data (blue), estimated noise data products used during inference (teal), theory-level calculations (purple), likelihoods (orange), and the spectral sirens method (red). The cross-correlation and spectral sirens methods are complementary, so they can be done separately and/or combined. \label{fig:flowchart}}
\end{figure*}

In \cref{sec:noise}, we discussed the high-\ac{SNR} limit, in which we can approximate our data as consisting of \ac{MAP} measurements of the parameters $\hat{\veclam}$ and associated Fisher matrix measurement errors $\Delta\veclam$. In particular, we show how to forward model localization errors into the likelihood on the observed cross-correlation estimators $\{\hat{\tilde{C}}_\ell^{[\alpha\beta]}\}$ rather than doing a Monte-Carlo integration over all possible values of the true parameters. 
We keep the same Gaussian distribution on the angular power spectrum estimators as in \cref{eq:Clhat_Gaussian_ll}:
\begin{equation} \label{eq:Clhat_Gaussian_ll_noise}
\begin{split}
    \mathcal{L}&\left(\left\{\hat{\tilde{C}}_\ell^{[\alpha\beta]}\right\} \mid \left\{\hat{C}_\ell^{[\alpha\beta]}\right\}\right) = 
    \prod_{\ell=\ell_{\min}}^{\ell_{\max}}
    \frac{1}{\sqrt{(2\pi)^{r}\det\mathbf{\Sigma}_\ell}} \\& 
    \times \exp\left[-\frac{1}{2} \sum_{\alpha \leq \beta} \sum_{\mu \leq \nu} \sum_{\ell} \Delta \hat{C}_{\ell}^{\alpha\beta} (\covS_\ell)^{-1}_{\alpha\beta, \mu\nu} \, \Delta \hat{C}_{\ell}^{\mu \nu}\right].
\end{split}
\end{equation}
The error-corrected theoretical expectation for the cross-correlation $\{{\hat{C}}_\ell^{[\alpha\beta]}\}$ can be computed by integrating the 3D power spectrum via \cref{eq:limber} with the error corrected effective window functions in \cref{eq:GW_eff_window_func,eq:g_eff_window_func}, and multiplying by the appropriate average beam window function per shell in \cref{eq:hatClGG_with_angloc,eq:hatClGg_with_angloc}. If the survey sky coverage is anisotropic for either tracer, then one must use an updated expression (\cref{eq:fsky_naive_estimator}) to compute the $\{\hat{\tilde{C}}_\ell^{[\alpha\beta]}\}$ in order for it to be an unbiased estimator of the $\{{\hat{C}}_\ell^{[\alpha\beta]}\}$, and update the covariance matrix $(\covS_\ell)_{\alpha\beta, \mu\nu}$ accordingly (\cref{eq:skymask_Clcovmat}). A flowchart summarizing our approach to the cross-correlation inference formalism can be found in \cref{fig:flowchart}.

We can now put everything together and write down a joint hierarchical likelihood given noisy galaxy and \ac{GW} data.
Taking \cref{eq:Ppop_xcorr_unified} with our error-corrected formulae as our population likelihood, we have
\begin{subequations} \label{eq:unified_jointll_no_sel}%
\begin{align}
    \mathcal{L}\left(\dG, \dg | \vecLam \right) &=
    \mathcal{L} \left(\left\{\hat{\tilde{C}}_\ell^{[\alpha\beta]}\right\} \bigg| \{\hat{z}^g, \Delta z^g\}, \{\hat{d}_L, \Delta\veclam\},\vecLam\right) \label{eq:unified_jointll_xcorr_term} \\
    & 
    \times  \mathcal{L} (N^G | \vecLam) 
    \times \prod_{i=1}^{N^G} \mathcal{L}_i \label{eq:unified_jointll_event_terms}
\end{align}
\end{subequations}
\cref{eq:unified_jointll_xcorr_term} is the Gaussian cross-correlation likelihood in \cref{eq:Clhat_Gaussian_ll_noise}; the dependence on $\{\hat{z}^g\}$ on $\{\hat{d}_L\}$ enters because these are used to estimate the shot noise per bin $1/\langle n^\alpha \rangle \approx 4\pi/N^\alpha$ (where $N^\alpha$ is the number of tracers observed in bin $\alpha$), while the galaxy redshift errors $\{\Delta z^g\}$ and \ac{GW} posterior measurement errors $\{\Delta\veclam\}$, $\veclam \in \{d_L, \vecOm, m_1, m_2\}$ are used to estimate the average measurement uncertainties, which are needed to compute the angular beam and effective radial window functions. These, along with the \ac{GW} and galaxy bias parameters and the 3D matter power spectrum, are sufficient to compute the error-corrected theoretical $\{\hat{C}_\ell^{\alpha\beta}\}$ as well as the corresponding covariance matrix on the $\{\hat{\tilde{C}}_\ell^{\alpha\beta}\}$. 

On the other hand, \cref{eq:unified_jointll_event_terms} is our usual inhomogeneous Poisson likelihood, with the usual product of independent event-level likelihoods
\begin{equation}
\begin{split}
\mathcal{L}_i = \int \di m_1 & \di m_2 \di d_L \,p(m_1, m_2, d_L | \vecLam) \\
&\mathcal{L}\left(\hat{m}_{1, i}, \hat{m}_{2, i}, \hat{d}_{L, i} | m_1, m_2, d_L, \Delta\veclam_i, \vecLam \right).
\end{split}
\end{equation}
Here, we are implicitly including a population prior on the \ac{GW} $d_L$ but not the $z^g$; in principle one could be added here to jointly model the rate evolution of the galaxies. We consider that to be outside the scope of this work.

Note that as written in \cref{eq:unified_jointll_no_sel}, the data enters both the cross-correlation estimator and the quantities on which it is conditioned (i.e. in $\{\hat{z}^g\}$, $\{\hat{d}_L\}$, and $\{\Delta\veclam\}$), which is technically not allowed as these are not completely independent. 
A fully self-consistent procedure would be to write down the likelihood on these quantities and marginalize over those, but here we opt not to because of the computational cost.
\changed{In deriving the noise-corrected likelihood, we have implicitly folded in the {\it realization variance} accounting for the possible realizations of the population and measurement noise.}
In conditioning on the Poisson-realized number of tracers per bin and realized distribution of measurement errors, rather than the Poisson rate and true distribution of measurement errors per bin, \changed{this realization variation is conditioned out and the actual input data then has less variability than the likelihood assumes}.
Thus, this procedure is conservative because we are assuming greater variability than is actually present. As discussed before, we require a large number of events per shell in order for the realization variance contribution to be small.

The final step is to incorporate selection effects.
Recall from \cref{subsec:darksiren} that writing down the likelihood with selection effects involves conditioning on the fact that the \ac{GW} events were detected. 
\cref{eq:unified_jointll_xcorr_term}, however, is already implicitly conditioned on the fact that the events are detected. The cross-correlation likelihood is a comparison of the observed map to the expected map, where the noisiness of the detections, sampling error (shot noise), and selection effects are already forward-modeled into the mean and covariance of the likelihood function. Radial (as well as angular, as discussed in \cref{subsec:aniso_sky}) selection effects are accounted for in the fact that overdensity fields are normalized to the expected number density in each shell and pixel. Put another way, the overall normalization (number of tracers per bin) does not enter in the intrinsic clustering patterns of the tracer. Although it affects the expected shot noise, as discussed at the end of \cref{sec:unified}, the shot noise can be directly estimated from the observed number density as long as the number of observations per bin is large, which decouples the cross-correlation from the rate likelihood.

Thus, because incorporating selection effects leaves the likelihood on the cross-correlation unchanged, we only need to impose selection effects on the \ac{GW} inhomogeneous Poisson likelihood in \cref{eq:unified_jointll_event_terms}, which we already reviewed in \cref{eq:poisson_ll}. The posterior on the hyperparameters $\vecLam$ given the galaxy and \ac{GW} data under selection effects is therefore given by
\begin{widetext}
\begin{equation} \label{eq:full_unified_hyperposterior}
\begin{split}
    \mathcal{P}(\vecLam |\{\vecd^G, D\}, \{\vecd^g\} ) &\propto \Pi(\vecLam) e^{-N_\text{exp}^G(\vecLam)} N_\text{tot}^{N^G} \left[\prod_{i=1}^{N^G} \int \di \veclam \,\Ppop(\veclam | \vecLam)\,\mathcal{L}(\vecd_i | \veclam, \vecLam)\right] \mathcal{L}\left(\left\{\hat{\tilde{C}}_\ell^{[\alpha\beta]}\right\}| \{\hat{z}^g\}, \{\hat{d}_L\}, \{\Delta z^g\}, \{\Delta\veclam\}, \vecLam\right)
\end{split}
\end{equation}
\end{widetext}
As before, $\{D\}$ denotes the fact that each \ac{GW} event was detected.
The spectral sirens and cross-correlation methods can therefore be carried out as two separate analyses: the joint posterior can be evaluated by using the posterior of one analysis as the prior for the other. The cross-correlation likelihood encodes the additional information not captured by the spectral sirens method, namely angular correlations of overdensities between \acp{GW} and galaxies and galaxy--galaxy 2-point correlations.

We will end this discussion with some final caveats on the relationship of the cross-correlation method with the standard galaxy catalog method. We argued in \cref{sec:unified} that the cross-correlation method is an extension of the standard galaxy catalog method---in particular the information from the conditional distribution $p(\vecOm | z, \vecLam)$---to 2-point statistics. In particular, the \changed{standard galaxy catalog method can be recovered from transforming the \ac{GRF} formalism to real space and conditioning on a fixed known underlying galaxy field.} 

In practice, \changed{the transformation from harmonic to real space cannot be done exactly because one must always choose a finite binning resolution.} Because we build measurement errors directly into the theory-level expectation, and because we work in harmonic space rather than real space, the resolution of both the density fields and the theoretical angular power spectrum are truncated at the measurement precision, and therefore one cannot impose exactly the requirement that \acp{GW} are hosted in galaxies as the galaxy catalog method does. In fact, much of our formalism relies on the central limit theorem to have many objects per voxel, which (along with the Limber approximation) breaks down if we choose radial bins that are too thin.
Another interesting point is that because galaxy clustering can be measured so well, this ``host requirement" ends up contributing very little to the constraining power of the cross-correlation method. In fact, we show upon implementing this method that it makes no difference to ignore it in our computation altogether. 
We discuss this at length in \cref{app:cross_shot}.

\section{Implementation on Mock Data}
\label{sec:implementation}

To test its validity and constraining power, we implement the cross-correlation framework presented above and test it on a set of mock \ac{GW} and galaxy catalogs, which we design to emulate next-generation galaxy survey and \ac{GW} detector networks. In this section, we will describe the details of the mock data generation and numerical implementation.

\subsection{Mock Data Generation}

The mock data generation pipeline can be broken down into four steps: (1) generation of a matter density field from a fiducial power spectrum used as a probability distribution to generate a Poisson-sampled galaxy field; (2) galaxy catalog generation by applying selection effects with a magnitude cut and photometric redshift errors; (3) \ac{GW} catalog generation via injection of \acp{BBH} drawn from a fiducial population model into galaxy hosts drawn with rate weighting from the galaxy catalogue, before applying selection effects with an \ac{SNR}; and (4) construction of the data products needed for the cross-correlation inference, namely the radial binning scheme, estimated localization errors (from simulated Fisher matrix posteriors) and shot noise per bin, and the sky map for each bin constructed from the \ac{MAP} positions, from which we compute the cross-correlation estimators $\{\hat{\tilde{C}}_\ell^{[\alpha\beta]}\}$. In all steps of the simulation pipeline, we adopt the flat Planck 2015 cosmology \cite{2016A&A...594A..13P} as our fiducial cosmology, which we aim to recover.

\subsubsection{Density field generation}

We use a linear matter power spectrum computed with the Boltzmann code \texttt{CLASS} \cite{2011arXiv1104.2932L,2011JCAP...07..034B} in our fiducial Planck 2015 cosmology. 
Because the theory presented in \cref{sec:theory} is valid only in the linear regime, we generate a simplified mock density field by drawing from a log-normal field \cite{1991MNRAS.248....1C,2017JCAP...10..003A,2024JCAP...12..013L}, which imposes non-negative densities such that $\delta > -1$.
The procedure we follow closely follows that of \cite{2024JCAP...12..013L}, so we will only summarize it briefly. The matter density field is generated by drawing a \ac{GRF} with a modified \ac{2PCF}
\begin{equation}
    \xi^\prime(r) = \ln[1 + \xi(r)]
\end{equation}
where $\xi$ is the \ac{2PCF} of the target power spectrum, while $\xi^\prime$ is the power spectrum from which we draw a \ac{GRF} $\delta^\prime$. (The \ac{2PCF} is the Fourier pair of the 3D power spectrum; we transform between the two using the package \texttt{mcfit} \cite{2019ascl.soft06017L}). We generate $\delta^\prime$ in Fourier space by drawing a complex field from white noise (enforcing the reality condition $\delta_{\veck}^* = \delta_{-\veck}$), reweighting it with the power spectrum, and taking the inverse discrete Fourier transform to transform the field back to real space. Then, we compute the field
\begin{equation}
    \delta_\vecr = \exp\left(\delta^\prime_\vecr - \frac{\sigma^2_{\delta^\prime}}{2}\right) - 1,
\end{equation}
where $\sigma^2_{\delta^\prime}$ is the variance of the density field $\delta^\prime$. The field respects $\delta_\vecr > -1$, and one can check that $\delta_\vecr$ has the 2-point statistics of the target power spectrum $\xi(r)$.

Because our goal is to generate a galaxy catalog, we draw a log-normal field with target power spectrum $P^g(k) = (b^g)^2 P^m(k)$ as our galaxy probability density field. Here, $b^g=1.1$ is our fiducial galaxy linear bias, and $P^m(k)$ is the linear matter power spectrum from \texttt{CLASS}. This field is generated on a cube of size $12$ Gpc divided into $2400^3 \sim 10^{10}$ cubic voxels of size $5$ Mpc. We are unable to generate a larger volume due to computational constraints; the voxel size is chosen to coincide with scales at which non-linearities become important and our simple linear formalism begins to break down.
Therefore, although a log-normal field is a very rough approximation to the true distribution of galaxies on non-linear scales, our mock galaxy density field is realistic on the scales that can be probed by this analysis. Furthermore, modeling non-linearities in the galaxy power spectrum for precision cosmology is an active area of research (see, e.g., \cite{2022MNRAS.516.5355A,2019PhRvD.100l3540W,2011MNRAS.415.3649V,2019OJAp....2E...4C}), and therefore extensions of this analysis to the non-linear regime will be left for future work.

To generate the Poisson-sampled galaxy field, we assume a uniform mean galaxy number density of $\langle n^g \rangle = 2 \cdot 10^{-3}\Mpc^{-3}$ \cite{2015MNRAS.454.2770T}. We then assign each voxel a galaxy number count by drawing from a Poisson distribution for each voxel. A complete galaxy catalog is then generated by assigning each galaxy a random position within the voxel and a $K$-band absolute magnitude drawn from a Schechter luminosity function \cite{1976ApJ...203..297S} with parameters $M_*=-23.39$, $\alpha=-1.09$, and luminosity cutoff $L > 0.01754 \, L_*$, following \cite{2023ApJ...949...76A,2001ApJ...560..566K}. Since we place the observer at the center of the box, we remove all galaxies outside the radius $\chi_{\max} = 6\Gpc$ ($z_{\max} = 2.52$, $d_{L, \max} = 21.1\Gpc$) to maintain isotropy. 

\subsubsection{Galaxy catalog}

\begin{figure}[t]
\centering
\includegraphics[width=0.47\textwidth]{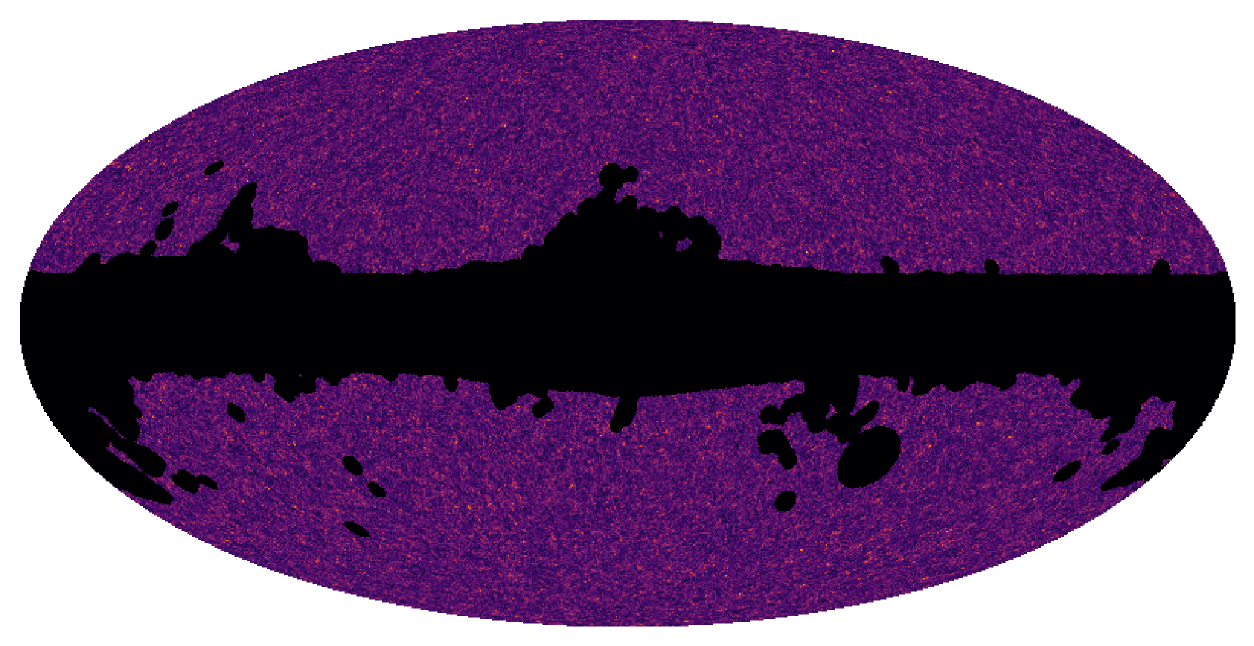}
\caption{A slice of the mock galaxy catalog, with measured redshifts $1.17 < \hat{z} < 1.22$. The galaxy sky mask, which we have taken from the WISE x SuperCosmos analysis, is clearly visible. \label{fig:gmap}}
\end{figure}

We generate the \textit{observed} galaxy catalog by computing the $K$-band apparent magnitude with the color and evolution corrections in \cite{2001ApJ...560..566K}. To simulate selection effects, we impose an apparent magnitude cut of $m < 26$, which is achievable (not necessarily in the K-band) by next-generation galaxy surveys such as Euclid \cite{2022A&A...662A.112E} and LSST \cite{2019ApJ...873..111I,2025ApJS..281...54C}. 

We also simulate galaxy survey limitations by masking out typical areas of stellar and galactic contamination using the galaxy mask from the WISE x SuperCOSMOS galaxy catalog data release \cite{2016ApJS..225....5B}, which masks out 32\% of the sky.
With these selection effects, we have an observed catalog with $\sim 5 \cdot 10^8$ galaxies out to redshift $3$. A number counts map of a representative slice of the mock galaxy catalog is shown in \cref{fig:gmap}.

Finally, we assign each galaxy a photometric redshift, assuming Gaussian measurement errors and a relative redshift error of $\sigma_{\ln z}=0.05$ \cite{2022A&A...662A..93E,2024ApJ...964..191B,2025arXiv250410482P}.

\subsubsection{BBH injections} \label{subsubsec:BBH_inj}

\acp{BBH} are drawn from a fiducial population model consistent with LVK population analyses \cite{2023PhRvX..13a1048A,2026ApJ..1005L..51A}, the full details of which we describe in \cref{app:BBH_pop}. 
Each event is assigned a host galaxy drawn from the mock (complete) galaxy catalog, weighted by the redshift-dependent rate in our fiducial model. 

To generate the observed catalog, we simulate a detector network of both \ac{ET} and {CE}. In particular, we consider 2 L-shaped \ac{ET} detectors each with 15 km arms offset in orientation by $45^\circ$, located in the candidate sites in Sardinia and in the Meuse-Rhine region \cite{2023JCAP...07..068B}; we choose this configuration over the triangular configuration as the latter has potentially highly degenerate localization modes \cite{2025PhRvD.112j3015S}. For \ac{CE}, we consider a $40$-km detector in the Idaho candidate site \cite{2021arXiv210909882E,CE_PSD}. 

We use \texttt{gwfast} \cite{2022ApJS..263....2I} with the \texttt{IMRPhenomHM} waveform model \cite{2018PhRvL.120p1102L} to generate approximate \acp{SNR} and Fisher-matrix based posteriors. For each event, we:
\begin{enumerate}
    \item Compute the Fisher matrix at the true \ac{BBH} parameters
    \item Draw a mock \ac{MAP} measurement from a multivariate Gaussian centered at the true parameters with the inverse Fisher matrix
    \item Compute the \ac{SNR} and covariance given by the inverse Fisher matrix at the \textit{measured} parameters, imposing a threshold of $\text{SNR} \geq 25$ to count the event as observed
\end{enumerate}
Crucially, our posterior measurements are not centered at the true parameters. Although this procedure does not exactly simulate a true Bayesian parameter estimation of noisy strain data, it is self-consistent if the Fisher matrix does not change significantly between the true and observed parameters, and in this high-\ac{SNR} regime it is good enough for verifying the validity of our formalism. 
We verify the self-consistency of this procedure by checking for biases in the recovered posteriors, which we do not see.

\begin{figure}[t]
\centering
\includegraphics[width=0.45\textwidth]{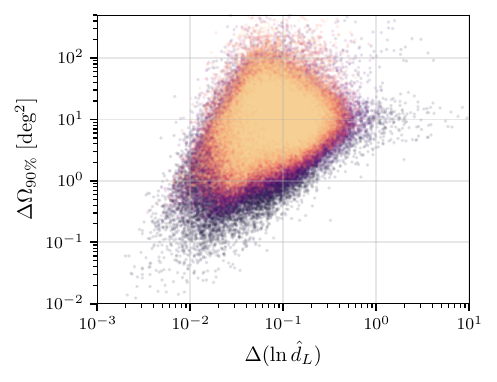}
\caption{90\% localization sky area and relative luminosity distance error $\Delta(\ln \hat{d}_L) = \Delta d_L / \hat{d}_L$ of the \acp{BBH} in the mock catalog. Colors correspond to the \ac{GW} radial bins in the top panel of \cref{fig:window}, i.e., the observed source distance $\hat{d}_L$. \label{fig:localizations}}
\end{figure}

The angular sky localizations and the relative luminosity distance errors of the mock \ac{BBH} catalog are shown in \cref{fig:localizations}. With a local merger rate of $20\Gpc^{-3}\,\mathrm{yr}^{-1}$ and a 2 year observing run, we have a total of $233499$ mock \acp{BBH} observed by our 2ET+1CE detector network. 

\subsubsection{Mock maps and data products} \label{subsubsec:data_products}

To construct the overdensity fields, we first choose a binning scheme based on the radial localization scale of each tracer. For both \ac{GW} and galaxies, we use tophat window functions in the observed coordinates, i.e., the \ac{MAP} luminosity distances for the \acp{GW} and photometric redshifts for the galaxies. For each tracer, we choose the number of bins by determining the highest number at which more than 75\% of the bins have mean redshift errors smaller than the bin half-width. Following \cite{2025arXiv250410482P,2021A&A...655A..44E}, the bin edges are chosen such that the bins are equally populated in order to maximize signal to noise. 

Recall that to compute the effective radial window functions that weight the power spectrum in \cref{eq:limber}, we need to convolve the tophat window functions with the radial localization errors, and then weight it by the radial distribution of the catalog, as in \cref{eq:eff_window_func,eq:g_eff_window_func,eq:GW_eff_window_func}. 
In \cref{subsec:radial_loc}, we proposed a strategy of empirically fitting the distribution of $P(\Delta \ln \hat{d}_L)$ for each bin. Since we have $\mathcal{O}(10^4)$ \acp{BBH} per radial bin, we can do so easily with a Gaussian \ac{KDE}. 
We also assume that we are able to obtain an estimate of the true radial distribution of observable tracers \textit{a priori}, a reasonable assumption given $\mathcal{O}(10^9)$ total galaxies and $\mathcal{O}(10^5)$ \acp{BBH}. Thus, we fit \acp{KDE} to the true galaxy redshifts and \ac{BBH} luminosity distances to pass into \cref{eq:eff_window_func}.

\begin{figure}[t]
\centering
\includegraphics[height=8cm]{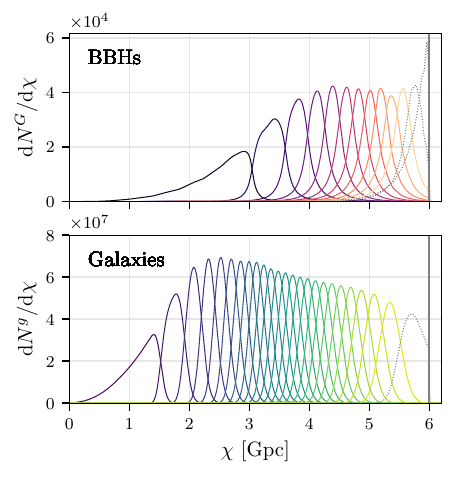}
\caption{Galaxy and \ac{BBH} effective window functions for our equally populated bins evaluated in the fiducial cosmology, normalized to the number count per bin. Dotted lines show bins that were removed due to being too close to the edge of the simulation volume at $\chi_\text{max}=6$ Gpc. \label{fig:window}}
\end{figure}

\cref{fig:window} shows the resulting effective window functions in comoving distance for the \acp{BBH} and galaxies, computed from the true cosmology; these window functions are computed on-the-fly during inference to convert between different distances. The overlapping tails are a result of convolving the tophat window functions with the localization error. Because we choose the binning of \acp{GW} and galaxies independently, they do not have a 1-to-1 correspondence; the distribution of \acp{GW} is visibly skewed towards higher redshifts relative to the galaxies. 
This is because the sensitive volume of \ac{GW} detectors is deeper than the galaxy survey, which is true for both current-generation and next-generation \ac{GW} and galaxy surveys.
Also note the hard boundary at $\chi_{\max}=6$ Gpc; since this artificially squishes the window functions near the boundary, we remove bins that are centered at $\chi > 0.9\chi_{\max}$, which are shown in \cref{fig:window} by dotted lines. In the end, we are left with 11 bins of $18,000$ \acp{BBH} each, and 24 bins of $2 \cdot 10^7$ galaxies each.

\begin{figure}[t]
\centering
\includegraphics[width=0.46\textwidth]{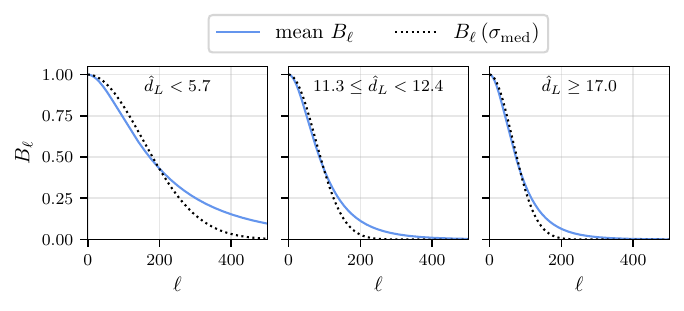}
\caption{Beam window functions for the 1st, 6th, and last of our 11 \ac{GW} bins. The solid blue line shows the average of $B_\ell$ over all events, while the dotted black line shows $B_\ell$ computed from the median angular \ac{RMS} localization error of each bin. \label{fig:beam}}
\end{figure}

For the \acp{BBH}, we account for their localization errors by computing the mean beam window function for each radial bin, as discussed in \cref{subsec:ang_loc}. As suggested in \cref{eq:mean_beam_approx}, we do this by computing $B_\ell(\sigma_i)$ for each event $i$ in the bin with the Gaussian beam window function in \cref{eq:Gaussian_beam} and taking the average over the events, where $\sigma$ is the \ac{RMS} localization error\footnote{Because events are not perfectly circular, this is computed as the geometric average of the square root of the eigenvalues of the $(\theta, \phi/\sin\theta)$ covariance matrix. All of the observed events have near-circular localization ellipses, with a median eccentricity of $0.98$.} of each event. Some resulting computed beams are shown in \cref{fig:beam}, compared with the result from approximating all events as having the same localization error set equal to the sample median.

Finally, we compute the \ac{GW} and galaxy overdensity maps and cross-correlations. Number count maps are constructed for each bin with the \ac{MAP} sky localization of the \acp{BBH} and true sky localization of the galaxies (we assume negligible localization error with respect to the pixel size). We use a \texttt{HEALPix} resolution of $\texttt{nside}=256$, corresponding to a conservative maximum resolution of $\ell_{\max} \approx 512$, well beyond what can be probed by the \ac{GW} localizations as seen in \cref{fig:beam}. Overdensity maps are created by normalizing the number count maps to the mean density. Finally, cross-correlation power spectra $\{\hat{\tilde{C}}_\ell^{\alpha\beta}\}$ are computed using \cref{eq:naive_estimator} for $\alpha\geq\beta$, where $\alpha, \beta$ index the $11+24$ total \ac{GW} and galaxy bins; the effect of the galaxy mask is then undone by inverting the mode-coupling matrix, which we compute with \texttt{NaMaster} \cite{2019MNRAS.484.4127A}. We use \texttt{healpy} \cite{2019JOSS....4.1298Z} for all spherical harmonic and angular power spectrum computations.

\subsection{Inference implementation}

We infer cosmological parameters $\vecLam_c = (H_0, \Omn)$; all other cosmological parameters are fixed at the fiducial Planck 2015 values. We use a simple linear bias $b=b^g=b^G$, assuming that the galaxy and \ac{GW} biases are the same and do not evolve with redshift, consistent with our assumptions. We also assume that the matter power spectrum is both fixed and known, such that we do not need to re-compute it on the fly during inference for different cosmologies. 
In a realistic analysis, none of these are a given and we would have further nuisance parameters to model these effects. Nonetheless, the goal of this work is to demonstrate the degree to which this method is precise and unbiased within the assumptions of our simplified mock catalogs, and thus we leave simulating a more sophisticated \ac{GW}--galaxy cross-correlation to future work.

Finally, we implement the Bayesian inference of the following posterior:
\begin{equation} \label{eq:xcorr_hyperposterior}
\begin{split}
    &\mathcal{P}(\vecLam |\{\vecd^G\}, \{\vecd^g\} ) \\
    & \qquad \quad \propto \Pi(\vecLam) \mathcal{L}\left(\left\{\hat{\tilde{C}}_\ell^{[\alpha\beta]}\right\}| \{\vecd\}, H_0, \Omn, b\right)
\end{split}
\end{equation}
where the likelihood on the observed angular power spectra $\hat{\tilde{C}}_\ell^{[\alpha\beta]}$ is our usual multivariate Gaussian in \cref{eq:Clhat_Gaussian_ll}. 
Here, $\{\vecd\}$ includes all the auxiliary data products described in the previous section, including the number density per bin for the shot noise computation, and the localization uncertainties encoded in the beam and radial window functions. These, along with our cosmology and bias parameter, are sufficient to compute the theoretical expected power spectrum and covariance matrix as follows:
\begin{enumerate}
    \item Using the cosmology, bias parameter, and the \ac{GW} and galaxy window functions, compute the theoretical angular power spectrum with the Limber approximation in \cref{eq:limber}
    \item Correct for \ac{GW} angular localization errors by multiplying by the appropriate beam window functions in \cref{eq:hatClGG_with_angloc,eq:hatClGg_with_angloc}
    \item Add the shot noise $\delta^{(K)}_{\alpha\beta}/\langle n^\alpha\rangle$
    \item Compute the covariance matrix with \cref{eq:Cl_covmat} from this error-corrected theoretical power spectra
    \item Divide all covariance matrix terms that involve a galaxy field by \fsky, as described at the end of \cref{subsec:aniso_sky}
\end{enumerate}
This is repeated at each step of the inference to evaluate the likelihood for each sampling of the cosmology and tracer bias.
As discussed at the end of \cref{sec:posterior} and at length in \cref{app:cross_shot}, we take the cross shot noise between \acp{GW} and galaxies to be negligible, which otherwise would be added between steps 1 and 2 above. 
We use uniform priors of $H_0 \in [65, 77]$, $\Omn \in [0.25, 0.35]$, and $b \in [0.7, 1.3]$.

We start the analysis at $\ell_\text{min}=20$ and set the cutoff $\ell_{\max}$ of the inference from the \ac{GW} localization resolution. For each \ac{GW} bin $\alpha$, we compute the maximum $\ell$ such that the beam $\langle B_\ell \rangle_\alpha \geq 1/e$; this goes from $\ell_{\max}^{[1]}=229$ for the first bin to $\ell_{\max}^{[11]}=93$ for the last bin. 
Upon computing the $\ell$ corresponding to the non-linear scale (i.e., Nyquist frequency) $k_\text{max}=0.1\Mpc^{-1}$ at the midpoint distance of each shell, we find $\ell_{\max, \text{nl}}$ starting from $262$ at the first shell and increasing onwards, and thus the $\ell_{\max}$ set by the localization resolution does not reach non-linear scales. We use $\ell_{\max}=229$ to analyze all power spectra in the analysis, as higher modes of the cross-correlation are not informative for any \ac{GW} bin.

\begin{figure*}[t]
    \centering
    \begin{subfigure}[b]{0.5\textwidth}
        \includegraphics[height=7.9cm]{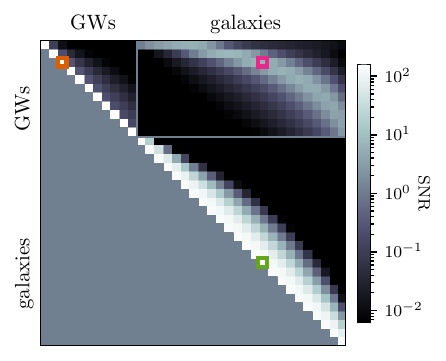}
        \caption{\label{subfig:cl_snr}}
    \end{subfigure}
    \hfill
    \begin{subfigure}[b]{0.49\textwidth}
        \includegraphics[height=7.6cm]{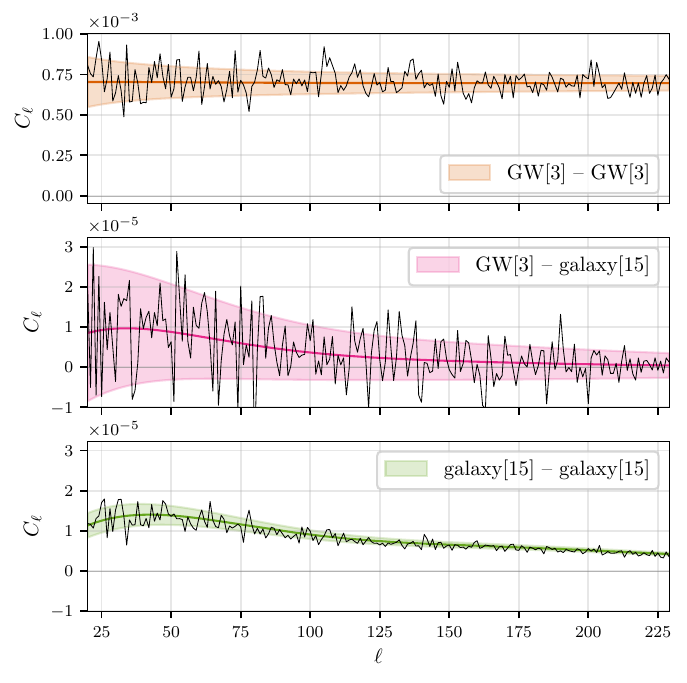}
        \caption{\label{subfig:cl_plots}}
    \end{subfigure}
    \caption{\textbf{(a)}: Theoretical SNR of the cross-correlation signal for each pair of radial bins $(\alpha, \beta)$ for $\alpha\geq\beta$. Highlighted pixels correspond to the power spectra plotted in (b). \textbf{(b)}: From top to bottom, select \ac{GW}--\ac{GW}, \ac{GW}--galaxy, and galaxy--galaxy power spectra; the bottom two panels are plotted on the same scale. The colored curves show the theoretical error-corrected power spectra and $1\sigma$ error bars, while the power spectra estimated from our mock data are plotted in black.}
    \label{fig:cl_plots}
\end{figure*}

\cref{fig:cl_plots} shows some theoretical power spectra computed using this framework for the true cosmology. In \cref{subfig:cl_snr}, we show a grid of error-corrected theoretical $C_\ell^{[\alpha\beta]}$; for each pair of bins $(\alpha, \beta)$ with $\alpha\geq\beta$, we plot the total \ac{SNR} across angular modes
\begin{equation}
\text{SNR} = \sqrt{\sum_{\ell=\ell_{\min}}^{\ell_{\max}} \frac{\left(C_\ell^{\alpha\beta}\right)^2}{\left\langle \left(C_\ell^{\alpha\beta}\right)^2 \right\rangle}}.
\end{equation}
The top left and bottom right triangles show the \ac{GW}--\ac{GW} and galaxy--galaxy auto-power spectra, which have the highest \ac{SNR} along the diagonal, and smaller but nonzero signals for nearby bins which overlap. The top right rectangle shows the grid of \ac{GW}--galaxy cross-correlations, in which a band showing the $d_L-z$ relation for the given cosmology is visible. 
This curve is not diagonal because we have chosen the \ac{GW} and galaxy bins separately, and thus are not spaced proportionately to each other as seen in \cref{fig:window}. 
One can also see the \ac{GW}--galaxy cross-correlations are only modestly well-measured, with a total \ac{SNR} of $30$ across all bins together, compared to an \ac{SNR} of $800$ for the galaxy--galaxy auto-power spectra.

To demonstrate this more clearly, we plot in \cref{subfig:cl_plots} the theoretical power spectra and corresponding error bars for the \ac{GW}--\ac{GW}, \ac{GW}--galaxy, and galaxy--galaxy power spectra in a representative \ac{GW} and galaxy bin, computed with the true cosmology and bias $b=1.1$. These three panels correspond to the highlighted pixels in \cref{subfig:cl_snr}; the \ac{GW} and galaxy bins are centered at $d_L \approx 8.2\Gpc$ and $z\approx1.2$, respectively, and thus their overlap generates a measurable cross-correlation. We also plot in black the power spectrum estimators computed from our mock data, verifying the goodness of fit of the theoretical prediction. 
Notice also the scale on the y-axes: $C_\ell^{GG}$ is large because it is shot noise dominated, while $C_\ell^{Gg}$ and $C_\ell^{gg}$ are signal dominated, i.e., $\overline{C}_\ell^{gg} \gg 1/\langle n^g \rangle$ for any galaxy bin. 
Thus, the \ac{GW}--\ac{GW} auto-correlation contributes little to no clustering information for the inference. 
Furthermore, the largeness of the \ac{GW} shot noise is the main contributor for the large uncertainty in the \ac{GW}--galaxy cross correlation.
The theoretical expectation for $C_\ell^{Gg}$ and $C_\ell^{gg}$ are on the same scale, differing by a factor of $B_\ell$ from the \ac{GW} localization uncertainty. 
A plot breaking down the contributions to the total cross-correlation from the shot noise and the signal $\overline{C}_\ell$ can be  found in \cref{app:cross_shot}.

To implement these computations, we use the architecture of \texttt{jax\_cosmo} \cite{2023OJAp....6E..15C}, a GPU-accelerated and differentiable cosmology library written in \texttt{jax}. 
In particular, we adapt their implementation of the Limber-approximated angular power spectra and computation of covariance matrices, and build upon their architecture of using container objects for the cosmology, bias parameters, and probes such that key \texttt{jax} features such as auto-differentiation and \texttt{vmap} (a wrapper that can vectorize any operation written in \texttt{jax}) are available to them.
Because of this, we are able to use \ac{HMC} sampling \cite{1987PhLB..195..216D,2011hmcm.book..113N,2017arXiv170102434B}, which efficiently explores the posterior space using gradient-based methods.
For our inference, we use the \texttt{NumPyro} \cite{2019arXiv191211554P,2025ascl.soft05005P} implementation of the No-U-Turn Sampler \cite{2011arXiv1111.4246H}, which is a flavor of \ac{HMC} that adaptively sets the path length.

\section{Results and Discussion} \label{sec:results}

\begin{figure}
\centering
\includegraphics[width=0.47\textwidth]{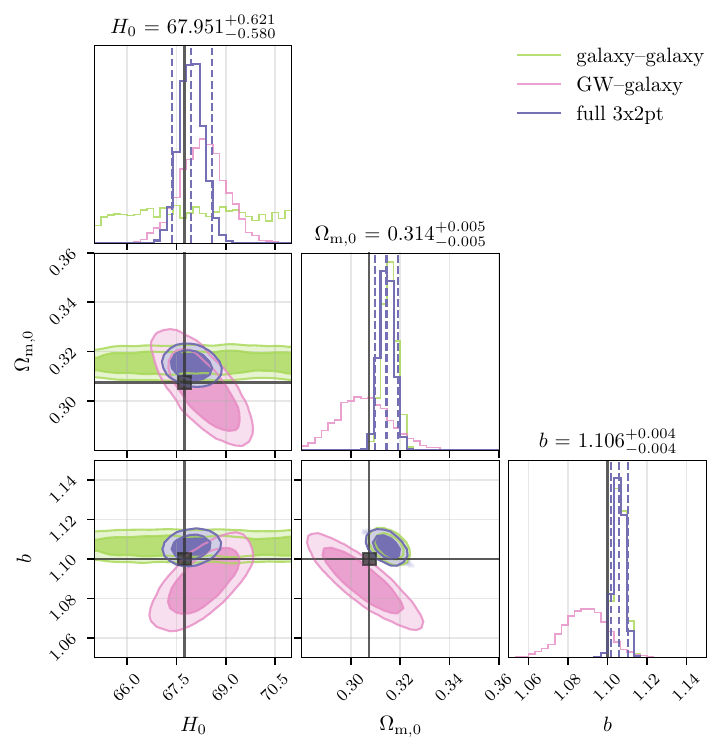}
\caption{Corner plot for $H_0$ (in km/s/Mpc), \Omn, and the tracer bias $b$. We overplot results from an analysis that uses only the galaxy--galaxy correlations, an analysis that uses only the \ac{GW}--galaxy cross-correlations, and the result from the full 3x2pt analysis.\label{fig:corner}}
\end{figure}

In \cref{fig:corner}, we present the posterior constraints on all of our population parameters --- $H_0$, \Omn, and $b$. 
Contours show the 50\% and 90\% constraints, while the 1D histograms show the 90\% credible intervals.

The results from the full 3x2pt analysis, corresponding to the posterior in \cref{eq:xcorr_hyperposterior}, are shown in dark purple. We also show the results from an analysis that uses only the galaxy--galaxy (green) or only the \ac{GW}--galaxy power spectra (pink), i.e., we marginalize over the other power spectra by slicing the covariance matrix in \cref{eq:Cl_covmat}. 
Note that because we marginalize over the galaxy power spectra instead of conditioning on the galaxy data, there is some overlapping information between the pink and green curves.
We can see that most of the constraints on $H_0$ itself come from the \ac{GW}--galaxy cross-correlation, but adding the galaxy--galaxy correlations tightens the constraint on \Omn, thus tightening the constraint on $H_0$ in the full analysis. This synergy is consistent with the Fisher-matrix based forecast in \cite{2025arXiv250410482P}.

In all three analyses, $H_0$ is recovered correctly and well within the $90\%$ credible intervals. Note also the precision: both the \ac{GW}--galaxy only and full 3x2pt analysis are able to achieve the percent level precision of modern cosmological experiments. In particular, the \ac{GW}--galaxy cross-correlation and full 3x2pt analysis measure $H_0$ to a $1\sigma$ precision of 1\% and 0.5\%, respectively. \Omn is also measured well, with a $1\sigma$ precision of 5\% and 1\%, respectively.

There is a small ($\sim 2\sigma$) bias in \Omn, driven by the galaxy--galaxy correlations preferring a higher value of \Omn. 
We show in \cref{app:lin_corr} that this bias is removed if we add a first-order correction in $b(k)$, indicating some kind of unmodeled error at small scales. Possible sources of this error include the spectral leakage and non-Gaussianities of the log-normal field discussed earlier, as well as the breakdown of the Limber approximation due to the relatively thin galaxy bin widths.
Because there are so many galaxies, and because galaxies have virtually no angular localization errors, the galaxy--galaxy power spectrum is extremely well measured at a total \ac{SNR} of $800$ across all bins, compared to $30$ for the \ac{GW}--galaxy cross power spectrum.
Because of this, the measurement precision can often approach or exceed unmodeled systematics. Since the number of independent modes scales linearly with $\ell$, small scales in particular contribute more information. Thus, one must have some caution when including information from galaxy--galaxy correlations, a lesson well known in the galaxy survey literature. Therefore, we do not find the bias in \Omn to be too concerning in this work.
Crucially, the posteriors on both $H_0$ and \Omn when we use the \ac{GW}--galaxy cross-correlation alone are unbiased, making the cross-correlation method a promising avenue for next-generation precision cosmology. 

Note that \cref{eq:xcorr_hyperposterior} does not include any information from a spectral sirens inference; recall that in the limit of large $N^G$, the spectral-sirens and cross-correlation likelihoods are decoupled. The goal of this example is to probe the effectiveness of the cross-correlation inference, and a study of the full joint posterior in \cref{eq:full_unified_hyperposterior} is left for future work.

Finally, we would like to highlight the speed of this analysis, which we see as among the primary advantages of this method in the \ac{3G} era of \ac{GW} detectors.
While the dimensionality of our posterior space does not necessarily require \ac{HMC},  GPU acceleration and \texttt{jax}'s XLA (accelerated linear algebra) compiler allows the sampling to be run in less than an hour.
Furthermore, we make a series of approximations that eliminates the need for keeping thousands of \ac{PE} samples per event at inference by binning and compressing the data into a few auxiliary data products.
Once this (relatively cheap) upfront computational cost is paid, the inference runtime is roughly constant with the number of \ac{GW} sources and galaxies.
Current methods for analyzing LVK data, involving many products of Monte-Carlo integrals, is known to scale very poorly as our catalog grows and measurement precision improves \cite{2023MNRAS.526.3495T}. Furthermore, current standard galaxy catalog methods are already quite computationally costly, since they involve pre-computing the entire redshift prior for $p^g(z, \vecOm)$  to be stored in memory at runtime, which also requires one to fix \Omn \cite{2023JCAP...12..023G,2020PhRvD.101l2001G,2023ApJ...949...76A,2024ApJ...964..191B,2025arXiv250904348T}. 
The cross-correlation method therefore provides a solution for cheap and accurate \ac{GW} cosmological inferences in the \ac{3G} era.

\section{Conclusions} \label{sec:conclusions}
We are entering an era in which it is increasingly possible to probe the cosmological properties of the Universe with multiple tracers alongside an ever-increasing catalog of galaxies. Significant upgrades of \ac{GW} detectors, in particular \ac{3G} detectors, would represent a huge leap in the power of multi-tracer cosmology, with tens of thousands of detected \acp{BBH} anticipated per year. Despite this, techniques for joint cosmological inference with \acp{GW} and galaxies are wide ranging, from the standard galaxy catalog method of statistically assigning each \ac{GW} a host galaxy to a variety of flavors of 2-point analyses, which are often at odds with standard \ac{GW} population analysis techniques.
In the era of precision cosmology, it is crucial to have a systematic understanding of these dark siren methods and their relationships.

Thus, we have set out in this paper to accomplish the following: 1) provide a pedagogical guide to the cross-correlation method by rigorously deriving the cross-correlation method from first principles (\cref{sec:theory,sec:unified}), including novel results on how to treat \ac{GW} measurement uncertainties (\cref{sec:noise}); and 2) describe the relationship between the cross-correlation method, the standard galaxy catalog method, and the spectral sirens method, and outline a path to a joint inference (\cref{sec:unified,sec:posterior}).

Our primary theoretical results are as follows:
\begin{itemize}
    \item The primary difference between the cross-correlation and standard galaxy catalog method is the information we condition upon. The standard formalism for analyzing a set of observed \ac{GW} sources assumes a fixed distribution of the source parameters ($d_L, \vecOm, m_1, m_2, \ldots$). In order to use this formalism, the standard galaxy catalog method implicitly conditions upon a known background galaxy distribution, usually by assuming a uniform-in-volume distribution for the unobserved galaxies. On the other hand, if we condition only upon the cosmology, the \acp{GW} are a doubly stochastic process: \acp{GW} are randomly drawn from a galaxy field which itself is a \ac{GRF}. To maximize the effectiveness of dark siren analyses with galaxies, one should either marginalize over all possible realizations of the missing galaxies, or more simply perform a joint analysis of the galaxies and \acp{GW} as \acp{GRF}, which is what the cross-correlation method does.
    \item The cross-correlation technique is generally thought to impose a weaker requirement than the standard galaxy catalog method, requiring only that the \acp{GW} trace the same density field as the galaxies instead of necessarily being hosted in galaxies. We showed \changed{in \cref{sec:unified,app:gcat_field_bridge}} that it is possible to impose this requirement exactly in the infinitesimal limit, by showing that the standard galaxy catalog method can be derived from the cross-correlation method under the assumption that the total galaxy field is known. In practice, however, we can only impose the host requirement on the scales that the events are measured, due to the localization errors of the tracers and the discretization of the field into radial bins and pixels. Thus, the constraining power of the cross-correlation method is largely derived from 2-point statistics independent of the constraining power of the standard galaxy catalog method.
    \item \changed{The cross-correlation method utilizes a summary statistic, which are the 2-point statistics between \acp{GW} and galaxies. In the infinitesimal limit, this compression is not lossy insofar as the fields are Gaussian and hence completely described by their 2-point statistics. Intuitively, all cosmological information is given by how correlated the two fields are summed over the entire sky at each given distance \cite{1997PhRvD..55.5895T}. Non-Gaussian information, as well as any information beyond the truncation resolution, are lost in this compression.}
    \item The cross-correlation method acts as a generalized angular distribution at a given radius $p(\{\vecOm\}, \{\vecOm^g\}|z)$. This is because the cross-correlation power spectra measure correlations between overdensity fields, which are by construction normalized to the rate at each distance. Therefore, they provide exactly complementary information to the spectral sirens method, which have no angular dependence. A corollary of this is that the cross-correlation population likelihood has radial selection effects folded in, and thus selection effects in a joint spectral sirens + cross-correlation analysis are only explicitly manifest in the former.
    \item Because of prohibitive computational costs of the cross-correlation method with Monte Carlo integration, we propose using an aggressive \ac{SNR} cut ($\gtrsim 25$) to analytically build localization errors into the theoretical prediction by modeling the likelihood as Gaussian in this limit.
    \item \ac{GW} angular localization errors are not the same process as a telescope point spread function and should not be treated as such. We show that if we model each event in a radial bin as being displaced from its true angular sky position by a draw from a homogeneous, isotropic, and Gaussian distribution, then the effective beam window function that multiplies the measured cross-correlation is the expectation value of an ensemble of Gaussian beams corresponding to the localization errors of events in the bin. In this model, contrary to current implementations of the cross-correlation method in the literature, the observed \ac{GW} overdensity field should be constructed from its maximum likelihood measurement, rather than the posterior draws themselves.
    \item The cross-correlation formalism relies on many assumptions, requiring at least $\sim$hundreds of events per bin with high \ac{SNR} and angular localizations exceeding $\ell \gtrsim 20$ ($\sim 3^\circ$ resolution) to satisfy the central limit theorem for both the shot noise and the likelihood of the power spectrum estimator, as well as to be able to estimate the distribution of localization errors directly from the data. We anticipate that these conditions will be difficult to meet before \ac{3G}, and encourage research directions of examining if or when these assumptions can be relaxed. To our knowledge, these caveats have not been addressed in the \ac{GW} cross-correlation literature, despite the usage of many elements of this formalism that rely on these assumptions.
\end{itemize}

We also validated our formalism by presenting an implementation on \ac{3G}-like data, forecasting its constraining power (\cref{sec:implementation,sec:results}).
We generated a mock galaxy catalog by drawing from a fiducial 3D matter power spectrum, and generated a mock \ac{BBH} catalog by computing Fisher matrix posteriors of \acp{BBH} drawn from these galaxies, and imposing selection effects for both. 
We then recovered the fiducial cosmology with a Bayesian inference on the mock data. 
To our knowledge, this is the first time that the cross-correlation method has been validated in a full Bayesian inference setting with realistic \ac{GW} localization errors, in which both the measured luminosity distances and angular sky locations are displaced from their true values \changed{and then self-consistently marginalized over during inference.}
With a 2 \ac{ET} + 1 \ac{CE} configuration and a 2 year observing run, we find that both the full 3x2pt analysis and an analysis using only the \ac{GW}--galaxy cross-correlation power spectra achieve percent-level precision in $H_0$; the former also achieves percent-level precision in \Omn. It is the \ac{GW}--galaxy cross-correlation that drives the measurement on $H_0$, whereas \Omn is mostly driven by the galaxy--galaxy power spectra.

Furthermore, by leveraging the optimal data compression of the cross-correlation method and a GPU-accelerated implementation, we were able to run the inference in less than an hour. Thus, we demonstrate an avenue for precise and scalable cosmological inferences in the \ac{3G} era. Unlike traditional population analysis techniques, the computational complexity of the cross-correlation method is flat with respect to the number of events.

Although we generated our mock galaxy catalog from a log-normal field, we do not expect the cross-correlation results to be different for a galaxy catalog generated from, e.g., hydrodynamical simulations, because \acp{GW} are not well localized enough to probe fully non-linear scales even in \ac{3G}.
Nonetheless, we acknowledge that this forecast is still relatively simplified. 
Because the aim of this paper was to focus on the information gained from modeling \acp{GW} and galaxies as \acp{GRF}, we ignored all effects beyond this model, including non-linearities, gravitational lensing, bias evolution, \acp{RSD} and \acp{LSD}. 
We also assumed that the linear matter power spectrum was known, but in general it is sensitive to the cosmology, and thus an analysis in which the 3D power spectrum itself is computed on the fly would be an interesting direction for future research. 
Furthermore, a more realistic analysis will not be able to assume \acp{GW} and galaxies have the same non-evolving bias; \ac{GW} host galaxies may not cluster the same way as the bright galaxies in our catalog, and therefore in full generality $b^G$ and $b^g$ would have to be inferred separately at each redshift.
A more detailed forecast that includes all of the effects is the most salient next step in advancing our understanding of cosmological inferences with \ac{GW}--galaxy cross-correlations.

\begin{acknowledgments}
The authors would like to thank Matteo Tagliazucchi, Haochen Wang, Sylvia Biscoveanu, and James Sunseri for their valuable comments and discussions. This work was made possible by the German-American Fulbright Commission. A.Q.C. is currently supported by the Lowell Wood Endowed Fellowship of the Fannie and John Hertz foundation.
\end{acknowledgments}

\section*{Data Availability}

All software used for this work is made available on Github.
This includes our branch of \texttt{jax\_cosmo} \cite{my_jax_cosmo}, which includes significant modifications to implement the specific details of the \ac{GW}--galaxy cross-correlation outlined in this paper, as well as all scripts and modules for the mock data generation and inference \cite{gwx}. Finally, the data products used in our mock inference, including the mock \ac{GW} and galaxy catalogs and derived cross-correlation estimators, can be found on Zenodo \cite{data_release}.

\bibliography{bib,bib_supplement}

\appendix

\section{Recovering the standard galaxy catalog method from a GRF} \label{app:gcat_field_bridge}

In this section, we check explicitly that we can recover the standard galaxy catalog population prior under the change of variables from event-level parameters $\{\vecr\}$ to an overdensity field $\{\alm\}$ described in \cref{subsubsec:field_ll}. That is, we show that the spatial distribution $p(\vecOm|z)$ constructed from the observed galaxies in the standard galaxy catalog method can be recovered from a Gaussian distribution on the \ac{GW} overdensity field. 

Recall that the standard galaxy catalog method conditions on the \textit{total} galaxy field as a fixed distribution.
Therefore, we are interested in the population prior of a \ac{GW} field on some thin shell\footnote{Here, we are ignoring for simplicity the coupling between shells via radial modes, but the same result can be shown in that case.} at redshift $\bar{z}$, conditioned on the \ac{GW} redshifts $\{z\}$ and total galaxy field
\begin{equation} \label{eq:p_almG_cond}
\begin{split}
    &\Ppop\left(\left\{\alm^G[\bar{z}]\right\} \Big| \{z\}, \{z\gtot\}, \left\{\alm\gtot[\bar{z}]\right\}, \vecLam\right) = \\
    & \qquad \prod_{\ell, m} \Ppop(\alm^G[\bar{z}] | n^G[\bar{z}], n\gtot[\bar{z}], \alm\gtot[\bar{z}], \vecLam)
\end{split}
\end{equation}
where $\{\alm\}$ refers to the set of $\alm$ over $(\ell, m)$, taking advantage of the separation of harmonic modes. 
We model the \ac{GW} field as a Poisson realization of the total galaxy field
\begin{equation} \label{eq:GW_poisson_g}
    \alm^G[\bar{z}] = \alm\gtot[\bar{z}] + s_{\ell m}^G[\bar{z}]
\end{equation}
where $s_{\ell m}^G[\bar{z}]$ is the shot noise of \acp{GW} being drawn from the galaxies of this shell, which averaged over all possible realizations of the universe is a scale-independent shot noise 
\begin{equation}
    \left\langle \left( s_{\ell m}^G[\bar{z}] \right)^2 \right\rangle = 1/\langle n^G[\bar{z}] \rangle.
\end{equation}
Both $\alm^G$ and $\alm\gtot$ refer to \textit{realized} tracer fields, and thus $\alm\gtot$ itself also implicitly contains shot noise, i.e.
\begin{equation}
    \left\langle  \left|\alm\gtot \right|^2 \right\rangle = C_\ell^{gg} = \overline{C}_\ell^{gg}[\bar{z}] + \frac{1}{\langle n\gtot[\bar{z}] \rangle}.
\end{equation}

To compute the distribution of the \ac{GW} overdensity field conditioned on the galaxy overdensity field, we can use the standard formulae for the conditional distribution of a multivariate Gaussian. For jointly distributed vectors $\mathbf{z} \equiv (\mathbf{x}, \mathbf{y})$ with mean $\langle \mathbf{z} \rangle \equiv \left( \langle \mathbf{x} \rangle, \langle \mathbf{y} \rangle \right)$ and covariance 
\begin{equation}
\left\langle \mathbf{z}^\dagger \mathbf{z}  \right\rangle = 
\begin{pmatrix}
\covS_{\mathrm{xx}} & \covS_{\mathrm{xy}}\\
\covS_{\mathrm{xy}}^\dagger & \covS_{\mathrm{yy}}
\end{pmatrix} \equiv 
\begin{pmatrix}
\left\langle \mathbf{x}^\dagger \mathbf{x}  \right\rangle  & \left\langle \mathbf{x}^\dagger \mathbf{y}  \right\rangle \\
\left\langle \mathbf{y}^\dagger \mathbf{x}  \right\rangle  & \left\langle \mathbf{y}^\dagger \mathbf{y}  \right\rangle 
\end{pmatrix},
\end{equation}
the distribution of $\mathbf{x}$ conditioned on $\mathbf{y}$ is a multivariate Gaussian with a modified mean and covariance matrix
\begin{equation} \label{eq:cond_gauss}
\begin{split}
    \langle \mathbf{x} \rangle_{\mathbf{y}} &= \langle \mathbf{x} \rangle + \covS_{\mathrm{xy}} \covS_{\mathrm{yy}}^{-1}\left( \mathbf{y} - \langle \mathbf{y} \rangle \right) \\
    \langle \mathbf{x}^\dagger \mathbf{x} \rangle_{\mathbf{y}} &= \covS_\mathrm{xx} - \covS_\mathrm{xy} \covS_\mathrm{yy}^{-1} \covS_\mathrm{xy}^\dagger.
\end{split}
\end{equation}
Hence, if we take $\mathbf{x}$ to be our \ac{GW} field $\{\alm^G[\bar{z}]\}$ and $\mathbf{y}$ to be our galaxy field $\{\alm\gtot[\bar{z}]\}$ at redshift $\bar{z}$, we need the theoretical power spectra $C_\ell^{G\,g,\mathrm{tot}}[\bar{z}]$ and $C_\ell^{GG}[\bar{z}]$ in order to compute \cref{eq:p_almG_cond}. Using \cref{eq:GW_poisson_g}, we have
\begin{equation} \label{eq:ClGgtot}
\begin{split}
    C_\ell^{G\,g,\mathrm{tot}}[\bar{z}] &= \left\langle \alm^G[\bar{z}]^* \alm\gtot[\bar{z}] \right\rangle \\ 
    &= \left\langle \left(\alm\gtot[\bar{z}] + s_{\ell m}^G[\bar{z}] \right)^* \alm\gtot[\bar{z}] \right\rangle \\
    &= C_\ell^{gg}[\bar{z}] = \overline{C}_\ell^{gg}[\bar{z}] + \frac{1}{\langle n\gtot[\bar{z}] \rangle}
\end{split}
\end{equation}
where we have used the standard approximation that the shot noise is decoupled from the underlying field from which it is drawn $\langle s_{\ell m}^G \alm\gtot \rangle =0$, valid for $\langle n^G \rangle \gg 1$.
A similar calculation yields
\begin{equation} \label{eq:ClGG}
    C_\ell^{GG}[\bar{z}] = \overline{C}_\ell^{gg}[\bar{z}] + \frac{1}{\langle n^G[\bar{z}] \rangle}
\end{equation}
Note that these theoretical power spectra are constructed by definition from the expectation value of the fields over all realizations of the universe; we condition upon a realization of the galaxy field later using \cref{eq:cond_gauss}.

\cref{eq:ClGgtot} is what we ought to expect; \acp{GW} have the same clustering patterns as the galaxies they are drawn from, and we can recognize the cross shot noise term in the last line (as in \cref{eq:alm_with_shot}).
\cref{eq:ClGG} is also what we expect from a Poisson point process which traces the galaxies. 
Furthermore, $\overline{C}_\ell^{gg} = \overline{C}_\ell^{Gg} = \overline{C}_\ell^{GG}$ because we have assumed  that \acp{GW} are uniformly drawn from the galaxies, and that observed and unobserved galaxies do not cluster differently. A more flexible model is easily accommodated by modeling a luminosity-dependent galaxy bias along with selection effects, or more simply by allowing \acp{GW} and galaxies to have separate biases.

\cref{eq:Gaussian_alm_ll} gives us a joint multivariate normal distribution on the \ac{GW} and galaxy overdensity fields.
Then, substituting \cref{eq:ClGgtot,eq:ClGG} into \cref{eq:cond_gauss} yields our desired conditional distribution
\begin{subequations} \label{eq:ppop_ang_condall_alm}
\begin{align} 
    \Ppop(\alm^G[\bar{z}] | n^G[\bar{z}], n\gtot[\bar{z}], \alm\gtot[\bar{z}], \vecLam) &= \mathcal{N}\left[\alm^G[\bar{z}]; \mu_{\ell m}, \sigma_{\ell m} \right] \\
    \mu_{\ell m} &= \alm\gtot[\bar{z}] \\
    \sigma_{\ell m} = \frac{1}{\langle n^G[\bar{z}] \rangle} - \frac{1}{\langle n^g[\bar{z}] \rangle} &\approx \frac{1}{\langle n^G[\bar{z}] \rangle}
\end{align}
\end{subequations}
assuming $n^G \changed{\ll} n^g$. \cref{eq:ppop_ang_condall_alm} is particularly simple because $C_\ell^{G\,g,\mathrm{tot}}=C_\ell^{gg}$.
Notice that the variance of the \ac{GW} field is now simply a scale-independent shot noise. Thus, we have now showed explicitly that conditioning on a fixed background of galaxies \changed{makes redundant information from} 2-point correlations \changed{of the probability density field}, as discussed in \cref{subsec:bridge}.

Transforming \cref{eq:ppop_ang_condall_alm} to real space, one can show that the probability of a \ac{GW} being found in a given infinitesimal shell $\bar{z}$ \changed{of width $\delta z$} and pixel $\bar{\vecOm}$ of \changed{size $\delta \vecOm$} is given by 
\begin{equation}
\begin{split}
    \Ppop[N^G(\bar{z},\bar{\vecOm}) = 1]  &=  \changed{\left\langle \frac{ \di n^G[z] }{\di z} \Bigg|_{\bar z} \delta z\right\rangle} [1 + \Delta\gtot(\bar{\vecOm} | \bar{z})]\,\delta\vecOm \\
    &= \deriv{N^G}{z}\bigg|_{\bar{z}}\frac{1}{4\pi}\left[ 1 + \Delta\gtot(\bar{\vecOm} | \bar{z}) \right] \delta z \, \delta\vecOm
\end{split}
\end{equation}
where $\delta z$ and $\delta \vecOm$ are the shell width and pixel size, respectively. 
Taking the infinitesimal limit gives us the continuous \ac{GW} spatial probability distribution of the standard galaxy catalog method, i.e., \cref{eq:gcat_basic}.
One can substitute in the expression for $\Delta\gtot$ in \cref{eq:pgal_ang_fields} and take $\Delta\gmiss = 0$, for example, for the ``homogeneous completion" implementation of the galaxy catalog method.
Thus, a spatial population prior informed by the galaxies $p(z, \vecOm | \vecLam, \dg)$ to be evaluated on an event-by-event basis, as in standard dark siren methods, can be recovered from the Gaussian population prior on the \ac{GW} fields, if we make some heuristic assumption about the missing galaxies to treat the total underlying galaxy field as known.

As discussed in the main text, the correct thing to do would be to condition only on the observed galaxies $\{\alm^g[\bar{z}]\}$, rather than the entire galaxy field $\{\alm\gtot[\bar{z}]\}$.
To compute this, we need the quantity $C_\ell^{Gg}[\bar{z}]$. Substituting \cref{eq:pgal_ang_fields}, we have 
\begin{equation} \label{eq:ClGg}
\begin{split}
    C_\ell^{Gg}[\bar{z}] &= \left\langle \alm^G[\bar{z}]^* \alm^g[\bar{z}] \right\rangle = \left\langle \left(\alm\gtot[\bar{z}] + s_{\ell m}^G[\bar{z}] \right)^* \alm^g[\bar{z}] \right\rangle \\
    &= p(D^g | \bar{z})C_\ell^{gg}[\bar{z}] + \big(1 - p(D^g | \bar{z})\big)C_\ell^{g\,g,\mathrm{miss}}[\bar{z}]\\
    &= p(D^g | \bar{z})\left(\overline{C}_\ell^{gg}[\bar{z}] + \frac{1}{n^g} \right) + \big(1 - p(D^g | \bar{z})\big)\overline{C}_\ell^{gg}[\bar{z}]\\
    &= \overline{C}_\ell^{gg}[\bar{z}] + \frac{p(D^g | \bar{z})}{n^g[\bar{z}]}.
\end{split}
\end{equation}
The second term in the last line can again be identified as the cross shot noise in \cref{eq:alm_with_shot}.
As above, we assume for simplicity that the missing and galaxies cluster with the same power spectrum $\overline{C}_\ell^{gg}$, which could be generalized with respective bias parameters.

Using \cref{eq:ClGG,eq:ClGg}, the distribution of the \ac{GW} overdensity field conditioned only on the observed galaxies is 
\begin{subequations} \label{eq:ppop_ang_condobs_alm}
\begin{align} 
    p(\alm^G[\bar{z}] | &\, n^G[\bar{z}], n^g[\bar{z}], \alm^g[\bar{z}], \vecLam) = \mathcal{N}\left[\alm^G; \mu_{\ell m}, \sigma_{\ell m} \right] \\
    \mu_{\ell m} &= \frac{\overline{C}_\ell^{gg} + \frac{p(D^g|z, \vecLam)}{\langle n^g[\bar{z}] \rangle}}{\overline{C}_\ell^{gg} + \frac{1}{\langle n^g[\bar{z}] \rangle}} \alm^g \\
    \sigma_{\ell m} &= \overline{C}_\ell^{gg} + \frac{1}{\langle n^G[\bar{z}] \rangle} - \frac{\left[ \overline{C}_\ell^{gg} + \frac{p(D^g|z, \vecLam)}{\langle n^g[\bar{z}] \rangle} \right]^2}{\overline{C}_\ell^{gg} + \frac{1}{\langle n^g[\bar{z}] \rangle}}.
\end{align}
\end{subequations}
With the usual assumption that $n^g \gg n^G$, this reduces to \cref{eq:ppop_ang_condall_alm} only in the case of galaxy catalog completeness, i.e., $p(D^g|z, \vecLam) = 1$. Thus, if all possible host galaxies are known, then 2-point correlations do not provide any new information for the possible \ac{GW} host locations.

\section{Beam window functions} \label{app:beam}

In this section, we derive the kernel for a Gaussian and isotropic localization error of an individual event. While this is a well known result in the context of, e.g., \ac{CMB} cosmology \cite{2020moco.book.....D}, we find it useful to derive this explicitly for the \ac{GW} measurement process, in which we model the \ac{MAP} sky localization as randomly displaced from its true position.

For a given realization of the universe, the probability density map under noise of the \ac{MAP} position of some observed \ac{GW} $i$ is given by $p^{(i)}(\vecOm)$, which we model as an isotropic Gaussian of width $\sigma_i$ centered at its true location $\vecOm_i$. Thus $p^{(i)}(\vecOm)$ is a function of $\vecOm, \vecOm_i$, and to make this manifest we will now use $p(\vecOm|\vecOm_i)=p^{(i)}(\vecOm)$ for the probability distribution of the observed position of the $i$th event.

By the orthonormality of the spherical harmonics, $p^{(i)}(\vecOm)$ admits a multipole expansion
\begin{equation}
    p(\vecOm|\vecOm_i) = \sum_{\ell, m} \sum_{\ell^\prime, m^\prime} D_{\ell m, \ell^\prime m^\prime} Y_{\ell m}(\vecOm) Y_{\ell^\prime m^\prime}^*(\vecOm_i).
\end{equation}
Let $d^{(i)}_{\ell m}$ be the \ac{SHT} of $p(\vecOm|\vecOm_i)$. The true number density field of this event and its spherical transform are given by
\begin{align}
    n_i(\vecOm) &= \delta^{(D)}(\vecOm - \vecOm_i) \\
    \alm^{(i)} &= Y_{\ell m}^*(\vecOm_i)
\end{align}
and thus $p(\vecOm|\vecOm_i)$ is written in harmonic space as
\begin{equation} \label{eq:aniso_beam_suppression}
d^{(i)}_{\ell m} = \sum_{\ell^\prime, m^\prime} D_{\ell m, \ell^\prime m^\prime} a_{\ell^\prime m^\prime}.
\end{equation}
This expression simplifies if $p(\vecOm | \vecOm_i)$ is isotropic and spatially invariant, i.e., it can be written as a function of $\cos\theta = \vecOm \cdot \vecOm_i$. We can do this by expanding $p(\vecOm | \vecOm_i)\equiv p(\cos \theta)$ in the complete basis of the Legendre polynomials $\mathcal{P}_\ell(\cos\theta)$
\begin{equation} \label{eq:beam_legendre_expansion}
\begin{split}
    p(\cos \theta) &= \sum_\ell B_\ell \frac{2\ell+1}{4\pi} \mathcal{P}_\ell(\cos\theta) \\
    &= \sum_{\ell, m} B_\ell Y_{\ell m}(\vecOm) Y_{\ell m}^*(\vecOm_i)
\end{split}
\end{equation}
which implies the result in \cref{eq:isotropic_beam_mult}
\begin{equation}
d^{(i)}_{\ell m} = B_\ell a_{\ell m}.
\end{equation}
Thus, $B_\ell$ serves to suppress features in the map on the scale of the localization error corresponding to the convolution; localization errors negligible with respect to the resolution of $\ell_{\max}$ have $B_\ell=1$, as is the case for galaxies.
One can solve for the $B_\ell$ corresponding to any localization error $P(\vecOm | \vecOm^\prime)$ by inverting \cref{eq:beam_legendre_expansion}:
\begin{equation}
    B_\ell = \frac{4\pi}{2\ell+1} \int \di\cos\theta p(\cos\theta) \mathcal{P}_\ell(\cos\theta).
\end{equation}
In particular, a Gaussian and isotropic localization error with \ac{RMS} width $\sigma$ can be written as
\begin{equation}
    p(\cos\theta) \propto \exp\left[ -\frac{1-\cos\theta}{\sigma^2} \right] \approx \exp\left[-\frac{\theta^2}{2\sigma^2} \right]
\end{equation}
where the small angle approximation is valid for $\sigma \lesssim 0.1$.
This has a corresponding expansion (to leading order in $\ell^2\sigma^2$) \cite{1992PhRvD..46.4198W}
\begin{equation}
    B_\ell \approx \exp\left[-\frac{1}{2}\ell(\ell+1)\sigma^2\right].
\end{equation}
valid for $\ell^2\sigma^2 \ll 1$. Becauase the signal is suppressed for $\ell^2\sigma^2 \gtrsim 1$ anyway, higher-order corrections are not needed. This yields the expression in \cref{eq:Gaussian_beam}.
Although we limit ourselves to isotropic and Gaussian localization errors in \cref{subsec:ang_loc}, our formalism can be extended to more general localization errors using \cref{eq:aniso_beam_suppression}.

\section{Cross shot noise} \label{app:cross_shot}

The fundamental assumption of the standard galaxy catalog method is that \ac{GW} sources are hosted in galaxies. In the cross-correlation method, this is encoded in the cross shot noise: the $1/n\gtot$ shot noise term in \cref{eq:ClGgtot}, for example, is a direct result from assuming that the \acp{GW} are drawn from the galaxies as a doubly stochastic process. Nonetheless, we claimed in the main text that, in practice, the cross shot noise does not actually contribute much to the cosmological constraints relative to the information gained from 2-point correlations, due to the binning of the data, a large number of galaxies per bin, and low galaxy completeness. In this section, we make this claim more concrete, as well as show for reference how one could in theory compute the expected cross shot noise for full consistency with the standard galaxy catalog method.

From \cref{sec:theory}, the cross shot noise between the \ac{GW} bin $\alpha$ and galaxy bin $\beta$ is
\begin{equation}
s_\ell^{G_\alpha g_\beta} = \frac{\langle n^{G_\alpha} n^{g_\beta} \rangle}{\langle n^{G_\alpha} \rangle \langle n^{g_\beta} \rangle}.
\end{equation}
$\langle n^{G_\alpha} n^{g_\beta} \rangle$ is the Poisson rate of coincidence, which encodes how often a \acp{GW} in bin $\alpha$ has a host galaxy in bin $\beta$. For two bins that are perfectly coincident, $\langle n^{G_\alpha} n^{g_\beta} \rangle = p(D^g|z_\beta) \langle n^{G_\alpha} \rangle$, the fraction of \ac{GW} host galaxies observed in the catalog, and thus the shot noise is something like $p(D^g|z_\beta)/n^g(z_\beta)$. This is already quite small if we have a large number of galaxies in each bin or if the galaxy catalog has low completeness, both of which we expect to be true. The cross shot noise is even smaller with arbitrary \ac{GW} and galaxy bins, which can be computed as
\begin{equation}
    \langle n^{G_\alpha}n^{g_\beta} \rangle = \langle n^{G_\alpha} \rangle \int \di z p(D^g|z)  \phi^{G_\alpha}(z) p^g(\beta|z)
\end{equation}
where $p^g(\beta|z)$ is the probability that a detected host galaxy at redshift $z$ is found in bin $\beta$, and $p(D^g|z)$ is again the host galaxy detection probability at redshift $z$, which can be computed following the standard galaxy catalog literature \cite{2020PhRvD.101l2001G,2023JCAP...12..023G,2024ApJ...964..191B} by integrating the luminosity function up to the k-corrected threshold magnitude.

For tophat radial bins, the probability $p^g(\beta|z)$ of a galaxy in the catalog with true redshift $z$ to be found in the bin with $\hat{z} \in [\hat{z}_{\min}^\beta, \hat{z}_{\max}^\beta]$ is
\begin{equation} \label{eq:p_bin_g}
    p^g(\beta|z) = \int_{\hat{z}_{\min}^\beta}^{\hat{z}_{\max}^\beta} \di \hat{z}\, \mathcal{L}(\hat{z}|z).
\end{equation}
Thus, the \ac{GW}--galaxy cross shot noise is given by
\begin{equation} \label{eq:ClGg_cross_shot}
s_\ell^{G_\alpha g_\beta} = \frac{1}{\langle n^{g_\beta} \rangle}\int \di z \,p(D^g | z) \phi^{G_\alpha}(z) \int_{\hat{z}_{\min}^\beta}^{\hat{z}_{\max}^\beta} \di \hat{z}\, \mathcal{L}(\hat{z}|z).
\end{equation}

\begin{figure}[t]
\centering
\includegraphics[height=8cm]{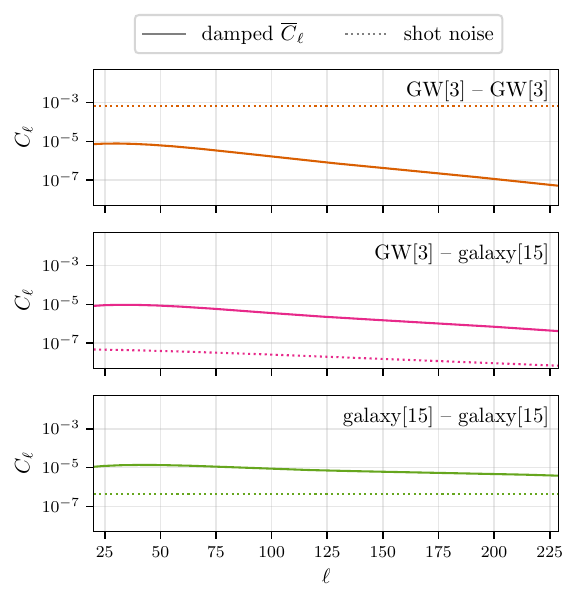}
\caption{Kernel-smoothed power spectrum of the probability density fields ($\overline{C}_\ell^{\alpha\beta}$) vs the shot noise for a \ac{GW}--\ac{GW}, \ac{GW}--galaxy, and galaxy--galaxy power spectrum (top to bottom). We show the same \ac{GW} and galaxy radial bins as in \cref{fig:cl_plots}. \label{fig:shot}}
\end{figure}

We can use this to compute the cross shot noise and compare it to the biased matter power spectrum in our mock universe in \cref{sec:implementation}. This is shown in \cref{fig:shot}, which plots the kernel-smoothed power spectrum of the probability density fields (i.e., the 2-point correlations that we can measure, suppressed at small scales by \ac{GW} angular localization errors) as well as the shot noise for the same power spectra that we showed in \cref{fig:cl_plots}. In particular, we can see that the \ac{GW}--galaxy cross-correlation in the middle panel is signal-dominated by a factor of $\sim100$. 
Furthermore, even though the signal is suppressed by the beam window function at large $\ell$ (it drops much faster with $\ell$ than the galaxy--galaxy power spectrum), the cross shot noise also drops by the same factor, as we showed in \cref{eq:hatClGg_with_angloc}.

With the large error bars in the cross-correlation (as seen in \cref{fig:cl_plots}), which are driven by the large shot noise of the \ac{GW} field, the cross shot noise ends up with negligible impact on the inference. We have repeated our inference in \cref{sec:implementation} including the computation of the cross shot noise outlined above and found no discernible differences in inference results. Thus, the constraints on $H_0$ presented in this paper arise purely from 2-point correlations as two tracers tracing the same Gaussian matter density field. 

The reason why host galaxies are not so informative are the same as why the standard galaxy catalog method is often not informative: galaxy catalogs are large and incomplete, and secondarily \ac{GW} localization errors are large. 
This is also made worse in the cross-correlation method because we discretize the data and build measurement errors into the theory prediction, such that the \ac{GW} host requirement can only be imposed up to a certain scale, which is a weaker requirement than the standard galaxy catalog method.
Thus, the information on cosmological parameters from matching a \ac{GW} to a host galaxy is negligible compared to that of the high-\ac{SNR} measurements of galaxy and \ac{GW} clustering. 
It is for these reasons that most multi-tracer applications ignore the cross shot noise altogether, even if the tracers can be coincident (e.g., \cite{2020PhRvD.102b3528R}).

\section{BBH population model} \label{app:BBH_pop}

\begin{table}[t]
    \centering
    \begin{tabular}{ccc}
        \toprule
         & ~Parameter~ & ~Value~ \\
        \midrule
        \multirow{4}{*}{ $\vecLam_r$ $[\Gpc^{-3}\,\mathrm{yr}^{-1}]$}
        & $\mathcal{R}_0$ & $20$ \\
        & $\gamma$ & $2.7$ \\
        & $\kappa$ & $3.0$ \\
        & $z_p$ & $2.0$ \\
        \midrule
        \multirow{8}{*}{$\vecLam_m$ $[M_\odot]$}
        & $\alpha$ & $3.4$ \\
        & $\mu_p$ & $34$ \\
        & $\sigma_p$ & $3.6$ \\
        & $\lambda_p$ & $0.04$ \\
        & $\delta_m$ & $4.8$ \\
        & $m_{\min}$ & $5.1$ \\
        & $m_{\max}$ & $87$ \\
        & $\beta$ & $1.1$ \\
        \bottomrule
    \end{tabular}
    \caption{True values of the \ac{BBH} rate ($\vecLam_r$) and mass ($\vecLam_m$) population model parameters used to generate the mock catalog. \label{tab:pop_params}}
\end{table}

In this section, we describe the details of the \ac{BBH} population model and the implementation of their injection into galaxies. The assumed values of the population model parameters of can be found in \cref{tab:pop_params}; recall that we use the Planck 2015 values for $H_0$ and \Omn as our fiducial cosmology. We will use $\vecLam_c$, $\vecLam_r$, and $\vecLam_m$ to abbreviate the cosmological, rate, and mass population parameters, respectively.

We use a Madau-Dickinson redshift model \cite{2014ARA&A..52..415M,2020ApJ...896L..32C} to model the rate evolution of the \acp{BBH}
\begin{equation} \label{eq:MD}
    \mathcal{R}(z | \vecLam_r)=\mathcal{C}\left(\gamma, \kappa, z_p\right) \frac{\mathcal{R}_0(1+z)^\gamma}{1+\left(\frac{1+z}{1+z_p}\right)^{\gamma+\kappa}}
\end{equation}
where $\mathcal{R}(z)$ is the source-frame rate per comoving volume at redshift $z$; the normalization constant $\mathcal{C}(\gamma, \kappa, z_p)=1+(1+z_p)^{-\gamma-\kappa}$ ensures that $\mathcal{R}_0$ corresponds to the present-day merger rate. 
Then, for a given redshift chunk of galaxies between $z_{\min}$ and $z_{\max}$, the total number of expected mergers during a detector-frame observing length of $T_\text{obs}$ is
\begin{equation} \label{eq:Nexp_per_chunk}
N_{\mathrm{exp}}(\vecLam_r, \vecLam_c, z_{\min}, z_{\max}) = T_\text{obs} \int_{z_{\min}}^{z_{\max}} \di z \frac{\mathcal{R}(z | \vecLam_r)}{1+z} \deriv{V_c}{z}(\vecLam_c)
\end{equation}
where the factor of $1+z$ is from the conversion of the source frame to the detector frame. We use an observing time of $T_\text{obs}=2$ years.

We divide the entire underlying set of galaxies into chunks in redshift. For each chunk $i$, we draw the number of \acp{BBH} in the chunk $N^i$ from a Poisson distribution with expectation $N_{\text{exp}}^i$ using \cref{eq:Nexp_per_chunk}. Then, we inject $N^i$ \acp{BBH} into randomly drawn galaxies in the chunk weighted by their redshift as in \cref{eq:MD}. Finally, for each of the $N_i$ \acp{BBH}, we draw the primary mass and mass ratio ($m_1, q$) via rejection sampling, and draw the aligned spins from a uniform distribution $\chi_{1, z}, \chi_{2, z} \in [-1, 1]$. These parameters, along with uniformly drawn extrinsic parameters (inclination angle, polarization angle, GPS time of arrival, coalescence phase), are sufficient to pass into \texttt{gwfast} for non-precessing waveforms (including \texttt{IMRPhenomHM}) as mock \acp{BBH}. This procedure is repeated for all chunks.

We now describe the mass population model. 
We model the primary mass distribution as a mixture model of a power law and a Gaussian distribution with a low-mass smoothing function \cite{2018ApJ...856..173T}
\begin{equation}
\begin{split}
\pi( & m_1 | \vecLam_m ) \\
\quad \propto & {\big[\left(1-\lambda_p\right) p_{\text{PL}}\left(m_1 | -\alpha, m_{\min}, m_{\max}\right)} \\
& \quad+\lambda_p \,\mathcal{N}\left[ m_1 ; \mu_p, \sigma_p\right] \big] 
\times S(m_1 \mid m_{\min }, \delta_m)
\end{split}
\end{equation}
where
\begin{equation}
\begin{split}
    &p_\text{PL}\left(m_1 | -\alpha, m_{\min}, m_{\max}\right) = \\
    &\quad\begin{cases}
        \frac{m_1^{-\alpha}}{m_{\max}^{1-\alpha} - m_{\min}^{1-\alpha}} & m_{\min} < m_1 < m_{\max} \\
        0 & \text{otherwise}
    \end{cases}
\end{split}
\end{equation}
is a truncated power law between $m_{\min}$ and $m_{\max}$ with exponent $-\alpha$, $\mu_p$ and $\sigma_p$ are the mean and width of the Gaussian $\mathcal{N}$, $\lambda_p$ is the mixing fraction of the peak, and
\begin{widetext}
\begin{equation}
    S(m_1 \mid m_{\min }, \delta_m) = 
    \begin{cases}
        0 & m_1 < m_{\min}\\
        \left[ 1 + \exp\left(\frac{\delta_m}{m - m_{\min}} + \frac{\delta_m}{m_1-m_{\min}-\delta_m}\right) \right]^{-1} & m_{\min} \leq m_1 < m_1 + \delta_m \\
        0 & m_1 \geq m_1 + \delta_m
    \end{cases}
\end{equation}
\end{widetext}
is a Planck taper window to smooth the distribution near the minimum mass. 

To model the mass ratio distribution, we use a truncated power law with exponent $\beta$
\begin{equation}
    p(q | m_1, \beta, m_{\max}) = p_\text{PL}\left(q \bigg| \beta, \frac{m_1}{m_{\max}}, 1\right).
\end{equation}
The observed \acp{SNR}, posterior measurements, etc. are then computed as described in \cref{subsubsec:BBH_inj}, giving the mock \ac{GW} catalog after applying our \ac{SNR}$\geq 25$ cut.

\section{Quasi-linear corrections to the power spectrum} \label{app:lin_corr}

\begin{figure*}[t]
\centering
\includegraphics[width=0.6\textwidth]{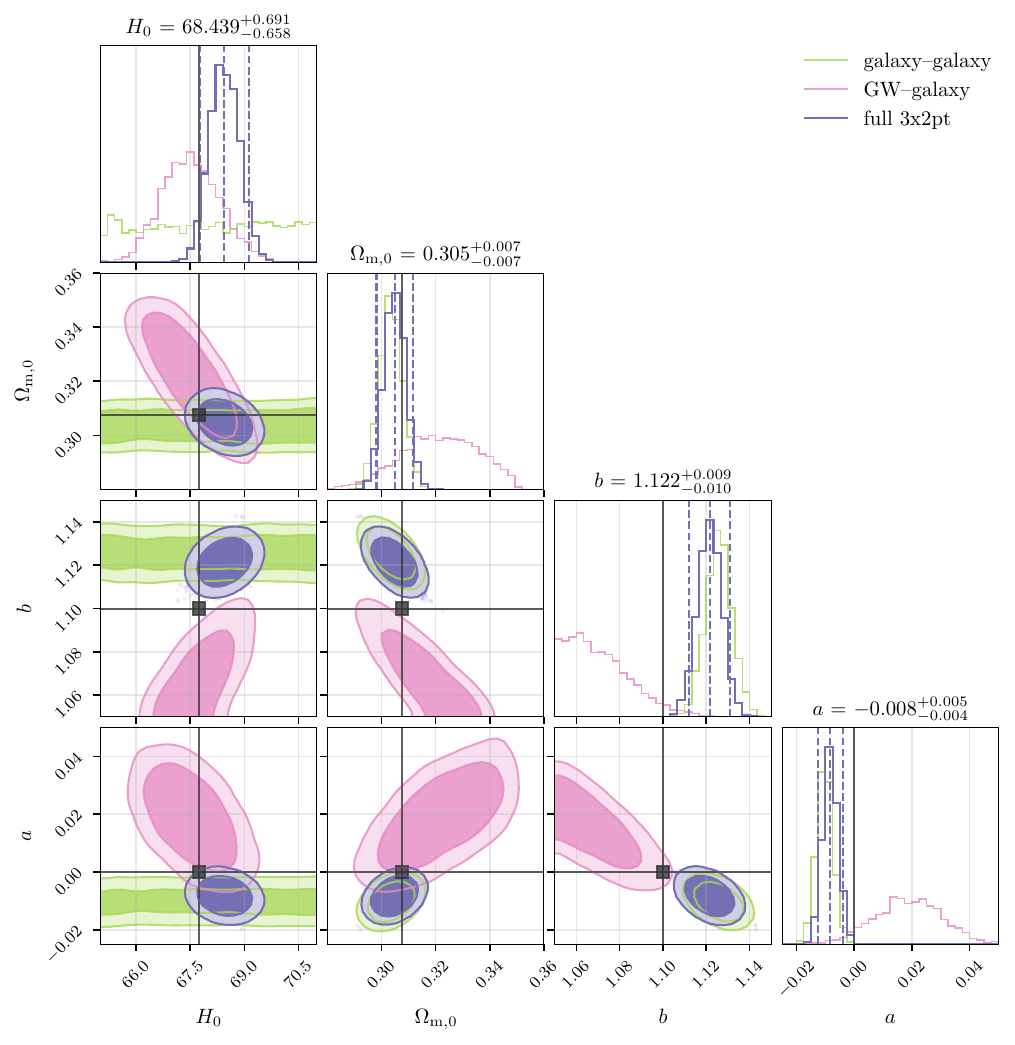}
\caption{Corner plot for an analysis that includes a first-order correction for non-linear scales, resulting in an unbiased measurement of \Omn from galaxy--galaxy clustering compared to the result in the main text, \cref{fig:corner}. \label{fig:corner_full_poly1bias}}
\end{figure*}

In \cref{eq:poly_bias}, we wrote down a simple series expansion for $b(k)$. This is a common technique to model and marginalize over unknown deviations from the linear power spectrum at quasi-linear scales (e.g., \cite{2012ApJ...761...14H}). 
A first order correction would be 
\begin{equation}
b(k, z) = b + a\left(\frac{k}{0.1\Mpc^{-1}}\right).
\end{equation}

\cref{fig:corner_full_poly1bias} shows the same corner plot as in \cref{fig:corner}, but with this added first-order correction parameter $a$; we used a uniform prior of $a \in [-0.2, 0.2]$.
Compared to the results in \cref{fig:corner}, we find that adding the nuisance parameter is sufficient to account for unmodeled systematics, removing the bias in \Omn from the galaxy--galaxy analysis.
Note that the true values of the nuisance parameters $a=0$ and $b=1.1$ are not recovered correctly, but for our purposes they are nuisance parameters that we marginalize over to encode uncertainty in the shape of the power spectrum. Inferring the shape of the power spectrum requires a more careful analysis that would, for example, include higher-order terms in the Limber approximation.

\end{document}

%% file: commands.tex
\newcommand\veck{\ensuremath{\mathbf{k}}\xspace}
\newcommand\vecr{\ensuremath{\mathbf{r}}\xspace}
\newcommand\vecd{\ensuremath{\bm{d}}\xspace}
\newcommand\vecOm{\ensuremath{\bm{\Omega}}\xspace}

\newcommand\vecLam{\ensuremath{\mathbf{\Lambda}}\xspace}
\newcommand\veclam{\ensuremath{\bm{\lambda}}\xspace}
\newcommand\vecF{\ensuremath{\bm{F}}\xspace}

\newcommand\alm{\ensuremath{a_{\ell m}}\xspace}

\newcommand\fsky{\ensuremath{f_\text{sky}}\xspace}
\newcommand\covS{\ensuremath{\mathbf{\Sigma}}\xspace}
\newcommand\covC{\ensuremath{\mathbf{C}}\xspace}

\newcommand\dg{\ensuremath{\{\bm{d}^g\}}\xspace}
\newcommand\dG{\ensuremath{\{\bm{d}^G\}}\xspace}

\newcommand\Ppop{\ensuremath{p}\xspace}
\newcommand\Pgal{\ensuremath{p^g}\xspace}

\newcommand\Hn{\ensuremath{H_0}\xspace}
\newcommand\Omn{\ensuremath{\Omega_{m, 0}}\xspace}

\newcommand\gmiss{\ensuremath{^{g, \text{miss}}}\xspace}
\newcommand\gobs{\ensuremath{^{g, \text{obs}}}\xspace}
\newcommand\gtot{\ensuremath{^{g, \text{tot}}}\xspace}
\newcommand\fgdet{\ensuremath{f^g_\text{obs}}\xspace}

\newcommand\Mpc{\ensuremath{~\mathrm{Mpc}}\xspace}
\newcommand\Gpc{\ensuremath{~\mathrm{Gpc}}\xspace}

\newcommand\di{\ensuremath{\mathrm{d}}\xspace}
\newcommand\deriv[2]{\ensuremath{\frac{\mathrm{d}#1}{\mathrm{d}#2}}\xspace}
\newcommand\E{\ensuremath{\mathbb{E}}\xspace}

\newcommand{\appropto}{\mathrel{\vcenter{
  \offinterlineskip\halign{\hfil$##$\cr
    \propto\cr\noalign{\kern2pt}\sim\cr\noalign{\kern-2pt}}}}}

\DeclareUnicodeCharacter{2500}{\rule[0.5ex]{1em}{0.4pt}}

%% file: acronyms.tex
\acrodef{GW}{gravitational wave}
\acrodef{BBH}{binary black hole}
\acrodef{PE}{parameter estimation}
\acrodef{GRF}{Gaussian Random Field}
\acrodef{LVK}{LIGO-Virgo-KAGRA}
\acrodef{CE}{Cosmic Explorer}
\acrodef{ET}{Einstein Telescope}
\acrodef{LSS}{large-scale structure}
\acrodef{SHT}{spherical harmonic transform}
\acrodef{BAO}{Baryonic Acoustic Oscillation}
\acrodef{CMB}{Cosmic Microwave Background}
\acrodef{MAP}{maximum a-posteriori}
\acrodef{SNR}{signal-to-noise ratio}
\acrodef{IID}{independent and identically distributed}
\acrodef{RMS}{root mean squared}
\acrodef{KDE}{kernel density estimation}
\acrodef{2PCF}{2-point correlation function}
\acrodef{HMC}{Hamiltonian Monte Carlo}
\acrodef{3G}{third generation}
\acrodef{RSD}{redshift space distortion}
\acrodef{LSD}{luminosity-distance space distortion}

%% file: bib.bib
@ARTICLE{2026ApJ..1005L..51A,
       author = {{Abac}, A.~G. and {Abouelfettouh}, I. and {Acernese}, F. and {Ackley}, K. and {Adamcewicz}, C. and {Adhicary}, S. and {Adhikari}, D. and {Adhikari}, N. and {Adhikari}, R.~X. and {Adkins}, V.~K. and others},
        title = "{GWTC-4.0: Population Properties of Merging Compact Binaries}",
      journal = {ApJL},
         year = 2026,
        month = jul,
       volume = {1005},
       number = {2},
          eid = {L51},
        pages = {L51},
          doi = {10.3847/2041-8213/ae771e},
archivePrefix = {arXiv},
       eprint = {2508.18083},
 primaryClass = {astro-ph.HE},
       adsurl = {https://ui.adsabs.harvard.edu/abs/2026ApJ..1005L..51A}
}

@ARTICLE{2026PhRvD.113j3532P,
       author = {{Pan}, Jiaming and {Huterer}, Dragan and {Avestruz}, Camille and {Cheung}, Damon H.~T. and {Trott}, Emery and {Dalal}, Neal and {Jeong}, Donghui},
        title = "{Determining the Hubble constant through cross-correlation of galaxies and gravitational waves}",
      journal = {PhRvD},
         year = 2026,
        month = may,
       volume = {113},
       number = {10},
          eid = {103532},
        pages = {103532},
          doi = {10.1103/2d96-fg7f},
archivePrefix = {arXiv},
       eprint = {2510.19931},
 primaryClass = {astro-ph.CO},
       adsurl = {https://ui.adsabs.harvard.edu/abs/2026PhRvD.113j3532P}
}

@ARTICLE{2026A&A...709A.197T,
       author = {{Tagliazucchi}, Matteo and {Moresco}, Michele and {Borghi}, Nicola and {Ciapetti}, Chiara},
        title = "{Mind the peak: Improving cosmological constraints from GWTC-4.0 spectral sirens using semiparametric mass models}",
      journal = {A\&A},
         year = 2026,
        month = may,
       volume = {709},
          eid = {A197},
        pages = {A197},
          doi = {10.1051/0004-6361/202558756},
archivePrefix = {arXiv},
       eprint = {2601.03347},
 primaryClass = {astro-ph.CO},
       adsurl = {https://ui.adsabs.harvard.edu/abs/2026A&A...709A.197T}
}

@ARTICLE{2026arXiv260527227T,
       author = {{The LIGO Scientific Collaboration} and {the Virgo Collaboration} and {the KAGRA Collaboration}},
        title = "{GWTC-5.0: Constraints on the Cosmic Expansion Rate and Modified Gravitational-wave Propagation}",
      journal = {arXiv},
         year = 2026,
        month = may,
          eid = {arXiv:2605.27227},
        pages = {arXiv:2605.27227},
          doi = {10.48550/arXiv.2605.27227},
archivePrefix = {arXiv},
       eprint = {2605.27227},
 primaryClass = {astro-ph.CO},
       adsurl = {https://ui.adsabs.harvard.edu/abs/2026arXiv260527227T}
}

@ARTICLE{2026JCAP...01..034D,
       author = {{Dalang}, Charles and {Fiorini}, Bartolomeo and {Baker}, Tessa},
        title = "{Large scale structure prior knowledge in the dark siren method}",
      journal = {JCAP},
         year = 2026,
        month = jan,
       volume = {2026},
       number = {1},
          eid = {034},
        pages = {034},
          doi = {10.1088/1475-7516/2026/01/034},
archivePrefix = {arXiv},
       eprint = {2410.03275},
 primaryClass = {astro-ph.CO},
       adsurl = {https://ui.adsabs.harvard.edu/abs/2026JCAP...01..034D}
}

@INPROCEEDINGS{2026enap....5..557P,
       author = {{Palmese}, Antonella and {Mastrogiovanni}, Simone},
        title = "{Gravitational wave cosmology}",
    booktitle = {Encyclopedia of Astrophysics, Volume 5},
         year = 2026,
       volume = {5},
        month = jan,
        pages = {557-576},
          doi = {10.1016/B978-0-443-21439-4.00107-3},
archivePrefix = {arXiv},
       eprint = {2502.00239},
 primaryClass = {astro-ph.CO},
       adsurl = {https://ui.adsabs.harvard.edu/abs/2026enap....5..557P}
}

@ARTICLE{2026JCAP...01..013L,
       author = {{Leyde}, Konstantin and {Baker}, Tessa and {Enzi}, Wolfgang},
        title = "{Cosmic cartography. Part II. Completing galaxy catalogs for gravitational-wave cosmology}",
      journal = {JCAP},
         year = 2026,
        month = jan,
       volume = {2026},
       number = {1},
          eid = {013},
        pages = {013},
          doi = {10.1088/1475-7516/2026/01/013},
archivePrefix = {arXiv},
       eprint = {2507.12171},
 primaryClass = {astro-ph.CO},
       adsurl = {https://ui.adsabs.harvard.edu/abs/2026JCAP...01..013L}
}

@ARTICLE{2025ApJS..281...54C,
       author = {{Crenshaw}, John Franklin and {Leistedt}, Boris and {Graham}, Melissa Lynn and {Payerne}, Constantin and {Connolly}, Andrew J. and {Gawiser}, Eric and {Karim}, Tanveer and {Malz}, Alex I. and {Newman}, Jeffrey A. and {Ricci}, Marina and others},
        title = "{Quantifying the Impact of LSST u-band Survey Strategy on Photometric Redshift Estimation and the Detection of Lyman-break Galaxies}",
      journal = {ApJS},
         year = 2025,
        month = dec,
       volume = {281},
       number = {2},
          eid = {54},
        pages = {54},
          doi = {10.3847/1538-4365/ae14f0},
archivePrefix = {arXiv},
       eprint = {2503.06016},
 primaryClass = {astro-ph.CO},
       adsurl = {https://ui.adsabs.harvard.edu/abs/2025ApJS..281...54C}
}

@ARTICLE{2025PASA...42..149C,
       author = {{Cross-Parkin}, Madeline L. and {Howlett}, Cullan and {Davis}, Tamara M. and {Khetan}, Nandita},
        title = "{Dark sirens and the impact of redshift precision}",
      journal = {PASA},
         year = 2025,
        month = nov,
       volume = {42},
          eid = {e149},
        pages = {e149},
          doi = {10.1017/pasa.2025.10111},
archivePrefix = {arXiv},
       eprint = {2502.17747},
 primaryClass = {astro-ph.CO},
       adsurl = {https://ui.adsabs.harvard.edu/abs/2025PASA...42..149C}
}

@ARTICLE{2025PhRvD.112j3015S,
       author = {{Santoliquido}, Filippo and {Tissino}, Jacopo and {Dupletsa}, Ulyana and {Branchesi}, Marica and {Harms}, Jan and {Arca Sedda}, Manuel and {Dax}, Maximilian and {Kofler}, Annalena and {Green}, Stephen R. and {Gupte}, Nihar and others},
        title = "{Fast and accurate parameter estimation of high-redshift sources with the Einstein Telescope}",
      journal = {PhRvD},
         year = 2025,
        month = nov,
       volume = {112},
       number = {10},
          eid = {103015},
        pages = {103015},
          doi = {10.1103/wf1k-p5cl},
archivePrefix = {arXiv},
       eprint = {2504.21087},
 primaryClass = {astro-ph.HE},
       adsurl = {https://ui.adsabs.harvard.edu/abs/2025PhRvD.112j3015S}
}

@ARTICLE{2025JCAP...09..008A,
       author = {{Adame}, A.~G. and {Aguilar}, J. and {Ahlen}, S. and {Alam}, S. and {Alexander}, D.~M. and {Alvarez}, M. and {Alves}, O. and {Anand}, A. and {Andrade}, U. and {Armengaud}, E. and others},
        title = "{DESI 2024 V: Full-Shape galaxy clustering from galaxies and quasars}",
      journal = {JCAP},
         year = 2025,
        month = sep,
       volume = {2025},
       number = {9},
          eid = {008},
        pages = {008},
          doi = {10.1088/1475-7516/2025/09/008},
archivePrefix = {arXiv},
       eprint = {2411.12021},
 primaryClass = {astro-ph.CO},
       adsurl = {https://ui.adsabs.harvard.edu/abs/2025JCAP...09..008A}
}

@ARTICLE{2025arXiv250904348T,
       author = {{The LIGO Scientific Collaboration} and {the Virgo Collaboration} and {the KAGRA Collaboration} and {Abac}, A.~G. and {Abouelfettouh}, I. and {Acernese}, F. and {Ackley}, K. and {Adamcewicz}, C. and {Adhicary}, S. and {Adhikari}, D. and others},
        title = "{GWTC-4.0: Constraints on the Cosmic Expansion Rate and Modified Gravitational-wave Propagation}",
      journal = {arXiv},
         year = 2025,
        month = sep,
          eid = {arXiv:2509.04348},
        pages = {arXiv:2509.04348},
          doi = {10.48550/arXiv.2509.04348},
archivePrefix = {arXiv},
       eprint = {2509.04348},
 primaryClass = {astro-ph.CO},
       adsurl = {https://ui.adsabs.harvard.edu/abs/2025arXiv250904348T}
}

@ARTICLE{2025A&A...698A.128P,
       author = {{Perna}, G. and {Mastrogiovanni}, S. and {Ricciardone}, A.},
        title = "{Investigating the impact of galaxies' compact binary hosting probability for gravitational wave cosmology}",
      journal = {A\&A},
         year = 2025,
        month = jun,
       volume = {698},
          eid = {A128},
        pages = {A128},
          doi = {10.1051/0004-6361/202450840},
archivePrefix = {arXiv},
       eprint = {2405.07904},
 primaryClass = {astro-ph.CO},
       adsurl = {https://ui.adsabs.harvard.edu/abs/2025A&A...698A.128P}
}

@ARTICLE{2025PhRvD.111j3012M,
       author = {{Mancarella}, Michele and {Gerosa}, Davide},
        title = "{Sampling the full hierarchical population posterior distribution in gravitational-wave astronomy}",
      journal = {PhRvD},
         year = 2025,
        month = may,
       volume = {111},
       number = {10},
          eid = {103012},
        pages = {103012},
          doi = {10.1103/PhysRevD.111.103012},
archivePrefix = {arXiv},
       eprint = {2502.12156},
 primaryClass = {gr-qc},
       adsurl = {https://ui.adsabs.harvard.edu/abs/2025PhRvD.111j3012M}
}

@software{2025ascl.soft05005P,
       author = {{Phan}, Du and {Pradhan}, Neeraj and {Jankowiak}, Martin and {Bingham}, Eli and {Chen}, Jonathan P. and {Obermeyer}, Fritz and {Karaletsos}, Theofanis and {Singh}, Rohit and {Szerlip}, Paul and {Horsfall}, Paul and others},
        title = "{NumPyro: Probabilistic programming with NumPy}",
 howpublished = {Astrophysics Source Code Library, record ascl:2505.005},
         year = 2025,
        month = may,
          eid = {ascl:2505.005},
archivePrefix = {ascl},
       eprint = {2505.005},
       adsurl = {https://ui.adsabs.harvard.edu/abs/2025ascl.soft05005P}
}

@ARTICLE{2025JCAP...04..008F,
       author = {{Ferri}, Jo{\~a}o and {Tashiro}, Ian L. and {Abramo}, L.~R. and {Matos}, Isabela and {Quartin}, Miguel and {Sturani}, Riccardo},
        title = "{A robust cosmic standard ruler from the cross-correlations of galaxies and dark sirens}",
      journal = {JCAP},
         year = 2025,
        month = apr,
       volume = {2025},
       number = {4},
          eid = {008},
        pages = {008},
          doi = {10.1088/1475-7516/2025/04/008},
archivePrefix = {arXiv},
       eprint = {2412.00202},
 primaryClass = {astro-ph.CO},
       adsurl = {https://ui.adsabs.harvard.edu/abs/2025JCAP...04..008F}
}

@ARTICLE{2025JCAP...04..056D,
       author = {{Dehghani}, Amir and {Kim}, J. Leo and {Hosseini}, Dorsa Sadat and {Krolewski}, Alex and {Mukherjee}, Suvodip and {Geshnizjani}, Ghazal},
        title = "{The gravitational wave bias parameter from angular power spectra: bridging between galaxies and binary black holes}",
      journal = {JCAP},
         year = 2025,
        month = apr,
       volume = {2025},
       number = {4},
          eid = {056},
        pages = {056},
          doi = {10.1088/1475-7516/2025/04/056},
archivePrefix = {arXiv},
       eprint = {2411.11965},
 primaryClass = {astro-ph.GA},
       adsurl = {https://ui.adsabs.harvard.edu/abs/2025JCAP...04..056D}
}

@ARTICLE{2025JCAP...04..012A,
       author = {{Adame}, A.~G. and {Aguilar}, J. and {Ahlen}, S. and {Alam}, S. and {Alexander}, D.~M. and {Alvarez}, M. and {Alves}, O. and {Anand}, A. and {Andrade}, U. and {Armengaud}, E. and others},
        title = "{DESI 2024 III: baryon acoustic oscillations from galaxies and quasars}",
      journal = {JCAP},
         year = 2025,
        month = apr,
       volume = {2025},
       number = {4},
          eid = {012},
        pages = {012},
          doi = {10.1088/1475-7516/2025/04/012},
archivePrefix = {arXiv},
       eprint = {2404.03000},
 primaryClass = {astro-ph.CO},
       adsurl = {https://ui.adsabs.harvard.edu/abs/2025JCAP...04..012A}
}

@ARTICLE{2025arXiv250410482P,
       author = {{Pedrotti}, Alessandro and {Mancarella}, Michele and {Bel}, Julien and {Santoni}, Michele and {Gerosa}, Davide},
        title = "{Cosmology with the angular cross-correlation of gravitational-wave and galaxy catalogs: forecasts for next-generation interferometers and the Euclid survey}",
      journal = {arXiv},
         year = 2025,
        month = apr,
          eid = {arXiv:2504.10482},
        pages = {arXiv:2504.10482},
          doi = {10.48550/arXiv.2504.10482},
archivePrefix = {arXiv},
       eprint = {2504.10482},
 primaryClass = {astro-ph.CO},
       adsurl = {https://ui.adsabs.harvard.edu/abs/2025arXiv250410482P}
}

@ARTICLE{2025ApJ...980...85M,
       author = {{Mali}, Utkarsh and {Essick}, Reed},
        title = "{Striking a Chord with Spectral Sirens: Multiple Features in the Compact Binary Population Correlate with H$_{0}$}",
      journal = {ApJ},
         year = 2025,
        month = feb,
       volume = {980},
       number = {1},
          eid = {85},
        pages = {85},
          doi = {10.3847/1538-4357/ad9de7},
archivePrefix = {arXiv},
       eprint = {2410.07416},
 primaryClass = {astro-ph.HE},
       adsurl = {https://ui.adsabs.harvard.edu/abs/2025ApJ...980...85M}
}

@ARTICLE{2025MNRAS.537.1912Z,
       author = {{Zazzera}, Stefano and {Fonseca}, Jos{\'e} and {Baker}, Tessa and {Clarkson}, Chris},
        title = "{Gravitational waves and galaxies cross-correlations: a forecast on GW biases for future detectors}",
      journal = {MNRAS},
         year = 2025,
        month = feb,
       volume = {537},
       number = {2},
        pages = {1912-1923},
          doi = {10.1093/mnras/staf150},
archivePrefix = {arXiv},
       eprint = {2412.01678},
 primaryClass = {astro-ph.CO},
       adsurl = {https://ui.adsabs.harvard.edu/abs/2025MNRAS.537.1912Z}
}

@ARTICLE{2025ApJ...978..153F,
       author = {{Farah}, Amanda M. and {Callister}, Thomas A. and {Ezquiaga}, Jose Mar{\'\i}a and {Zevin}, Michael and {Holz}, Daniel E.},
        title = "{No Need to Know: Toward Astrophysics-free Gravitational-wave Cosmology}",
      journal = {ApJ},
         year = 2025,
        month = jan,
       volume = {978},
       number = {2},
          eid = {153},
        pages = {153},
          doi = {10.3847/1538-4357/ad9253},
archivePrefix = {arXiv},
       eprint = {2404.02210},
 primaryClass = {astro-ph.CO},
       adsurl = {https://ui.adsabs.harvard.edu/abs/2025ApJ...978..153F}
}

@ARTICLE{2025ApJ...979....9H,
       author = {{Hanselman}, Alexandra G. and {Vijaykumar}, Aditya and {Fishbach}, Maya and {Holz}, Daniel E.},
        title = "{Gravitational-wave Dark Siren Cosmology Systematics from Galaxy Weighting}",
      journal = {ApJ},
         year = 2025,
        month = jan,
       volume = {979},
       number = {1},
          eid = {9},
        pages = {9},
          doi = {10.3847/1538-4357/ad9393},
archivePrefix = {arXiv},
       eprint = {2405.14818},
 primaryClass = {astro-ph.CO},
       adsurl = {https://ui.adsabs.harvard.edu/abs/2025ApJ...979....9H}
}

@ARTICLE{2024JCAP...12..013L,
       author = {{Leyde}, Konstantin and {Baker}, Tessa and {Enzi}, Wolfgang},
        title = "{Cosmic cartography: Bayesian reconstruction of the galaxy density informed by large-scale structure}",
      journal = {JCAP},
         year = 2024,
        month = dec,
       volume = {2024},
       number = {12},
          eid = {013},
        pages = {013},
          doi = {10.1088/1475-7516/2024/12/013},
archivePrefix = {arXiv},
       eprint = {2409.20531},
 primaryClass = {astro-ph.CO},
       adsurl = {https://ui.adsabs.harvard.edu/abs/2024JCAP...12..013L}
}

@ARTICLE{2024OJAp....7E.110G,
       author = {{Gagnon}, E.~L. and {Anbajagane}, D. and {Prat}, J. and {Chang}, C. and {Frieman}, J.},
        title = "{Cosmological Constraints from Combining Photometric Galaxy Surveys and Gravitational Wave Observatories}",
      journal = {OJAp},
         year = 2024,
        month = dec,
       volume = {7},
          eid = {110},
        pages = {110},
          doi = {10.33232/001c.127131},
archivePrefix = {arXiv},
       eprint = {2312.16289},
 primaryClass = {astro-ph.CO},
       adsurl = {https://ui.adsabs.harvard.edu/abs/2024OJAp....7E.110G}
}

@ARTICLE{2024ApJ...975..189M,
       author = {{Mukherjee}, Suvodip and {Krolewski}, Alex and {Wandelt}, Benjamin D. and {Silk}, Joseph},
        title = "{Cross-correlating Dark Sirens and Galaxies: Constraints on H $_{0}$ from GWTC-3 of LIGO─Virgo─KAGRA}",
      journal = {ApJ},
         year = 2024,
        month = nov,
       volume = {975},
       number = {2},
          eid = {189},
        pages = {189},
          doi = {10.3847/1538-4357/ad7d90},
archivePrefix = {arXiv},
       eprint = {2203.03643},
 primaryClass = {astro-ph.CO},
       adsurl = {https://ui.adsabs.harvard.edu/abs/2024ApJ...975..189M}
}

@ARTICLE{2024JCAP...10..087B,
       author = {{Begnoni}, Andrea and {Dall'Armi}, Lorenzo Valbusa and {Bertacca}, Daniele and {Raccanelli}, Alvise},
        title = "{Gravitational wave luminosity distance-weighted anisotropies}",
      journal = {JCAP},
         year = 2024,
        month = oct,
       volume = {2024},
       number = {10},
          eid = {087},
        pages = {087},
          doi = {10.1088/1475-7516/2024/10/087},
archivePrefix = {arXiv},
       eprint = {2404.12351},
 primaryClass = {astro-ph.CO},
       adsurl = {https://ui.adsabs.harvard.edu/abs/2024JCAP...10..087B}
}

@ARTICLE{2024JCAP...10..074B,
       author = {{Beltrame}, Matteo and {Bonici}, Marco and {Carbone}, Carmelita},
        title = "{Cosmological forecasts from the combination of Stage-IV photometric galaxy surveys and the magnification from forthcoming GW observatories}",
      journal = {JCAP},
         year = 2024,
        month = oct,
       volume = {2024},
       number = {10},
          eid = {074},
        pages = {074},
          doi = {10.1088/1475-7516/2024/10/074},
archivePrefix = {arXiv},
       eprint = {2403.03859},
 primaryClass = {astro-ph.CO},
       adsurl = {https://ui.adsabs.harvard.edu/abs/2024JCAP...10..074B}
}

@INPROCEEDINGS{2024IAUGA..32P1303G,
       author = {{Ghosh}, Tathagata and {More}, Surhud and {Bera}, Sayantani and {Bose}, Sukanta},
        title = "{Bayesian Inference of the Hubble constant from Cross-Correlation of Individual Gravitational Wave Events with Galaxies}",
    booktitle = {32nd General Assembly International Union (IAUGA 2024)},
         year = 2024,
        month = aug,
          eid = {1303},
        pages = {1303},
          doi = {10.48550/arXiv.2312.16305},
archivePrefix = {arXiv},
       eprint = {2312.16305},
 primaryClass = {astro-ph.CO},
       adsurl = {https://ui.adsabs.harvard.edu/abs/2024IAUGA..32P1303G}
}

@ARTICLE{2024JCAP...05..095Z,
       author = {{Zazzera}, Stefano and {Fonseca}, Jos{\'e} and {Baker}, Tessa and {Clarkson}, Chris},
        title = "{Magnification and evolution bias of transient sources: GWs and SNIa}",
      journal = {JCAP},
         year = 2024,
        month = may,
       volume = {2024},
       number = {5},
          eid = {095},
        pages = {095},
          doi = {10.1088/1475-7516/2024/05/095},
archivePrefix = {arXiv},
       eprint = {2309.04391},
 primaryClass = {astro-ph.CO},
       adsurl = {https://ui.adsabs.harvard.edu/abs/2024JCAP...05..095Z}
}

@ARTICLE{2024ApJ...964..191B,
       author = {{Borghi}, Nicola and {Mancarella}, Michele and {Moresco}, Michele and {Tagliazucchi}, Matteo and {Iacovelli}, Francesco and {Cimatti}, Andrea and {Maggiore}, Michele},
        title = "{Cosmology and Astrophysics with Standard Sirens and Galaxy Catalogs in View of Future Gravitational Wave Observations}",
      journal = {ApJ},
         year = 2024,
        month = apr,
       volume = {964},
       number = {2},
          eid = {191},
        pages = {191},
          doi = {10.3847/1538-4357/ad20eb},
archivePrefix = {arXiv},
       eprint = {2312.05302},
 primaryClass = {astro-ph.CO},
       adsurl = {https://ui.adsabs.harvard.edu/abs/2024ApJ...964..191B}
}

@ARTICLE{2024JCAP...02..024D,
       author = {{Dalang}, Charles and {Baker}, Tessa},
        title = "{The clustering of dark sirens' invisible host galaxies}",
      journal = {JCAP},
         year = 2024,
        month = feb,
       volume = {2024},
       number = {2},
          eid = {024},
        pages = {024},
          doi = {10.1088/1475-7516/2024/02/024},
archivePrefix = {arXiv},
       eprint = {2310.08991},
 primaryClass = {astro-ph.CO},
       adsurl = {https://ui.adsabs.harvard.edu/abs/2024JCAP...02..024D}
}

@ARTICLE{2024AnP...53600180M,
       author = {{Mastrogiovanni}, Simone and {Karathanasis}, Christos and {Gair}, Jonathan and {Ashton}, Gregory and {Rinaldi}, Stefano and {Huang}, Hsiang-Yu and {D{\'a}lya}, Gergely},
        title = "{Cosmology with Gravitational Waves: A Review}",
      journal = {AnP},
         year = 2024,
        month = feb,
       volume = {536},
       number = {2},
          eid = {2200180},
        pages = {2200180},
          doi = {10.1002/andp.202200180},
       adsurl = {https://ui.adsabs.harvard.edu/abs/2024AnP...53600180M}
}

@ARTICLE{2024MNRAS.528.3249A,
       author = {{Alfradique}, V. and {Bom}, C.~R. and {Palmese}, A. and {Teixeira}, G. and {Santana-Silva}, L. and {Drlica-Wagner}, A. and {Riley}, A.~H. and {Mart{\'\i}nez-V{\'a}zquez}, C.~E. and {Sand}, D.~J. and {Stringfellow}, G.~S. and others},
        title = "{A dark siren measurement of the Hubble constant using gravitational wave events from the first three LIGO/Virgo observing runs and DELVE}",
      journal = {MNRAS},
         year = 2024,
        month = feb,
       volume = {528},
       number = {2},
        pages = {3249-3259},
          doi = {10.1093/mnras/stae086},
archivePrefix = {arXiv},
       eprint = {2310.13695},
 primaryClass = {astro-ph.CO},
       adsurl = {https://ui.adsabs.harvard.edu/abs/2024MNRAS.528.3249A}
}

@ARTICLE{2024ApJ...962..169E,
       author = {{Essick}, Reed and {Fishbach}, Maya},
        title = "{Ensuring Consistency between Noise and Detection in Hierarchical Bayesian Inference}",
      journal = {ApJ},
         year = 2024,
        month = feb,
       volume = {962},
       number = {2},
          eid = {169},
        pages = {169},
          doi = {10.3847/1538-4357/ad1604},
archivePrefix = {arXiv},
       eprint = {2310.02017},
 primaryClass = {gr-qc},
       adsurl = {https://ui.adsabs.harvard.edu/abs/2024ApJ...962..169E}
}

@ARTICLE{2023ApJ...959...35K,
       author = {{Kumar}, Sumit},
        title = "{Probing Cosmology with Baryon Acoustic Oscillations Using Gravitational Waves}",
      journal = {ApJ},
         year = 2023,
        month = dec,
       volume = {959},
       number = {1},
          eid = {35},
        pages = {35},
          doi = {10.3847/1538-4357/acf618},
archivePrefix = {arXiv},
       eprint = {2203.04273},
 primaryClass = {astro-ph.CO},
       adsurl = {https://ui.adsabs.harvard.edu/abs/2023ApJ...959...35K}
}

@ARTICLE{2023MNRAS.526.3495T,
       author = {{Talbot}, Colm and {Golomb}, Jacob},
        title = "{Growing pains: understanding the impact of likelihood uncertainty on hierarchical Bayesian inference for gravitational-wave astronomy}",
      journal = {MNRAS},
         year = 2023,
        month = dec,
       volume = {526},
       number = {3},
        pages = {3495-3503},
          doi = {10.1093/mnras/stad2968},
archivePrefix = {arXiv},
       eprint = {2304.06138},
 primaryClass = {astro-ph.IM},
       adsurl = {https://ui.adsabs.harvard.edu/abs/2023MNRAS.526.3495T}
}

@ARTICLE{2023JCAP...12..023G,
       author = {{Gray}, Rachel and {Beirnaert}, Freija and {Karathanasis}, Christos and {Revenu}, Beno{\^\i}t and {Turski}, Cezary and {Chen}, Anson and {Baker}, Tessa and {Vallejo}, Sergio and {Romano}, Antonio Enea and {Ghosh}, Tathagata and others},
        title = "{Joint cosmological and gravitational-wave population inference using dark sirens and galaxy catalogues}",
      journal = {JCAP},
         year = 2023,
        month = dec,
       volume = {2023},
       number = {12},
          eid = {023},
        pages = {023},
          doi = {10.1088/1475-7516/2023/12/023},
archivePrefix = {arXiv},
       eprint = {2308.02281},
 primaryClass = {astro-ph.CO},
       adsurl = {https://ui.adsabs.harvard.edu/abs/2023JCAP...12..023G}
}

@ARTICLE{2023RNAAS...7..250B,
       author = {{Ballard}, W. and {Palmese}, A. and {Hernandez}, I. Maga{\~n}a and {BenZvi}, S. and {Moon}, J. and {Ross}, A.~J. and {Rossi}, G. and {Aguilar}, J. and {Ahlen}, S. and {Blum}, R. and others},
        title = "{A Dark Siren Measurement of the Hubble Constant with the LIGO/Virgo Gravitational Wave Event GW190412 and DESI Galaxies}",
      journal = {RNAAS},
         year = 2023,
        month = nov,
       volume = {7},
       number = {11},
          eid = {250},
        pages = {250},
          doi = {10.3847/2515-5172/ad0eda},
archivePrefix = {arXiv},
       eprint = {2311.13062},
 primaryClass = {astro-ph.CO},
       adsurl = {https://ui.adsabs.harvard.edu/abs/2023RNAAS...7..250B}
}

@ARTICLE{2023PhRvD.108j3017V,
       author = {{Vijaykumar}, Aditya and {Saketh}, M.~V.~S. and {Kumar}, Sumit and {Ajith}, Parameswaran and {Choudhury}, Tirthankar Roy},
        title = "{Probing the large scale structure using gravitational-wave observations of binary black holes}",
      journal = {PhRvD},
         year = 2023,
        month = nov,
       volume = {108},
       number = {10},
          eid = {103017},
        pages = {103017},
          doi = {10.1103/PhysRevD.108.103017},
archivePrefix = {arXiv},
       eprint = {2005.01111},
 primaryClass = {astro-ph.CO},
       adsurl = {https://ui.adsabs.harvard.edu/abs/2023PhRvD.108j3017V}
}

@ARTICLE{2023JCAP...08..050F,
       author = {{Fonseca}, Jos{\'e} and {Zazzera}, Stefano and {Baker}, Tessa and {Clarkson}, Chris},
        title = "{The observed number counts in luminosity distance space}",
      journal = {JCAP},
         year = 2023,
        month = aug,
       volume = {2023},
       number = {8},
          eid = {050},
        pages = {050},
          doi = {10.1088/1475-7516/2023/08/050},
archivePrefix = {arXiv},
       eprint = {2304.14253},
 primaryClass = {astro-ph.CO},
       adsurl = {https://ui.adsabs.harvard.edu/abs/2023JCAP...08..050F}
}

@ARTICLE{2023PhRvD.108d2002M,
       author = {{Mastrogiovanni}, Simone and {Laghi}, Danny and {Gray}, Rachel and {Santoro}, Giada Caneva and {Ghosh}, Archisman and {Karathanasis}, Christos and {Leyde}, Konstantin and {Steer}, Dani{\`e}le A. and {Perri{\`e}s}, St{\'e}phane and {Pierra}, Gr{\'e}goire},
        title = "{Joint population and cosmological properties inference with gravitational waves standard sirens and galaxy surveys}",
      journal = {PhRvD},
         year = 2023,
        month = aug,
       volume = {108},
       number = {4},
          eid = {042002},
        pages = {042002},
          doi = {10.1103/PhysRevD.108.042002},
archivePrefix = {arXiv},
       eprint = {2305.10488},
 primaryClass = {astro-ph.CO},
       adsurl = {https://ui.adsabs.harvard.edu/abs/2023PhRvD.108d2002M}
}

@ARTICLE{2023JCAP...07..068B,
       author = {{Branchesi}, Marica and {Maggiore}, Michele and {Alonso}, David and {Badger}, Charles and {Banerjee}, Biswajit and {Beirnaert}, Freija and {Belgacem}, Enis and {Bhagwat}, Swetha and {Boileau}, Guillaume and {Borhanian}, Ssohrab and others},
        title = "{Science with the Einstein Telescope: a comparison of different designs}",
      journal = {JCAP},
         year = 2023,
        month = jul,
       volume = {2023},
       number = {7},
          eid = {068},
        pages = {068},
          doi = {10.1088/1475-7516/2023/07/068},
archivePrefix = {arXiv},
       eprint = {2303.15923},
 primaryClass = {gr-qc},
       adsurl = {https://ui.adsabs.harvard.edu/abs/2023JCAP...07..068B}
}

@ARTICLE{2023AJ....166...22G,
       author = {{Gair}, Jonathan R. and {Ghosh}, Archisman and {Gray}, Rachel and {Holz}, Daniel E. and {Mastrogiovanni}, Simone and {Mukherjee}, Suvodip and {Palmese}, Antonella and {Tamanini}, Nicola and {Baker}, Tessa and {Beirnaert}, Freija and others},
        title = "{The Hitchhiker's Guide to the Galaxy Catalog Approach for Dark Siren Gravitational-wave Cosmology}",
      journal = {AJ},
         year = 2023,
        month = jul,
       volume = {166},
       number = {1},
          eid = {22},
        pages = {22},
          doi = {10.3847/1538-3881/acca78},
archivePrefix = {arXiv},
       eprint = {2212.08694},
 primaryClass = {gr-qc},
       adsurl = {https://ui.adsabs.harvard.edu/abs/2023AJ....166...22G}
}

@ARTICLE{2023ApJ...949...76A,
       author = {{Abbott}, R. and {Abe}, H. and {Acernese}, F. and {Ackley}, K. and {Adhikari}, N. and {Adhikari}, R.~X. and {Adkins}, V.~K. and {Adya}, V.~B. and {Affeldt}, C. and {Agarwal}, D. and others},
        title = "{Constraints on the Cosmic Expansion History from GWTC─3}",
      journal = {ApJ},
         year = 2023,
        month = jun,
       volume = {949},
       number = {2},
          eid = {76},
        pages = {76},
          doi = {10.3847/1538-4357/ac74bb},
archivePrefix = {arXiv},
       eprint = {2111.03604},
 primaryClass = {astro-ph.CO},
       adsurl = {https://ui.adsabs.harvard.edu/abs/2023ApJ...949...76A}
}

@ARTICLE{2023OJAp....6E..15C,
       author = {{Campagne}, Jean-Eric and {Lanusse}, Fran{\c{c}}ois and {Zuntz}, Joe and {Boucaud}, Alexandre and {Casas}, Santiago and {Karamanis}, Minas and {Kirkby}, David and {Lanzieri}, Denise and {Peel}, Austin and {Li}, Yin},
        title = "{JAX-COSMO: An End-to-End Differentiable and GPU Accelerated Cosmology Library}",
      journal = {OJAp},
         year = 2023,
        month = apr,
       volume = {6},
          eid = {15},
        pages = {15},
          doi = {10.21105/astro.2302.05163},
archivePrefix = {arXiv},
       eprint = {2302.05163},
 primaryClass = {astro-ph.CO},
       adsurl = {https://ui.adsabs.harvard.edu/abs/2023OJAp....6E..15C}
}

@ARTICLE{2023MNRAS.519.2736G,
       author = {{Gair}, Jonathan R. and {Antonelli}, Andrea and {Barbieri}, Riccardo},
        title = "{A Fisher matrix for gravitational-wave population inference}",
      journal = {MNRAS},
         year = 2023,
        month = feb,
       volume = {519},
       number = {2},
        pages = {2736-2753},
          doi = {10.1093/mnras/stac3560},
archivePrefix = {arXiv},
       eprint = {2205.07893},
 primaryClass = {gr-qc},
       adsurl = {https://ui.adsabs.harvard.edu/abs/2023MNRAS.519.2736G}
}

@ARTICLE{2023ApJ...943...56P,
       author = {{Palmese}, A. and {Bom}, C.~R. and {Mucesh}, S. and {Hartley}, W.~G.},
        title = "{A Standard Siren Measurement of the Hubble Constant Using Gravitational-wave Events from the First Three LIGO/Virgo Observing Runs and the DESI Legacy Survey}",
      journal = {ApJ},
         year = 2023,
        month = jan,
       volume = {943},
       number = {1},
          eid = {56},
        pages = {56},
          doi = {10.3847/1538-4357/aca6e3},
archivePrefix = {arXiv},
       eprint = {2111.06445},
 primaryClass = {astro-ph.CO},
       adsurl = {https://ui.adsabs.harvard.edu/abs/2023ApJ...943...56P}
}

@ARTICLE{2023PhRvX..13a1048A,
       author = {{Abbott}, R. and {Abbott}, T.~D. and {Acernese}, F. and {Ackley}, K. and {Adams}, C. and {Adhikari}, N. and {Adhikari}, R.~X. and {Adya}, V.~B. and {Affeldt}, C. and {Agarwal}, D. and others},
        title = "{Population of Merging Compact Binaries Inferred Using Gravitational Waves through GWTC-3}",
      journal = {PhRvX},
         year = 2023,
        month = jan,
       volume = {13},
       number = {1},
          eid = {011048},
        pages = {011048},
          doi = {10.1103/PhysRevX.13.011048},
archivePrefix = {arXiv},
       eprint = {2111.03634},
 primaryClass = {astro-ph.HE},
       adsurl = {https://ui.adsabs.harvard.edu/abs/2023PhRvX..13a1048A}
}

@ARTICLE{2022MNRAS.516.5355A,
       author = {{Amon}, Alexandra and {Efstathiou}, George},
        title = "{A non-linear solution to the S$_{8}$ tension?}",
      journal = {MNRAS},
         year = 2022,
        month = nov,
       volume = {516},
       number = {4},
        pages = {5355-5366},
          doi = {10.1093/mnras/stac2429},
archivePrefix = {arXiv},
       eprint = {2206.11794},
 primaryClass = {astro-ph.CO},
       adsurl = {https://ui.adsabs.harvard.edu/abs/2022MNRAS.516.5355A}
}

@ARTICLE{2022ApJS..263....2I,
       author = {{Iacovelli}, Francesco and {Mancarella}, Michele and {Foffa}, Stefano and {Maggiore}, Michele},
        title = "{GWFAST: A Fisher Information Matrix Python Code for Third-generation Gravitational-wave Detectors}",
      journal = {ApJS},
         year = 2022,
        month = nov,
       volume = {263},
       number = {1},
          eid = {2},
        pages = {2},
          doi = {10.3847/1538-4365/ac9129},
archivePrefix = {arXiv},
       eprint = {2207.06910},
 primaryClass = {astro-ph.IM},
       adsurl = {https://ui.adsabs.harvard.edu/abs/2022ApJS..263....2I}
}

@ARTICLE{2022JCAP...09..012L,
       author = {{Leyde}, K. and {Mastrogiovanni}, S. and {Steer}, D.~A. and {Chassande-Mottin}, E. and {Karathanasis}, C.},
        title = "{Current and future constraints on cosmology and modified gravitational wave friction from binary black holes}",
      journal = {JCAP},
         year = 2022,
        month = sep,
       volume = {2022},
       number = {9},
          eid = {012},
        pages = {012},
          doi = {10.1088/1475-7516/2022/09/012},
archivePrefix = {arXiv},
       eprint = {2203.11680},
 primaryClass = {gr-qc},
       adsurl = {https://ui.adsabs.harvard.edu/abs/2022JCAP...09..012L}
}

@ARTICLE{2022JCAP...08..073A,
       author = {{Abramo}, L. Raul and {Dinarte Ferri}, Jo{\~a}o Vitor and {Tashiro}, Ian Lucas and {Loureiro}, Arthur},
        title = "{Fisher matrix for the angular power spectrum of multi-tracer galaxy surveys}",
      journal = {JCAP},
         year = 2022,
        month = aug,
       volume = {2022},
       number = {8},
          eid = {073},
        pages = {073},
          doi = {10.1088/1475-7516/2022/08/073},
archivePrefix = {arXiv},
       eprint = {2204.05057},
 primaryClass = {astro-ph.CO},
       adsurl = {https://ui.adsabs.harvard.edu/abs/2022JCAP...08..073A}
}

@ARTICLE{2022PhRvL.129f1102E,
       author = {{Ezquiaga}, Jose Mar{\'\i}a and {Holz}, Daniel E.},
        title = "{Spectral Sirens: Cosmology from the Full Mass Distribution of Compact Binaries}",
      journal = {PhRvL},
         year = 2022,
        month = aug,
       volume = {129},
       number = {6},
          eid = {061102},
        pages = {061102},
          doi = {10.1103/PhysRevLett.129.061102},
archivePrefix = {arXiv},
       eprint = {2202.08240},
 primaryClass = {astro-ph.CO},
       adsurl = {https://ui.adsabs.harvard.edu/abs/2022PhRvL.129f1102E}
}

@ARTICLE{2022ApJ...934L...7R,
       author = {{Riess}, Adam G. and {Yuan}, Wenlong and {Macri}, Lucas M. and {Scolnic}, Dan and {Brout}, Dillon and {Casertano}, Stefano and {Jones}, David O. and {Murakami}, Yukei and {Anand}, Gagandeep S. and {Breuval}, Louise and others},
        title = "{A Comprehensive Measurement of the Local Value of the Hubble Constant with 1 km s$^{-1}$ Mpc$^{-1}$ Uncertainty from the Hubble Space Telescope and the SH0ES Team}",
      journal = {ApJL},
         year = 2022,
        month = jul,
       volume = {934},
       number = {1},
          eid = {L7},
        pages = {L7},
          doi = {10.3847/2041-8213/ac5c5b},
archivePrefix = {arXiv},
       eprint = {2112.04510},
 primaryClass = {astro-ph.CO},
       adsurl = {https://ui.adsabs.harvard.edu/abs/2022ApJ...934L...7R}
}

@ARTICLE{2022A&A...662A..93E,
       author = {{Euclid Collaboration} and {Lepori}, F. and {Tutusaus}, I. and {Viglione}, C. and {Bonvin}, C. and {Camera}, S. and {Castander}, F.~J. and {Durrer}, R. and {Fosalba}, P. and {Jelic-Cizmek}, G. and others},
        title = "{Euclid preparation. XIX. Impact of magnification on photometric galaxy clustering}",
      journal = {A\&A},
         year = 2022,
        month = jun,
       volume = {662},
          eid = {A93},
        pages = {A93},
          doi = {10.1051/0004-6361/202142419},
archivePrefix = {arXiv},
       eprint = {2110.05435},
 primaryClass = {astro-ph.CO},
       adsurl = {https://ui.adsabs.harvard.edu/abs/2022A&A...662A..93E}
}

@ARTICLE{2022A&A...662A.112E,
       author = {{Euclid Collaboration} and {Scaramella}, R. and {Amiaux}, J. and {Mellier}, Y. and {Burigana}, C. and {Carvalho}, C.~S. and {Cuillandre}, J.-C. and {Da Silva}, A. and {Derosa}, A. and {Dinis}, J. and others},
        title = "{Euclid preparation. I. The Euclid Wide Survey}",
      journal = {A\&A},
         year = 2022,
        month = jun,
       volume = {662},
          eid = {A112},
        pages = {A112},
          doi = {10.1051/0004-6361/202141938},
archivePrefix = {arXiv},
       eprint = {2108.01201},
 primaryClass = {astro-ph.CO},
       adsurl = {https://ui.adsabs.harvard.edu/abs/2022A&A...662A.112E}
}

@ARTICLE{2022MNRAS.511.2782C,
       author = {{Cigarr{\'a}n D{\'\i}az}, Cristina and {Mukherjee}, Suvodip},
        title = "{Mapping the cosmic expansion history from LIGO-Virgo-KAGRA in synergy with DESI and SPHEREx}",
      journal = {MNRAS},
         year = 2022,
        month = apr,
       volume = {511},
       number = {2},
        pages = {2782-2795},
          doi = {10.1093/mnras/stac208},
archivePrefix = {arXiv},
       eprint = {2107.12787},
 primaryClass = {astro-ph.CO},
       adsurl = {https://ui.adsabs.harvard.edu/abs/2022MNRAS.511.2782C}
}

@ARTICLE{2022JCAP...04..013A,
       author = {{Abramo}, L. Raul and {Ferri}, Jo{\~a}o Vitor Dinarte and {Tashiro}, Ian Lucas},
        title = "{Fisher matrix for multiple tracers: the information in the cross-spectra}",
      journal = {JCAP},
         year = 2022,
        month = apr,
       volume = {2022},
       number = {4},
          eid = {013},
        pages = {013},
          doi = {10.1088/1475-7516/2022/04/013},
archivePrefix = {arXiv},
       eprint = {2112.01812},
 primaryClass = {astro-ph.CO},
       adsurl = {https://ui.adsabs.harvard.edu/abs/2022JCAP...04..013A}
}

@INCOLLECTION{2022hgwa.bookE..45V,
       author = {{Vitale}, Salvatore and {Gerosa}, Davide and {Farr}, Will M. and {Taylor}, Stephen R.},
        title = "{Inferring the Properties of a Population of Compact Binaries in Presence of Selection Effects}",
    booktitle = {Handbook of Gravitational Wave Astronomy},
         year = 2022,
       editor = {{Bambi}, Cosimo and {Katsanevas}, Stavros and {Kokkotas}, Konstantinos D.},
          eid = {45},
        pages = {45},
          doi = {10.1007/978-981-15-4702-7_45-1},
       adsurl = {https://ui.adsabs.harvard.edu/abs/2022hgwa.bookE..45V}
}

@ARTICLE{2021A&A...655A..44E,
       author = {{Euclid Collaboration} and {Pocino}, A. and {Tutusaus}, I. and {Castander}, F.~J. and {Fosalba}, P. and {Crocce}, M. and {Porredon}, A. and {Camera}, S. and {Cardone}, V. and {Casas}, S. and others},
        title = "{Euclid preparation. XII. Optimizing the photometric sample of the Euclid survey for galaxy clustering and galaxy-galaxy lensing analyses}",
      journal = {A\&A},
         year = 2021,
        month = nov,
       volume = {655},
          eid = {A44},
        pages = {A44},
          doi = {10.1051/0004-6361/202141061},
archivePrefix = {arXiv},
       eprint = {2104.05698},
 primaryClass = {astro-ph.CO},
       adsurl = {https://ui.adsabs.harvard.edu/abs/2021A&A...655A..44E}
}

@ARTICLE{2021arXiv210909882E,
       author = {{Evans}, Matthew and {Adhikari}, Rana X and {Afle}, Chaitanya and {Ballmer}, Stefan W. and {Biscoveanu}, Sylvia and {Borhanian}, Ssohrab and {Brown}, Duncan A. and {Chen}, Yanbei and {Eisenstein}, Robert and {Gruson}, Alexandra and others},
        title = "{A Horizon Study for Cosmic Explorer: Science, Observatories, and Community}",
      journal = {arXiv},
         year = 2021,
        month = sep,
          eid = {arXiv:2109.09882},
        pages = {arXiv:2109.09882},
          doi = {10.48550/arXiv.2109.09882},
archivePrefix = {arXiv},
       eprint = {2109.09882},
 primaryClass = {astro-ph.IM},
       adsurl = {https://ui.adsabs.harvard.edu/abs/2021arXiv210909882E}
}

@ARTICLE{2021PhRvD.104f2009M,
       author = {{Mastrogiovanni}, S. and {Leyde}, K. and {Karathanasis}, C. and {Chassande-Mottin}, E. and {Steer}, D.~A. and {Gair}, J. and {Ghosh}, A. and {Gray}, R. and {Mukherjee}, S. and {Rinaldi}, S.},
        title = "{On the importance of source population models for gravitational-wave cosmology}",
      journal = {PhRvD},
         year = 2021,
        month = sep,
       volume = {104},
       number = {6},
          eid = {062009},
        pages = {062009},
          doi = {10.1103/PhysRevD.104.062009},
archivePrefix = {arXiv},
       eprint = {2103.14663},
 primaryClass = {gr-qc},
       adsurl = {https://ui.adsabs.harvard.edu/abs/2021PhRvD.104f2009M}
}

@ARTICLE{2021CQGra..38o3001D,
       author = {{Di Valentino}, Eleonora and {Mena}, Olga and {Pan}, Supriya and {Visinelli}, Luca and {Yang}, Weiqiang and {Melchiorri}, Alessandro and {Mota}, David F. and {Riess}, Adam G. and {Silk}, Joseph},
        title = "{In the realm of the Hubble tension-a review of solutions}",
      journal = {CQGra},
         year = 2021,
        month = jul,
       volume = {38},
       number = {15},
          eid = {153001},
        pages = {153001},
          doi = {10.1088/1361-6382/ac086d},
archivePrefix = {arXiv},
       eprint = {2103.01183},
 primaryClass = {astro-ph.CO},
       adsurl = {https://ui.adsabs.harvard.edu/abs/2021CQGra..38o3001D}
}

@ARTICLE{2021JCAP...02..035L,
       author = {{Libanore}, S. and {Artale}, M.~C. and {Karagiannis}, D. and {Liguori}, M. and {Bartolo}, N. and {Bouffanais}, Y. and {Giacobbo}, N. and {Mapelli}, M. and {Matarrese}, S.},
        title = "{Gravitational Wave mergers as tracers of Large Scale Structures}",
      journal = {JCAP},
         year = 2021,
        month = feb,
       volume = {2021},
       number = {2},
          eid = {035},
        pages = {035},
          doi = {10.1088/1475-7516/2021/02/035},
archivePrefix = {arXiv},
       eprint = {2007.06905},
 primaryClass = {astro-ph.CO},
       adsurl = {https://ui.adsabs.harvard.edu/abs/2021JCAP...02..035L}
}

@ARTICLE{2021ApJ...908L...4C,
       author = {{Chen}, Hsin-Yu and {Cowperthwaite}, Philip S. and {Metzger}, Brian D. and {Berger}, Edo},
        title = "{A Program for Multimessenger Standard Siren Cosmology in the Era of LIGO A+, Rubin Observatory, and Beyond}",
      journal = {ApJL},
         year = 2021,
        month = feb,
       volume = {908},
       number = {1},
          eid = {L4},
        pages = {L4},
          doi = {10.3847/2041-8213/abdab0},
archivePrefix = {arXiv},
       eprint = {2011.01211},
 primaryClass = {astro-ph.CO},
       adsurl = {https://ui.adsabs.harvard.edu/abs/2021ApJ...908L...4C}
}

@ARTICLE{2021PhRvD.103d3520M,
       author = {{Mukherjee}, Suvodip and {Wandelt}, Benjamin D. and {Nissanke}, Samaya M. and {Silvestri}, Alessandra},
        title = "{Accurate precision cosmology with redshift unknown gravitational wave sources}",
      journal = {PhRvD},
         year = 2021,
        month = feb,
       volume = {103},
       number = {4},
          eid = {043520},
        pages = {043520},
          doi = {10.1103/PhysRevD.103.043520},
archivePrefix = {arXiv},
       eprint = {2007.02943},
 primaryClass = {astro-ph.CO},
       adsurl = {https://ui.adsabs.harvard.edu/abs/2021PhRvD.103d3520M}
}

@ARTICLE{2021JCAP...01..036N,
       author = {{Namikawa}, Toshiya},
        title = "{Analyzing clustering of astrophysical gravitational-wave sources: luminosity-distance space distortions}",
      journal = {JCAP},
         year = 2021,
        month = jan,
       volume = {2021},
       number = {1},
          eid = {036},
        pages = {036},
          doi = {10.1088/1475-7516/2021/01/036},
archivePrefix = {arXiv},
       eprint = {2007.04359},
 primaryClass = {astro-ph.CO},
       adsurl = {https://ui.adsabs.harvard.edu/abs/2021JCAP...01..036N}
}

@ARTICLE{2020ApJ...900L..33P,
       author = {{Palmese}, A. and {deVicente}, J. and {Pereira}, M.~E.~S. and {Annis}, J. and {Hartley}, W. and {Herner}, K. and {Soares-Santos}, M. and {Crocce}, M. and {Huterer}, D. and {Maga{\~n}a Hernandez}, I. and others},
        title = "{A Statistical Standard Siren Measurement of the Hubble Constant from the LIGO/Virgo Gravitational Wave Compact Object Merger GW190814 and Dark Energy Survey Galaxies}",
      journal = {ApJL},
         year = 2020,
        month = sep,
       volume = {900},
       number = {2},
          eid = {L33},
        pages = {L33},
          doi = {10.3847/2041-8213/abaeff},
archivePrefix = {arXiv},
       eprint = {2006.14961},
 primaryClass = {astro-ph.CO},
       adsurl = {https://ui.adsabs.harvard.edu/abs/2020ApJ...900L..33P}
}

@ARTICLE{2020A&A...641A...6P,
       author = {{Planck Collaboration} and {Aghanim}, N. and {Akrami}, Y. and {Ashdown}, M. and {Aumont}, J. and {Baccigalupi}, C. and {Ballardini}, M. and {Banday}, A.~J. and {Barreiro}, R.~B. and {Bartolo}, N. and others},
        title = "{Planck 2018 results. VI. Cosmological parameters}",
      journal = {A\&A},
         year = 2020,
        month = sep,
       volume = {641},
          eid = {A6},
        pages = {A6},
          doi = {10.1051/0004-6361/201833910},
archivePrefix = {arXiv},
       eprint = {1807.06209},
 primaryClass = {astro-ph.CO},
       adsurl = {https://ui.adsabs.harvard.edu/abs/2020A&A...641A...6P}
}

@ARTICLE{2020PhRvD.102b3528R,
       author = {{Rafiei-Ravandi}, Masoud and {Smith}, Kendrick M. and {Masui}, Kiyoshi W.},
        title = "{Characterizing fast radio bursts through statistical cross-correlations}",
      journal = {PhRvD},
         year = 2020,
        month = jul,
       volume = {102},
       number = {2},
          eid = {023528},
        pages = {023528},
          doi = {10.1103/PhysRevD.102.023528},
archivePrefix = {arXiv},
       eprint = {1912.09520},
 primaryClass = {astro-ph.CO},
       adsurl = {https://ui.adsabs.harvard.edu/abs/2020PhRvD.102b3528R}
}

@ARTICLE{2020PhRvD.101l2001G,
       author = {{Gray}, Rachel and {Hernandez}, Ignacio Maga{\~n}a and {Qi}, Hong and {Sur}, Ankan and {Brady}, Patrick R. and {Chen}, Hsin-Yu and {Farr}, Will M. and {Fishbach}, Maya and {Gair}, Jonathan R. and {Ghosh}, Archisman and others},
        title = "{Cosmological inference using gravitational wave standard sirens: A mock data analysis}",
      journal = {PhRvD},
         year = 2020,
        month = jun,
       volume = {101},
       number = {12},
          eid = {122001},
        pages = {122001},
          doi = {10.1103/PhysRevD.101.122001},
archivePrefix = {arXiv},
       eprint = {1908.06050},
 primaryClass = {gr-qc},
       adsurl = {https://ui.adsabs.harvard.edu/abs/2020PhRvD.101l2001G}
}

@ARTICLE{2020ApJ...896L..32C,
       author = {{Callister}, Tom and {Fishbach}, Maya and {Holz}, Daniel E. and {Farr}, Will M.},
        title = "{Shouts and Murmurs: Combining Individual Gravitational-wave Sources with the Stochastic Background to Measure the History of Binary Black Hole Mergers}",
      journal = {ApJL},
         year = 2020,
        month = jun,
       volume = {896},
       number = {2},
          eid = {L32},
        pages = {L32},
          doi = {10.3847/2041-8213/ab9743},
archivePrefix = {arXiv},
       eprint = {2003.12152},
 primaryClass = {astro-ph.HE},
       adsurl = {https://ui.adsabs.harvard.edu/abs/2020ApJ...896L..32C}
}

@ARTICLE{2020MNRAS.494.1956M,
       author = {{Mukherjee}, Suvodip and {Wandelt}, Benjamin D. and {Silk}, Joseph},
        title = "{Probing the theory of gravity with gravitational lensing of gravitational waves and galaxy surveys}",
      journal = {MNRAS},
         year = 2020,
        month = may,
       volume = {494},
       number = {2},
        pages = {1956-1970},
          doi = {10.1093/mnras/staa827},
archivePrefix = {arXiv},
       eprint = {1908.08951},
 primaryClass = {astro-ph.CO},
       adsurl = {https://ui.adsabs.harvard.edu/abs/2020MNRAS.494.1956M}
}

@BOOK{2020moco.book.....D,
       author = {{Dodelson}, Scott and {Schmidt}, Fabian},
        title = "{Modern Cosmology}",
         year = 2020,
          doi = {10.1016/C2017-0-01943-2},
       adsurl = {https://ui.adsabs.harvard.edu/abs/2020moco.book.....D}
}

@ARTICLE{2019arXiv191211554P,
       author = {{Phan}, Du and {Pradhan}, Neeraj and {Jankowiak}, Martin},
        title = "{Composable Effects for Flexible and Accelerated Probabilistic Programming in NumPyro}",
      journal = {arXiv},
         year = 2019,
        month = dec,
          eid = {arXiv:1912.11554},
        pages = {arXiv:1912.11554},
          doi = {10.48550/arXiv.1912.11554},
archivePrefix = {arXiv},
       eprint = {1912.11554},
 primaryClass = {stat.ML},
       adsurl = {https://ui.adsabs.harvard.edu/abs/2019arXiv191211554P}
}

@ARTICLE{2019PhRvD.100l3540W,
       author = {{Winther}, Hans A. and {Casas}, Santiago and {Baldi}, Marco and {Koyama}, Kazuya and {Li}, Baojiu and {Lombriser}, Lucas and {Zhao}, Gong-Bo},
        title = "{Emulators for the nonlinear matter power spectrum beyond {\ensuremath{\Lambda}} CDM}",
      journal = {PhRvD},
         year = 2019,
        month = dec,
       volume = {100},
       number = {12},
          eid = {123540},
        pages = {123540},
          doi = {10.1103/PhysRevD.100.123540},
archivePrefix = {arXiv},
       eprint = {1903.08798},
 primaryClass = {astro-ph.CO},
       adsurl = {https://ui.adsabs.harvard.edu/abs/2019PhRvD.100l3540W}
}

@ARTICLE{2019NatAs...3..891V,
       author = {{Verde}, Licia and {Treu}, Tommaso and {Riess}, Adam G.},
        title = "{Tensions between the early and late Universe}",
      journal = {NatAs},
         year = 2019,
        month = sep,
       volume = {3},
        pages = {891-895},
          doi = {10.1038/s41550-019-0902-0},
archivePrefix = {arXiv},
       eprint = {1907.10625},
 primaryClass = {astro-ph.CO},
       adsurl = {https://ui.adsabs.harvard.edu/abs/2019NatAs...3..891V}
}

@software{2019ascl.soft06017L,
       author = {{Li}, Yin},
        title = "{mcfit: Multiplicatively Convolutional Fast Integral Transforms}",
 howpublished = {Astrophysics Source Code Library, record ascl:1906.017},
         year = 2019,
        month = jun,
          eid = {ascl:1906.017},
archivePrefix = {ascl},
       eprint = {1906.017},
       adsurl = {https://ui.adsabs.harvard.edu/abs/2019ascl.soft06017L}
}

@ARTICLE{2019OJAp....2E...4C,
       author = {{Chisari}, Nora Elisa and {Mead}, Alexander J. and {Joudaki}, Shahab and {Ferreira}, Pedro G. and {Schneider}, Aurel and {Mohr}, Joseph and {Tr{\"o}ster}, Tilman and {Alonso}, David and {McCarthy}, Ian G. and {Martin-Alvarez}, Sergio and others},
        title = "{Modelling baryonic feedback for survey cosmology}",
      journal = {OJAp},
         year = 2019,
        month = jun,
       volume = {2},
       number = {1},
          eid = {4},
        pages = {4},
          doi = {10.21105/astro.1905.06082},
archivePrefix = {arXiv},
       eprint = {1905.06082},
 primaryClass = {astro-ph.CO},
       adsurl = {https://ui.adsabs.harvard.edu/abs/2019OJAp....2E...4C}
}

@ARTICLE{2019MNRAS.486.1086M,
       author = {{Mandel}, Ilya and {Farr}, Will M. and {Gair}, Jonathan R.},
        title = "{Extracting distribution parameters from multiple uncertain observations with selection biases}",
      journal = {MNRAS},
         year = 2019,
        month = jun,
       volume = {486},
       number = {1},
        pages = {1086-1093},
          doi = {10.1093/mnras/stz896},
archivePrefix = {arXiv},
       eprint = {1809.02063},
 primaryClass = {physics.data-an},
       adsurl = {https://ui.adsabs.harvard.edu/abs/2019MNRAS.486.1086M}
}

@ARTICLE{2019ApJ...876L...7S,
       author = {{Soares-Santos}, M. and {Palmese}, A. and {Hartley}, W. and {Annis}, J. and {Garcia-Bellido}, J. and {Lahav}, O. and {Doctor}, Z. and {Fishbach}, M. and {Holz}, D.~E. and {Lin}, H. and others},
        title = "{First Measurement of the Hubble Constant from a Dark Standard Siren using the Dark Energy Survey Galaxies and the LIGO/Virgo Binary-Black-hole Merger GW170814}",
      journal = {ApJL},
         year = 2019,
        month = may,
       volume = {876},
       number = {1},
          eid = {L7},
        pages = {L7},
          doi = {10.3847/2041-8213/ab14f1},
archivePrefix = {arXiv},
       eprint = {1901.01540},
 primaryClass = {astro-ph.CO},
       adsurl = {https://ui.adsabs.harvard.edu/abs/2019ApJ...876L...7S}
}

@ARTICLE{2019MNRAS.484.4127A,
       author = {{Alonso}, David and {Sanchez}, Javier and {Slosar}, An{\v{z}}e and {LSST Dark Energy Science Collaboration}},
        title = "{A unified pseudo-C$_{{\ensuremath{\ell}}}$ framework}",
      journal = {MNRAS},
         year = 2019,
        month = apr,
       volume = {484},
       number = {3},
        pages = {4127-4151},
          doi = {10.1093/mnras/stz093},
archivePrefix = {arXiv},
       eprint = {1809.09603},
 primaryClass = {astro-ph.CO},
       adsurl = {https://ui.adsabs.harvard.edu/abs/2019MNRAS.484.4127A}
}

@ARTICLE{2019JOSS....4.1298Z,
       author = {{Zonca}, Andrea and {Singer}, Leo and {Lenz}, Daniel and {Reinecke}, Martin and {Rosset}, Cyrille and {Hivon}, Eric and {Gorski}, Krzysztof},
        title = "{healpy: equal area pixelization and spherical harmonics transforms for data on the sphere in Python}",
      journal = {JOSS},
         year = 2019,
        month = mar,
       volume = {4},
       number = {35},
          eid = {1298},
        pages = {1298},
          doi = {10.21105/joss.01298},
       adsurl = {https://ui.adsabs.harvard.edu/abs/2019JOSS....4.1298Z}
}

@ARTICLE{2019PASA...36...10T,
       author = {{Thrane}, Eric and {Talbot}, Colm},
        title = "{An introduction to Bayesian inference in gravitational-wave astronomy: Parameter estimation, model selection, and hierarchical models}",
      journal = {PASA},
         year = 2019,
        month = mar,
       volume = {36},
          eid = {e010},
        pages = {e010},
          doi = {10.1017/pasa.2019.2},
archivePrefix = {arXiv},
       eprint = {1809.02293},
 primaryClass = {astro-ph.IM},
       adsurl = {https://ui.adsabs.harvard.edu/abs/2019PASA...36...10T}
}

@ARTICLE{2019ApJ...873..111I,
       author = {{Ivezi{\'c}}, {\v{Z}}eljko and {Kahn}, Steven M. and {Tyson}, J. Anthony and {Abel}, Bob and {Acosta}, Emily and {Allsman}, Robyn and {Alonso}, David and {AlSayyad}, Yusra and {Anderson}, Scott F. and {Andrew}, John and others},
        title = "{LSST: From Science Drivers to Reference Design and Anticipated Data Products}",
      journal = {ApJ},
         year = 2019,
        month = mar,
       volume = {873},
       number = {2},
          eid = {111},
        pages = {111},
          doi = {10.3847/1538-4357/ab042c},
archivePrefix = {arXiv},
       eprint = {0805.2366},
 primaryClass = {astro-ph},
       adsurl = {https://ui.adsabs.harvard.edu/abs/2019ApJ...873..111I}
}

@ARTICLE{2019ApJ...871L..13F,
       author = {{Fishbach}, M. and {Gray}, R. and {Maga{\~n}a Hernandez}, I. and {Qi}, H. and {Sur}, A. and {Acernese}, F. and {Aiello}, L. and {Allocca}, A. and {Aloy}, M.~A. and {Amato}, A. and others},
        title = "{A Standard Siren Measurement of the Hubble Constant from GW170817 without the Electromagnetic Counterpart}",
      journal = {ApJL},
         year = 2019,
        month = jan,
       volume = {871},
       number = {1},
          eid = {L13},
        pages = {L13},
          doi = {10.3847/2041-8213/aaf96e},
archivePrefix = {arXiv},
       eprint = {1807.05667},
 primaryClass = {astro-ph.CO},
       adsurl = {https://ui.adsabs.harvard.edu/abs/2019ApJ...871L..13F}
}

@ARTICLE{2018PhRvD..98h3524O,
       author = {{Osato}, Ken},
        title = "{Exploring the distance-redshift relation with gravitational wave standard sirens and tomographic weak lensing}",
      journal = {PhRvD},
         year = 2018,
        month = oct,
       volume = {98},
       number = {8},
          eid = {083524},
        pages = {083524},
          doi = {10.1103/PhysRevD.98.083524},
archivePrefix = {arXiv},
       eprint = {1807.00016},
 primaryClass = {astro-ph.CO},
       adsurl = {https://ui.adsabs.harvard.edu/abs/2018PhRvD..98h3524O}
}

@ARTICLE{2018arXiv181011915Z,
       author = {{Zhang}, Pengjie},
        title = "{The large scale structure in the 3D luminosity-distance space and its cosmological applications}",
      journal = {arXiv},
         year = 2018,
        month = oct,
          eid = {arXiv:1810.11915},
        pages = {arXiv:1810.11915},
          doi = {10.48550/arXiv.1810.11915},
archivePrefix = {arXiv},
       eprint = {1810.11915},
 primaryClass = {astro-ph.CO},
       adsurl = {https://ui.adsabs.harvard.edu/abs/2018arXiv181011915Z}
}

@ARTICLE{2018Natur.562..545C,
       author = {{Chen}, Hsin-Yu and {Fishbach}, Maya and {Holz}, Daniel E.},
        title = "{A two per cent Hubble constant measurement from standard sirens within five years}",
      journal = {Natur},
         year = 2018,
        month = oct,
       volume = {562},
       number = {7728},
        pages = {545-547},
          doi = {10.1038/s41586-018-0606-0},
archivePrefix = {arXiv},
       eprint = {1712.06531},
 primaryClass = {astro-ph.CO},
       adsurl = {https://ui.adsabs.harvard.edu/abs/2018Natur.562..545C}
}

@ARTICLE{2018ApJ...856..173T,
       author = {{Talbot}, Colm and {Thrane}, Eric},
        title = "{Measuring the Binary Black Hole Mass Spectrum with an Astrophysically Motivated Parameterization}",
      journal = {ApJ},
         year = 2018,
        month = apr,
       volume = {856},
       number = {2},
          eid = {173},
        pages = {173},
          doi = {10.3847/1538-4357/aab34c},
archivePrefix = {arXiv},
       eprint = {1801.02699},
 primaryClass = {astro-ph.HE},
       adsurl = {https://ui.adsabs.harvard.edu/abs/2018ApJ...856..173T}
}

@ARTICLE{2018PhRvL.120p1102L,
       author = {{London}, Lionel and {Khan}, Sebastian and {Fauchon-Jones}, Edward and {Garc{\'\i}a}, Cecilio and {Hannam}, Mark and {Husa}, Sascha and {Jim{\'e}nez-Forteza}, Xisco and {Kalaghatgi}, Chinmay and {Ohme}, Frank and {Pannarale}, Francesco},
        title = "{First Higher-Multipole Model of Gravitational Waves from Spinning and Coalescing Black-Hole Binaries}",
      journal = {PhRvL},
         year = 2018,
        month = apr,
       volume = {120},
       number = {16},
          eid = {161102},
        pages = {161102},
          doi = {10.1103/PhysRevLett.120.161102},
archivePrefix = {arXiv},
       eprint = {1708.00404},
 primaryClass = {gr-qc},
       adsurl = {https://ui.adsabs.harvard.edu/abs/2018PhRvL.120p1102L}
}

@ARTICLE{2017Natur.551...85A,
       author = {{Abbott}, B.~P. and {Abbott}, R. and {Abbott}, T.~D. and {Acernese}, F. and {Ackley}, K. and {Adams}, C. and {Adams}, T. and {Addesso}, P. and {Adhikari}, R.~X. and {Adya}, V.~B. and others},
        title = "{A gravitational-wave standard siren measurement of the Hubble constant}",
      journal = {Natur},
         year = 2017,
        month = nov,
       volume = {551},
       number = {7678},
        pages = {85-88},
          doi = {10.1038/nature24471},
archivePrefix = {arXiv},
       eprint = {1710.05835},
 primaryClass = {astro-ph.CO},
       adsurl = {https://ui.adsabs.harvard.edu/abs/2017Natur.551...85A}
}

@ARTICLE{2017JCAP...10..003A,
       author = {{Agrawal}, Aniket and {Makiya}, Ryu and {Chiang}, Chi-Ting and {Jeong}, Donghui and {Saito}, Shun and {Komatsu}, Eiichiro},
        title = "{Generating log-normal mock catalog of galaxies in redshift space}",
      journal = {JCAP},
         year = 2017,
        month = oct,
       volume = {2017},
       number = {10},
          eid = {003},
        pages = {003},
          doi = {10.1088/1475-7516/2017/10/003},
archivePrefix = {arXiv},
       eprint = {1706.09195},
 primaryClass = {astro-ph.CO},
       adsurl = {https://ui.adsabs.harvard.edu/abs/2017JCAP...10..003A}
}

@ARTICLE{2017ApJ...835...31C,
       author = {{Chen}, Hsin-Yu and {Essick}, Reed and {Vitale}, Salvatore and {Holz}, Daniel E. and {Katsavounidis}, Erik},
        title = "{Observational Selection Effects with Ground-based Gravitational Wave Detectors}",
      journal = {ApJ},
         year = 2017,
        month = jan,
       volume = {835},
       number = {1},
          eid = {31},
        pages = {31},
          doi = {10.3847/1538-4357/835/1/31},
archivePrefix = {arXiv},
       eprint = {1608.00164},
 primaryClass = {astro-ph.HE},
       adsurl = {https://ui.adsabs.harvard.edu/abs/2017ApJ...835...31C}
}

@ARTICLE{2017arXiv170102434B,
       author = {{Betancourt}, Michael},
        title = "{A Conceptual Introduction to Hamiltonian Monte Carlo}",
      journal = {arXiv},
         year = 2017,
        month = jan,
          eid = {arXiv:1701.02434},
        pages = {arXiv:1701.02434},
          doi = {10.48550/arXiv.1701.02434},
archivePrefix = {arXiv},
       eprint = {1701.02434},
 primaryClass = {stat.ME},
       adsurl = {https://ui.adsabs.harvard.edu/abs/2017arXiv170102434B}
}

@ARTICLE{2016arXiv161100036D,
       author = {{DESI Collaboration} and {Aghamousa}, Amir and {Aguilar}, Jessica and {Ahlen}, Steve and {Alam}, Shadab and {Allen}, Lori E. and {Allende Prieto}, Carlos and {Annis}, James and {Bailey}, Stephen and {Balland}, Christophe and others},
        title = "{The DESI Experiment Part I: Science,Targeting, and Survey Design}",
      journal = {arXiv},
         year = 2016,
        month = oct,
          eid = {arXiv:1611.00036},
        pages = {arXiv:1611.00036},
          doi = {10.48550/arXiv.1611.00036},
archivePrefix = {arXiv},
       eprint = {1611.00036},
 primaryClass = {astro-ph.IM},
       adsurl = {https://ui.adsabs.harvard.edu/abs/2016arXiv161100036D}
}

@ARTICLE{2016A&A...594A..13P,
       author = {{Planck Collaboration} and {Ade}, P.~A.~R. and {Aghanim}, N. and {Arnaud}, M. and {Ashdown}, M. and {Aumont}, J. and {Baccigalupi}, C. and {Banday}, A.~J. and {Barreiro}, R.~B. and {Bartlett}, J.~G. and others},
        title = "{Planck 2015 results. XIII. Cosmological parameters}",
      journal = {A\&A},
         year = 2016,
        month = sep,
       volume = {594},
          eid = {A13},
        pages = {A13},
          doi = {10.1051/0004-6361/201525830},
archivePrefix = {arXiv},
       eprint = {1502.01589},
 primaryClass = {astro-ph.CO},
       adsurl = {https://ui.adsabs.harvard.edu/abs/2016A&A...594A..13P}
}

@ARTICLE{2016ApJS..225....5B,
       author = {{Bilicki}, Maciej and {Peacock}, John A. and {Jarrett}, Thomas H. and {Cluver}, Michelle E. and {Maddox}, Natasha and {Brown}, Michael J.~I. and {Taylor}, Edward N. and {Hambly}, Nigel C. and {Solarz}, Aleksandra and {Holwerda}, Benne W. and others},
        title = "{WISE {\texttimes} SuperCOSMOS Photometric Redshift Catalog: 20 Million Galaxies over 3/pi Steradians}",
      journal = {ApJS},
         year = 2016,
        month = jul,
       volume = {225},
       number = {1},
          eid = {5},
        pages = {5},
          doi = {10.3847/0067-0049/225/1/5},
archivePrefix = {arXiv},
       eprint = {1607.01182},
 primaryClass = {astro-ph.CO},
       adsurl = {https://ui.adsabs.harvard.edu/abs/2016ApJS..225....5B}
}

@ARTICLE{2016PhRvD..94b4013N,
       author = {{Namikawa}, Toshiya and {Nishizawa}, Atsushi and {Taruya}, Atsushi},
        title = "{Detecting black-hole binary clustering via the second-generation gravitational-wave detectors}",
      journal = {PhRvD},
         year = 2016,
        month = jul,
       volume = {94},
       number = {2},
          eid = {024013},
        pages = {024013},
          doi = {10.1103/PhysRevD.94.024013},
archivePrefix = {arXiv},
       eprint = {1603.08072},
 primaryClass = {astro-ph.CO},
       adsurl = {https://ui.adsabs.harvard.edu/abs/2016PhRvD..94b4013N}
}

@ARTICLE{2016PhRvD..93h3511O,
       author = {{Oguri}, Masamune},
        title = "{Measuring the distance-redshift relation with the cross-correlation of gravitational wave standard sirens and galaxies}",
      journal = {PhRvD},
         year = 2016,
        month = apr,
       volume = {93},
       number = {8},
          eid = {083511},
        pages = {083511},
          doi = {10.1103/PhysRevD.93.083511},
archivePrefix = {arXiv},
       eprint = {1603.02356},
 primaryClass = {astro-ph.CO},
       adsurl = {https://ui.adsabs.harvard.edu/abs/2016PhRvD..93h3511O}
}

@ARTICLE{2016PhRvL.116l1302N,
       author = {{Namikawa}, Toshiya and {Nishizawa}, Atsushi and {Taruya}, Atsushi},
        title = "{Anisotropies of Gravitational-Wave Standard Sirens as a New Cosmological Probe without Redshift Information}",
      journal = {PhRvL},
         year = 2016,
        month = mar,
       volume = {116},
       number = {12},
          eid = {121302},
        pages = {121302},
          doi = {10.1103/PhysRevLett.116.121302},
archivePrefix = {arXiv},
       eprint = {1511.04638},
 primaryClass = {astro-ph.CO},
       adsurl = {https://ui.adsabs.harvard.edu/abs/2016PhRvL.116l1302N}
}

@ARTICLE{2015MNRAS.454.2770T,
       author = {{Torrey}, Paul and {Wellons}, Sarah and {Machado}, Francisco and {Griffen}, Brendan and {Nelson}, Dylan and {Rodriguez-Gomez}, Vicente and {McKinnon}, Ryan and {Pillepich}, Annalisa and {Ma}, Chung-Pei and {Vogelsberger}, Mark and others},
        title = "{An analysis of the evolving comoving number density of galaxies in hydrodynamical simulations}",
      journal = {MNRAS},
         year = 2015,
        month = dec,
       volume = {454},
       number = {3},
        pages = {2770-2786},
          doi = {10.1093/mnras/stv1986},
archivePrefix = {arXiv},
       eprint = {1507.01942},
 primaryClass = {astro-ph.GA},
       adsurl = {https://ui.adsabs.harvard.edu/abs/2015MNRAS.454.2770T}
}

@ARTICLE{2014ARA&A..52..415M,
       author = {{Madau}, Piero and {Dickinson}, Mark},
        title = "{Cosmic Star-Formation History}",
      journal = {ARA\&A},
         year = 2014,
        month = aug,
       volume = {52},
        pages = {415-486},
          doi = {10.1146/annurev-astro-081811-125615},
archivePrefix = {arXiv},
       eprint = {1403.0007},
 primaryClass = {astro-ph.CO},
       adsurl = {https://ui.adsabs.harvard.edu/abs/2014ARA&A..52..415M}
}

@ARTICLE{2012MNRAS.427.1891A,
       author = {{Asorey}, Jacobo and {Crocce}, Martin and {Gazta{\~n}aga}, Enrique and {Lewis}, Antony},
        title = "{Recovering 3D clustering information with angular correlations}",
      journal = {MNRAS},
         year = 2012,
        month = dec,
       volume = {427},
       number = {3},
        pages = {1891-1902},
          doi = {10.1111/j.1365-2966.2012.21972.x},
archivePrefix = {arXiv},
       eprint = {1207.6487},
 primaryClass = {astro-ph.CO},
       adsurl = {https://ui.adsabs.harvard.edu/abs/2012MNRAS.427.1891A}
}

@ARTICLE{2012ApJ...761...14H,
       author = {{Ho}, Shirley and {Cuesta}, Antonio and {Seo}, Hee-Jong and {de Putter}, Roland and {Ross}, Ashley J. and {White}, Martin and {Padmanabhan}, Nikhil and {Saito}, Shun and {Schlegel}, David J. and {Schlafly}, Eddie and others},
        title = "{Clustering of Sloan Digital Sky Survey III Photometric Luminous Galaxies: The Measurement, Systematics, and Cosmological Implications}",
      journal = {ApJ},
         year = 2012,
        month = dec,
       volume = {761},
       number = {1},
          eid = {14},
        pages = {14},
          doi = {10.1088/0004-637X/761/1/14},
archivePrefix = {arXiv},
       eprint = {1201.2137},
 primaryClass = {astro-ph.CO},
       adsurl = {https://ui.adsabs.harvard.edu/abs/2012ApJ...761...14H}
}

@ARTICLE{2012ApJ...761..152T,
       author = {{Takahashi}, Ryuichi and {Sato}, Masanori and {Nishimichi}, Takahiro and {Taruya}, Atsushi and {Oguri}, Masamune},
        title = "{Revising the Halofit Model for the Nonlinear Matter Power Spectrum}",
      journal = {ApJ},
         year = 2012,
        month = dec,
       volume = {761},
       number = {2},
          eid = {152},
        pages = {152},
          doi = {10.1088/0004-637X/761/2/152},
archivePrefix = {arXiv},
       eprint = {1208.2701},
 primaryClass = {astro-ph.CO},
       adsurl = {https://ui.adsabs.harvard.edu/abs/2012ApJ...761..152T}
}

@ARTICLE{2012PhRvD..86d3011D,
       author = {{Del Pozzo}, Walter},
        title = "{Inference of cosmological parameters from gravitational waves: Applications to second generation interferometers}",
      journal = {PhRvD},
         year = 2012,
        month = aug,
       volume = {86},
       number = {4},
          eid = {043011},
        pages = {043011},
          doi = {10.1103/PhysRevD.86.043011},
archivePrefix = {arXiv},
       eprint = {1108.1317},
 primaryClass = {astro-ph.CO},
       adsurl = {https://ui.adsabs.harvard.edu/abs/2012PhRvD..86d3011D}
}

@ARTICLE{2012PhRvD..86b3502T,
       author = {{Taylor}, Stephen R. and {Gair}, Jonathan R.},
        title = "{Cosmology with the lights off: Standard sirens in the Einstein Telescope era}",
      journal = {PhRvD},
         year = 2012,
        month = jul,
       volume = {86},
       number = {2},
          eid = {023502},
        pages = {023502},
          doi = {10.1103/PhysRevD.86.023502},
archivePrefix = {arXiv},
       eprint = {1204.6739},
 primaryClass = {astro-ph.CO},
       adsurl = {https://ui.adsabs.harvard.edu/abs/2012PhRvD..86b3502T}
}

@ARTICLE{2012PhRvD..85b3535T,
       author = {{Taylor}, Stephen R. and {Gair}, Jonathan R. and {Mandel}, Ilya},
        title = "{Cosmology using advanced gravitational-wave detectors alone}",
      journal = {PhRvD},
         year = 2012,
        month = jan,
       volume = {85},
       number = {2},
          eid = {023535},
        pages = {023535},
          doi = {10.1103/PhysRevD.85.023535},
archivePrefix = {arXiv},
       eprint = {1108.5161},
 primaryClass = {gr-qc},
       adsurl = {https://ui.adsabs.harvard.edu/abs/2012PhRvD..85b3535T}
}

@ARTICLE{2011arXiv1111.4246H,
       author = {{Hoffman}, Matthew D. and {Gelman}, Andrew},
        title = "{The No-U-Turn Sampler: Adaptively Setting Path Lengths in Hamiltonian Monte Carlo}",
      journal = {arXiv},
         year = 2011,
        month = nov,
          eid = {arXiv:1111.4246},
        pages = {arXiv:1111.4246},
          doi = {10.48550/arXiv.1111.4246},
archivePrefix = {arXiv},
       eprint = {1111.4246},
 primaryClass = {stat.CO},
       adsurl = {https://ui.adsabs.harvard.edu/abs/2011arXiv1111.4246H}
}

@ARTICLE{2011PhRvD..84f3505B,
       author = {{Bonvin}, Camille and {Durrer}, Ruth},
        title = "{What galaxy surveys really measure}",
      journal = {PhRvD},
         year = 2011,
        month = sep,
       volume = {84},
       number = {6},
          eid = {063505},
        pages = {063505},
          doi = {10.1103/PhysRevD.84.063505},
archivePrefix = {arXiv},
       eprint = {1105.5280},
 primaryClass = {astro-ph.CO},
       adsurl = {https://ui.adsabs.harvard.edu/abs/2011PhRvD..84f3505B}
}

@ARTICLE{2011MNRAS.415.3649V,
       author = {{van Daalen}, Marcel P. and {Schaye}, Joop and {Booth}, C.~M. and {Dalla Vecchia}, Claudio},
        title = "{The effects of galaxy formation on the matter power spectrum: a challenge for precision cosmology}",
      journal = {MNRAS},
         year = 2011,
        month = aug,
       volume = {415},
       number = {4},
        pages = {3649-3665},
          doi = {10.1111/j.1365-2966.2011.18981.x},
archivePrefix = {arXiv},
       eprint = {1104.1174},
 primaryClass = {astro-ph.CO},
       adsurl = {https://ui.adsabs.harvard.edu/abs/2011MNRAS.415.3649V}
}

@ARTICLE{2011JCAP...07..034B,
       author = {{Blas}, Diego and {Lesgourgues}, Julien and {Tram}, Thomas},
        title = "{The Cosmic Linear Anisotropy Solving System (CLASS). Part II: Approximation schemes}",
      journal = {JCAP},
         year = 2011,
        month = jul,
       volume = {2011},
       number = {7},
          eid = {034},
        pages = {034},
          doi = {10.1088/1475-7516/2011/07/034},
archivePrefix = {arXiv},
       eprint = {1104.2933},
 primaryClass = {astro-ph.CO},
       adsurl = {https://ui.adsabs.harvard.edu/abs/2011JCAP...07..034B}
}

@INCOLLECTION{2011hmcm.book..113N,
       author = {{Neal}, Radford},
        title = "{MCMC Using Hamiltonian Dynamics}",
    booktitle = {Handbook of Markov Chain Monte Carlo},
         year = 2011,
        pages = {113-162},
          doi = {10.1201/b10905},
       adsurl = {https://ui.adsabs.harvard.edu/abs/2011hmcm.book..113N}
}

@ARTICLE{2011arXiv1104.2932L,
       author = {{Lesgourgues}, Julien},
        title = "{The Cosmic Linear Anisotropy Solving System (CLASS) I: Overview}",
      journal = {arXiv},
         year = 2011,
        month = apr,
          eid = {arXiv:1104.2932},
        pages = {arXiv:1104.2932},
          doi = {10.48550/arXiv.1104.2932},
archivePrefix = {arXiv},
       eprint = {1104.2932},
 primaryClass = {astro-ph.IM},
       adsurl = {https://ui.adsabs.harvard.edu/abs/2011arXiv1104.2932L}
}

@ARTICLE{2010MNRAS.406...60J,
       author = {{Jasche}, Jens and {Kitaura}, Francisco S. and {Wandelt}, Benjamin D. and {En{\ss}lin}, Torsten A.},
        title = "{Bayesian power-spectrum inference for large-scale structure data}",
      journal = {MNRAS},
         year = 2010,
        month = jul,
       volume = {406},
       number = {1},
        pages = {60-85},
          doi = {10.1111/j.1365-2966.2010.16610.x},
archivePrefix = {arXiv},
       eprint = {0911.2493},
 primaryClass = {astro-ph.CO},
       adsurl = {https://ui.adsabs.harvard.edu/abs/2010MNRAS.406...60J}
}

@ARTICLE{2009arXiv0912.0201L,
       author = {{LSST Science Collaboration} and {Abell}, Paul A. and {Allison}, Julius and {Anderson}, Scott F. and {Andrew}, John R. and {Angel}, J. Roger P. and {Armus}, Lee and {Arnett}, David and {Asztalos}, S.~J. and {Axelrod}, Tim S. and others},
        title = "{LSST Science Book, Version 2.0}",
      journal = {arXiv},
         year = 2009,
        month = dec,
          eid = {arXiv:0912.0201},
        pages = {arXiv:0912.0201},
          doi = {10.48550/arXiv.0912.0201},
archivePrefix = {arXiv},
       eprint = {0912.0201},
 primaryClass = {astro-ph.IM},
       adsurl = {https://ui.adsabs.harvard.edu/abs/2009arXiv0912.0201L}
}

@ARTICLE{2008PhRvD..78l3506L,
       author = {{LoVerde}, Marilena and {Afshordi}, Niayesh},
        title = "{Extended Limber approximation}",
      journal = {PhRvD},
         year = 2008,
        month = dec,
       volume = {78},
       number = {12},
          eid = {123506},
        pages = {123506},
          doi = {10.1103/PhysRevD.78.123506},
archivePrefix = {arXiv},
       eprint = {0809.5112},
 primaryClass = {astro-ph},
       adsurl = {https://ui.adsabs.harvard.edu/abs/2008PhRvD..78l3506L}
}

@ARTICLE{2008PhRvD..77d2001V,
       author = {{Vallisneri}, Michele},
        title = "{Use and abuse of the Fisher information matrix in the assessment of gravitational-wave parameter-estimation prospects}",
      journal = {PhRvD},
         year = 2008,
        month = feb,
       volume = {77},
       number = {4},
          eid = {042001},
        pages = {042001},
          doi = {10.1103/PhysRevD.77.042001},
archivePrefix = {arXiv},
       eprint = {gr-qc/0703086},
 primaryClass = {gr-qc},
       adsurl = {https://ui.adsabs.harvard.edu/abs/2008PhRvD..77d2001V}
}

@ARTICLE{2007MNRAS.381.1347C,
       author = {{Cabr{\'e}}, Anna and {Fosalba}, Pablo and {Gazta{\~n}aga}, Enrique and {Manera}, Marc},
        title = "{Error analysis in cross-correlation of sky maps: application to the Integrated Sachs-Wolfe detection}",
      journal = {MNRAS},
         year = 2007,
        month = nov,
       volume = {381},
       number = {4},
        pages = {1347-1368},
          doi = {10.1111/j.1365-2966.2007.12280.x},
archivePrefix = {arXiv},
       eprint = {astro-ph/0701393},
 primaryClass = {astro-ph},
       adsurl = {https://ui.adsabs.harvard.edu/abs/2007MNRAS.381.1347C}
}

@ARTICLE{2007A&A...473..711S,
       author = {{Simon}, P.},
        title = "{How accurate is Limber's equation?}",
      journal = {A\&A},
         year = 2007,
        month = oct,
       volume = {473},
       number = {3},
        pages = {711-714},
          doi = {10.1051/0004-6361:20066352},
archivePrefix = {arXiv},
       eprint = {astro-ph/0609165},
 primaryClass = {astro-ph},
       adsurl = {https://ui.adsabs.harvard.edu/abs/2007A&A...473..711S}
}

@ARTICLE{2007MNRAS.378..852P,
       author = {{Padmanabhan}, Nikhil and {Schlegel}, David J. and {Seljak}, Uro{\v{s}} and {Makarov}, Alexey and {Bahcall}, Neta A. and {Blanton}, Michael R. and {Brinkmann}, Jonathan and {Eisenstein}, Daniel J. and {Finkbeiner}, Douglas P. and {Gunn}, James E. and others},
        title = "{The clustering of luminous red galaxies in the Sloan Digital Sky Survey imaging data}",
      journal = {MNRAS},
         year = 2007,
        month = jul,
       volume = {378},
       number = {3},
        pages = {852-872},
          doi = {10.1111/j.1365-2966.2007.11593.x},
archivePrefix = {arXiv},
       eprint = {astro-ph/0605302},
 primaryClass = {astro-ph},
       adsurl = {https://ui.adsabs.harvard.edu/abs/2007MNRAS.378..852P}
}

@ARTICLE{2006PhRvD..74f3006D,
       author = {{Dalal}, Neal and {Holz}, Daniel E. and {Hughes}, Scott A. and {Jain}, Bhuvnesh},
        title = "{Short GRB and binary black hole standard sirens as a probe of dark energy}",
      journal = {PhRvD},
         year = 2006,
        month = sep,
       volume = {74},
       number = {6},
          eid = {063006},
        pages = {063006},
          doi = {10.1103/PhysRevD.74.063006},
archivePrefix = {arXiv},
       eprint = {astro-ph/0601275},
 primaryClass = {astro-ph},
       adsurl = {https://ui.adsabs.harvard.edu/abs/2006PhRvD..74f3006D}
}

@ARTICLE{2005ApJ...629...15H,
       author = {{Holz}, Daniel E. and {Hughes}, Scott A.},
        title = "{Using Gravitational-Wave Standard Sirens}",
      journal = {ApJ},
         year = 2005,
        month = aug,
       volume = {629},
       number = {1},
        pages = {15-22},
          doi = {10.1086/431341},
archivePrefix = {arXiv},
       eprint = {astro-ph/0504616},
 primaryClass = {astro-ph},
       adsurl = {https://ui.adsabs.harvard.edu/abs/2005ApJ...629...15H}
}

@ARTICLE{2003MNRAS.341.1311S,
       author = {{Smith}, R.~E. and {Peacock}, J.~A. and {Jenkins}, A. and {White}, S.~D.~M. and {Frenk}, C.~S. and {Pearce}, F.~R. and {Thomas}, P.~A. and {Efstathiou}, G. and {Couchman}, H.~M.~P.},
        title = "{Stable clustering, the halo model and non-linear cosmological power spectra}",
      journal = {MNRAS},
         year = 2003,
        month = jun,
       volume = {341},
       number = {4},
        pages = {1311-1332},
          doi = {10.1046/j.1365-8711.2003.06503.x},
archivePrefix = {arXiv},
       eprint = {astro-ph/0207664},
 primaryClass = {astro-ph},
       adsurl = {https://ui.adsabs.harvard.edu/abs/2003MNRAS.341.1311S}
}

@ARTICLE{2002PhR...372....1C,
       author = {{Cooray}, Asantha and {Sheth}, Ravi},
        title = "{Halo models of large scale structure}",
      journal = {PhR},
         year = 2002,
        month = dec,
       volume = {372},
       number = {1},
        pages = {1-129},
          doi = {10.1016/S0370-1573(02)00276-4},
archivePrefix = {arXiv},
       eprint = {astro-ph/0206508},
 primaryClass = {astro-ph},
       adsurl = {https://ui.adsabs.harvard.edu/abs/2002PhR...372....1C}
}

@ARTICLE{2002ApJ...572..140D,
       author = {{Dodelson}, Scott and {Narayanan}, Vijay K. and {Tegmark}, Max and {Scranton}, Ryan and {Budav{\'a}ri}, Tamas and {Connolly}, Andrew and {Csabai}, Istvan and {Eisenstein}, Daniel and {Frieman}, Joshua A. and {Gunn}, James E. and others},
        title = "{The Three-dimensional Power Spectrum from Angular Clustering of Galaxies in Early Sloan Digital Sky Survey Data}",
      journal = {ApJ},
         year = 2002,
        month = jun,
       volume = {572},
       number = {1},
        pages = {140-156},
          doi = {10.1086/340225},
archivePrefix = {arXiv},
       eprint = {astro-ph/0107421},
 primaryClass = {astro-ph},
       adsurl = {https://ui.adsabs.harvard.edu/abs/2002ApJ...572..140D}
}

@ARTICLE{2002ApJ...567....2H,
       author = {{Hivon}, Eric and {G{\'o}rski}, Krzysztof M. and {Netterfield}, C. Barth and {Crill}, Brendan P. and {Prunet}, Simon and {Hansen}, Frode},
        title = "{MASTER of the Cosmic Microwave Background Anisotropy Power Spectrum: A Fast Method for Statistical Analysis of Large and Complex Cosmic Microwave Background Data Sets}",
      journal = {ApJ},
         year = 2002,
        month = mar,
       volume = {567},
       number = {1},
        pages = {2-17},
          doi = {10.1086/338126},
archivePrefix = {arXiv},
       eprint = {astro-ph/0105302},
 primaryClass = {astro-ph},
       adsurl = {https://ui.adsabs.harvard.edu/abs/2002ApJ...567....2H}
}

@ARTICLE{2001ApJ...560..566K,
       author = {{Kochanek}, C.~S. and {Pahre}, M.~A. and {Falco}, E.~E. and {Huchra}, J.~P. and {Mader}, J. and {Jarrett}, T.~H. and {Chester}, T. and {Cutri}, R. and {Schneider}, S.~E.},
        title = "{The K-Band Galaxy Luminosity Function}",
      journal = {ApJ},
         year = 2001,
        month = oct,
       volume = {560},
       number = {2},
        pages = {566-579},
          doi = {10.1086/322488},
archivePrefix = {arXiv},
       eprint = {astro-ph/0011456},
 primaryClass = {astro-ph},
       adsurl = {https://ui.adsabs.harvard.edu/abs/2001ApJ...560..566K}
}

@ARTICLE{1998ApJ...498L...1S,
       author = {{Szalay}, Alexander S. and {Matsubara}, Takahiko and {Landy}, Stephen D.},
        title = "{Redshift-Space Distortions of the Correlation Function in Wide-Angle Galaxy Surveys}",
      journal = {ApJL},
         year = 1998,
        month = may,
       volume = {498},
       number = {1},
        pages = {L1-L4},
          doi = {10.1086/311293},
archivePrefix = {arXiv},
       eprint = {astro-ph/9712007},
 primaryClass = {astro-ph},
       adsurl = {https://ui.adsabs.harvard.edu/abs/1998ApJ...498L...1S}
}

@ARTICLE{1998PhRvD..57.2117B,
       author = {{Bond}, J.~R. and {Jaffe}, A.~H. and {Knox}, L.},
        title = "{Estimating the power spectrum of the cosmic microwave background}",
      journal = {PhRvD},
         year = 1998,
        month = feb,
       volume = {57},
       number = {4},
        pages = {2117-2137},
          doi = {10.1103/PhysRevD.57.2117},
archivePrefix = {arXiv},
       eprint = {astro-ph/9708203},
 primaryClass = {astro-ph},
       adsurl = {https://ui.adsabs.harvard.edu/abs/1998PhRvD..57.2117B}
}

@ARTICLE{1997ApJ...480...22T,
       author = {{Tegmark}, Max and {Taylor}, Andy N. and {Heavens}, Alan F.},
        title = "{Karhunen-Lo{\`e}ve Eigenvalue Problems in Cosmology: How Should We Tackle Large Data Sets?}",
      journal = {ApJ},
         year = 1997,
        month = may,
       volume = {480},
       number = {1},
        pages = {22-35},
          doi = {10.1086/303939},
archivePrefix = {arXiv},
       eprint = {astro-ph/9603021},
 primaryClass = {astro-ph},
       adsurl = {https://ui.adsabs.harvard.edu/abs/1997ApJ...480...22T}
}

@ARTICLE{1997PhRvD..55.5895T,
       author = {{Tegmark}, Max},
        title = "{How to measure CMB power spectra without losing information}",
      journal = {PhRvD},
         year = 1997,
        month = may,
       volume = {55},
       number = {10},
        pages = {5895-5907},
          doi = {10.1103/PhysRevD.55.5895},
archivePrefix = {arXiv},
       eprint = {astro-ph/9611174},
 primaryClass = {astro-ph},
       adsurl = {https://ui.adsabs.harvard.edu/abs/1997PhRvD..55.5895T}
}

@ARTICLE{1995MNRAS.275..483H,
       author = {{Heavens}, A.~F. and {Taylor}, A.~N.},
        title = "{A spherical harmonic analysis of redshift space}",
      journal = {MNRAS},
         year = 1995,
        month = jul,
       volume = {275},
       number = {2},
        pages = {483-497},
          doi = {10.1093/mnras/275.2.483},
archivePrefix = {arXiv},
       eprint = {astro-ph/9409027},
 primaryClass = {astro-ph},
       adsurl = {https://ui.adsabs.harvard.edu/abs/1995MNRAS.275..483H}
}

@ARTICLE{1994MNRAS.266..219F,
       author = {{Fisher}, Karl B. and {Scharf}, Caleb A. and {Lahav}, Ofer},
        title = "{A spherical harmonic approach to redshift distortion and a measurement of Omega(0) from the 1.2-Jy IRAS Redshift Survey}",
      journal = {MNRAS},
         year = 1994,
        month = jan,
       volume = {266},
        pages = {219},
          doi = {10.1093/mnras/266.1.219},
archivePrefix = {arXiv},
       eprint = {astro-ph/9309027},
 primaryClass = {astro-ph},
       adsurl = {https://ui.adsabs.harvard.edu/abs/1994MNRAS.266..219F}
}

@INPROCEEDINGS{1993AAS...183.6705F,
       author = {{Finn}, L.~S. and {Chernoff}, D.~F.},
        title = "{Gravitational Radiation, Inspiraling Binaries, and Cosmology}",
    booktitle = {American Astronomical Society Meeting Abstracts},
         year = 1993,
       series = {American Astronomical Society Meeting Abstracts},
       volume = {183},
        month = dec,
          eid = {67.05},
        pages = {67.05},
       adsurl = {https://ui.adsabs.harvard.edu/abs/1993AAS...183.6705F}
}

@ARTICLE{1992PhRvD..46.4198W,
       author = {{White}, Martin},
        title = "{Contribution of long-wavelength gravitational waves to the cosmic microwave background anisotropy}",
      journal = {PhRvD},
         year = 1992,
        month = nov,
       volume = {46},
       number = {10},
        pages = {4198-4205},
          doi = {10.1103/PhysRevD.46.4198},
archivePrefix = {arXiv},
       eprint = {hep-ph/9207239},
 primaryClass = {hep-ph},
       adsurl = {https://ui.adsabs.harvard.edu/abs/1992PhRvD..46.4198W}
}

@ARTICLE{1991MNRAS.248....1C,
       author = {{Coles}, Peter and {Jones}, Bernard},
        title = "{A lognormal model for the cosmological mass distribution.}",
      journal = {MNRAS},
         year = 1991,
        month = jan,
       volume = {248},
        pages = {1-13},
          doi = {10.1093/mnras/248.1.1},
       adsurl = {https://ui.adsabs.harvard.edu/abs/1991MNRAS.248....1C}
}

@ARTICLE{1987PhLB..195..216D,
       author = {{Duane}, Simon and {Kennedy}, A.~D. and {Pendleton}, Brian J. and {Roweth}, Duncan},
        title = "{Hybrid Monte Carlo}",
      journal = {PhLB},
         year = 1987,
        month = sep,
       volume = {195},
       number = {2},
        pages = {216-222},
          doi = {10.1016/0370-2693(87)91197-X},
       adsurl = {https://ui.adsabs.harvard.edu/abs/1987PhLB..195..216D}
}

@ARTICLE{1986Natur.323..310S,
       author = {{Schutz}, B.~F.},
        title = "{Determining the Hubble constant from gravitational wave observations}",
      journal = {Natur},
         year = 1986,
        month = sep,
       volume = {323},
       number = {6086},
        pages = {310-311},
          doi = {10.1038/323310a0},
       adsurl = {https://ui.adsabs.harvard.edu/abs/1986Natur.323..310S}
}

@BOOK{1981lssu.book.....P,
       author = {{Peebles}, P.~J.~E.},
        title = "{The Large-Scale Structure of the Universe}",
         year = 1981,
       adsurl = {https://ui.adsabs.harvard.edu/abs/1981lssu.book.....P}
}

@ARTICLE{1976ApJ...203..297S,
       author = {{Schechter}, P.},
        title = "{An analytic expression for the luminosity function for galaxies.}",
      journal = {ApJ},
         year = 1976,
        month = jan,
       volume = {203},
        pages = {297-306},
          doi = {10.1086/154079},
       adsurl = {https://ui.adsabs.harvard.edu/abs/1976ApJ...203..297S}
}

@ARTICLE{1973ApJ...185..413P,
       author = {{Peebles}, P.~J.~E.},
        title = "{Statistical Analysis of Catalogs of Extragalactic Objects. I. Theory}",
      journal = {ApJ},
         year = 1973,
        month = oct,
       volume = {185},
        pages = {413-440},
          doi = {10.1086/152431},
       adsurl = {https://ui.adsabs.harvard.edu/abs/1973ApJ...185..413P}
}

@ARTICLE{1953ApJ...117..134L,
       author = {{Limber}, D. Nelson},
        title = "{The Analysis of Counts of the Extragalactic Nebulae in Terms of a Fluctuating Density Field.}",
      journal = {ApJ},
         year = 1953,
        month = jan,
       volume = {117},
        pages = {134},
          doi = {10.1086/145672},
       adsurl = {https://ui.adsabs.harvard.edu/abs/1953ApJ...117..134L}
}


%% file: bib_supplement.bib
@misc{CE_PSD,
       author = {Kuns, Kevin and Fulda, Paul and Barsotti, Lisa and Evans, Matthew},
        title = "{Cosmic Explorer Strain and Displacement Sensitivity}",
         year = 2024,
       howpublished = "\url{https://dcc.cosmicexplorer.org/CE-T2000017/public}",
}

@misc{gwx,
  author = {{Cheng}, A.~Q.},
  howpublished = "\url{https://github.com/aqcheng/gw_xcosmo}",
  year = 2026
}

@misc{my_jax_cosmo,
  author = {{Cheng}, A.~Q. and {Differentiable Universe Initiative}},
  howpublished = "\url{https://github.com/aqcheng/jax_cosmo}",
  year = 2026
}

@dataset{data_release,
  author    = {{Cheng}, A.~Q.},
  title     = {A unified harmonic framework for dark siren cosmology - simulation data},
  month     = Apr,
  year      = 2026,
  publisher = {Zenodo},
  version   = {v1.0},
  doi       = {10.5281/zenodo.19673580.},
  url       = {https://zenodo.org/records/19673581}
}
